\documentclass[11pt]{amsart}
\usepackage{epsfig}
\usepackage{graphicx,color}
\usepackage{amsmath, amssymb,mathabx,mathrsfs}
\usepackage[toc,page]{appendix}
\usepackage{comment}
\usepackage{hyperref}
\usepackage{float}
\usepackage{subcaption}

\theoremstyle{definition}

\theoremstyle{remark}

\numberwithin{equation}{section}
\newcommand{\eps}{\varepsilon}
\def\N{\mathbb{N}}
\def\R{\mathbb{R}}

\begin{document}
\title[Why TDA Detects Financial Bubbles?]{Why Topological Data Analysis Detects Financial Bubbles?}

\author[S. W. Akingbade]{Samuel W. Akingbade}
\address{Department of Mathematical Sciences, Yeshiva University, USA}
\email{sakingba@mail.yu.edu}

\author[M. Gidea]{Marian Gidea}
\address{Department of Mathematical Sciences, Yeshiva University, USA}
\email{marian.gidea@yu.edu}

\author[M. Manzi]{Matteo Manzi}
\address{CrunchDAO, France}
\email{matteo.manzi@crunchdao.com}

\author[V. Nateghi]{Vahid Nateghi}
\address{Aerospace Sciences and Technology Department, Politecnico di Milano, Italy}
\email{vahid.nateghi@mail.polimi.it}

\begin{abstract}
We present a heuristic argument for the propensity of  Topological Data Analysis (TDA) to detect early warning signals of critical transitions in financial time series.
Our argument is based on the Log-Periodic Power Law Singularity  (LPPLS) model,  which characterizes financial bubbles as super-exponential growth (or decay) of an asset price superimposed with oscillations increasing in frequency and decreasing in amplitude when approaching a critical transition (tipping point).  We show that whenever the LPPLS model is fitting with the data, TDA  generates early warning signals. As an application, we illustrate this approach on a sample of positive and negative bubbles in the Bitcoin historical price.
\end{abstract}

\maketitle
\section{Introduction}

Topological Data Analysis (TDA) has emerged in recent years as a powerful methodology in time-series analysis and signal
processing. A prevalent approach relies on time-delay coordinate embedding, which is used to construct, from the time-series, a point-cloud in some high-dimensional Euclidean space \cite{takens1981detecting}.
The dynamics on the point-cloud allows one to recover the evolution laws driving the time-series.
The shape of the point-cloud can be characterized via  Persistent Homology -- a fundamental tool in TDA -- which measures the number of holes of a geometric  approximation of  the point-cloud, as well as their `visibility' at any given `resolution' level.
Applying a sliding window to a time series allows one to measure how the shape of the point-cloud associated to each window changes over time.
The robustness of persistent homology of the point-cloud under small perturbations makes this approach particularly suitable for analyzing noisy time series.
The method can be used, for example, for  detection of periodic behavior,  attractors, and bifurcations. More broadly, it can be used for detection of critical transitions (tipping points). These represent abrupt shifts of a system
from one steady state to another, due to  small changes in external
conditions.
Critical transitions are relevant to many disciplines, including
finance,  climate, ecology, and biomedicine; see, e.g., \cite{scheffer2009early,lenton2011early,thompson2011climate,dakos2012methods,bury2021deep}.
Some applications of TDA  to detect critical transitions have been developed in \cite{berwald2014automatic,berwald2014critical,saggar2018towards,islambekov2020harnessing,dindin2020topological}.

An important class of applications of TDA concerns critical transitions in financial time series, particularly market crashes and financial bubbles.
A  positive bubble represents a rapid increase of the price of an  asset  -- which does not reflect the asset’s intrinsic value -- followed by a crash.
A negative bubble  represents a rapid decay followed by a quick rebound. In this context, the asset price peak in a positive bubble prior to a crash, and the
trough in a negative bubble prior to a rebound, represent tipping points of the system.
There is a practical interest in the detection of early warning signals of critical transitions, which would allow  minimizing  the losses from a bubble crash, or maximize the profit from a rebound.
Some related references include:  \cite{phillips2011dating,yan2012diagnosis,fry2016negative,gidea2017topological,gidea2018topological,gidea2020topological,katz2021time,
rudkin2021uncertainty,dlotko2021financial}.

It is important to distinguish between crashes of an endogenous origin, essentially caused by speculative, unsustainable, accelerating bubbles, and crashes of an exogenous  origin, caused by external factors (e.g., natural cataclysms, pandemics, fraud,
political changes).  Exogenous crashes are less likely to yield early warning signals in comparison to the endogenous ones \cite{song20222020,shu2021covid}.


Most of the evidence, so far, on the ability of TDA to detect financial bubbles has been empirical. A key question remains: what features of a time series yield to significant changes in the shape of the associated point cloud,  more precisely, in terms of the holes that appear in the point-cloud?
Some previous works attempted to explain this by drawing analogies between a market  undergoing a crash
and   bifurcations in a complex dynamical system, or in terms of changes in drift and volatility of the strong non-stationary process
described by a market near a crash
\cite{gidea2017topological,gidea2018topological,gidea2020topological,aromi2021topological,katz2022topological}

In this paper, we propose a heuristic argument  for why TDA is able to detect financial bubbles.
As we are interested in early warning signals, we focus on the time period of a bubble before the tipping point.
The ansatz is that  the   time series during such a period follows the  Log-Periodic Power Law Singularity  (LPPLS) model, which characterizes positive bubbles by super-exponential growth superimposed with oscillations of increasing frequency and decreasing amplitude when approaching the tipping point. A similar characterization holds for negative bubbles. The LPPLS model is based on the theory of rational expectation, and provides a signature for endogenous bubbles. See, e.g., \cite{blanchard1979speculative,blanchard1982bubbles,feigenbaum1996discrete,johansen1996discrete,
johansen2000crashes,yan2012detection,lin2013diagnostics,seyrich2016micro,sornette2017stock,gerlach2018dissection}.

From a topological view-point,  the time-delay reconstruction of LPPLS oscillatory signals  yields loops (surrounding  holes) in the point-clouds, and changes in the frequency and/or amplitude of the oscillations yield changes in the structure of  holes in the point-cloud, which can be measured via persistent homology.
When approaching the tipping point of a bubble, there is a significant change in the features of the oscillatory signal and consequently in the TDA output.

The upshot of our work is that whenever the LPPLS model for the detection of a positive (negative) bubble
applies to some data set,  the TDA method  on that data set  provides early warning signals of critical transition.
As far as we know, this is the first time when the TDA method is justified in terms of a deterministic model for the expected dynamics of financial bubbles.

As an illustration, we apply both the LPPLS model and the TDA to Bitcoin -- the largest cryptocurrency by market cap. Bitcoin has been experiencing numerous  phases of extreme price growth and  massive crashes, hence it provides an excellent test bed  for detection of financial bubbles.
Some related references include \cite{gerlach2018dissection,akcora2018bitcoin,rivera2019topological,gidea2020topological,akcora2020bitcoinheist,shu2020real,9309855}.

Our experiments show a good agreement between the TDA method and the LPPLS model. Moreover, even when the LPPLS model fits poorly with some of the data,
the TDA method can show relatively strong signal prior to the tipping point. While the TDA method is robust, the LPPLS appears to be very sensitive to noise \cite{bree2013prediction}.

\section{Background}

\subsection{Log-Periodic Power Law Singularity (LPPLS) Model}
Below we summarize the  LPPLS model following \cite{johansen1996discrete,johansen2000crashes,yan2012detection,seyrich2016micro,sornette2017stock,gerlach2018dissection}.
This is a model for both positive
bubbles 
as well as for  negative bubbles. 
In either case the market reaches a point when it  no longer reflects the real underlying value of the asset (or it violates the market
efficiency with respect to its time-scale \cite{sohn2017bubbles}),
which triggers a critical transition.

The LPPLS model can be derived from a network  description of the market, which assumes that the trading of the asset is driven by two  types of agents:
a group consisting of traders with rational expectations, and another group formed
by noise traders. Noise traders are likely to deviate from fundamental valuation in an accelerating mode, due to positive feedback and herding behaviour as a
group.  The  collective behavior of the agents eventually leads to market instability by pushing the market away from an equilibrium between supply and demand.   See \cite{johansen2000crashes,suchecki2005log,seyrich2016micro,sornette2017stock}.

It is important to keep in mind that the LPPLS model is concerned only with bubble generated endogenously by the dynamics of the market,
and does not apply to bubbles of exogenous origins produced by external shocks.

According to the model, the asset price $p(t)$ during a  (positive or negative) bubble evolves according to the  stochastic differential equation
\begin{equation}
\begin{split}
 \frac{\textrm{d}p}{p}=&B'(t_c-t)^{m-1}+C'(t_c-t)^{m-1}\cos(\omega\ln (t_c-t)-\phi')+\sigma(t) \textrm{d}W(t)\\
\end{split}\label{eqn:LPPLS_SDE}
\end{equation}
where $t_c$ represents the critical time, $B',C',\phi'\in\R$, $0<m<1$ are parameters, $\sigma(t)$ is the volatility, and $dW(t)$	is the increment of a standard
Wiener process with zero mean and unit variance.
Hence the expectation of the logarithm of the asset price satisfies
\begin{equation}
\begin{split}
E[\ln p(t)]=&A+B(t_c-t)^m+C(t_c-t)^m\cos(\omega\ln (t_c-t)-\phi)\\
=&A+B(t_c-t)^m+C_1(t_c-t)^m\cos(\omega\ln (t_c-t))\\&\qquad \qquad\qquad\,\,\,+C_2(t_c-t)^m\sin(\omega\ln (t_c-t)).
\end{split}
\label{eqn:LPPLS_expected}
\end{equation}

Here $A>0$, $0<m<1$, $B,C$ and $\phi$ are parameters, whose role is explained below.
Also, $C_1=C\cos(\phi)$, $C_2=-C\sin(\phi)$, and $C=\sqrt{C_1^2+C_2^2}$.

The function in \eqref{eqn:LPPLS_expected} has a singularity at  $t=t_c$. Since
\[\lim_{t\to t_c} (t_c-t)^m \cos(\omega\ln(t_c-t))=0,\] the function can be extended by continuity at $t=t_c$ by $E[\ln p(t_c)]=A$.
The critical time $t_c$ marks the termination of the
bubble  regime and the transition time to a different regime, which could be a
be a major crash or rebound.
The parameter $A$ represents the expected value of the log price $\ln p(t)$ at the critical time $t_c$.

The condition $m < 1$ ensures that there is a singularity at the critical time $t_c$, while $m > 0$
ensures that the expected price remains finite  at $t_c$.  The parameter $m$ is associated to the non-linearity of the trend,
that is,  $m\approx 0$ corresponds to a  strong  super-linear trend of $\ln p(t)$, while $m\approx 1$ corresponds to a  nearly linear trend.
The parameter $m$ is also affecting, jointly with the parameters $C_1$ and $C_2$, the changing amplitude of the oscillations.

The case when $B < 0$ corresponds to a super-exponential growth of  the price $p(t)$ as the time $t$ approaches $t_c$, while
$B>0$ corresponds to a super-exponential decay. 

Thus, the main features of the model \eqref{eqn:LPPLS_SDE} are the following
\begin{itemize}
\item the price of the asset oscillates about  a trend-line that is characterized by super-exponential growth or decay;
\item the amplitude of the oscillations decreases when approaching the critical time $t_c$;
\item the frequency of the oscillations increases when approaching the critical time $t_c$;
\item the oscillations are superimposed with Gaussian noise.
\end{itemize}

\begin{figure}
\centering
$\begin{array}{cc}
\includegraphics[width=0.32\textwidth]{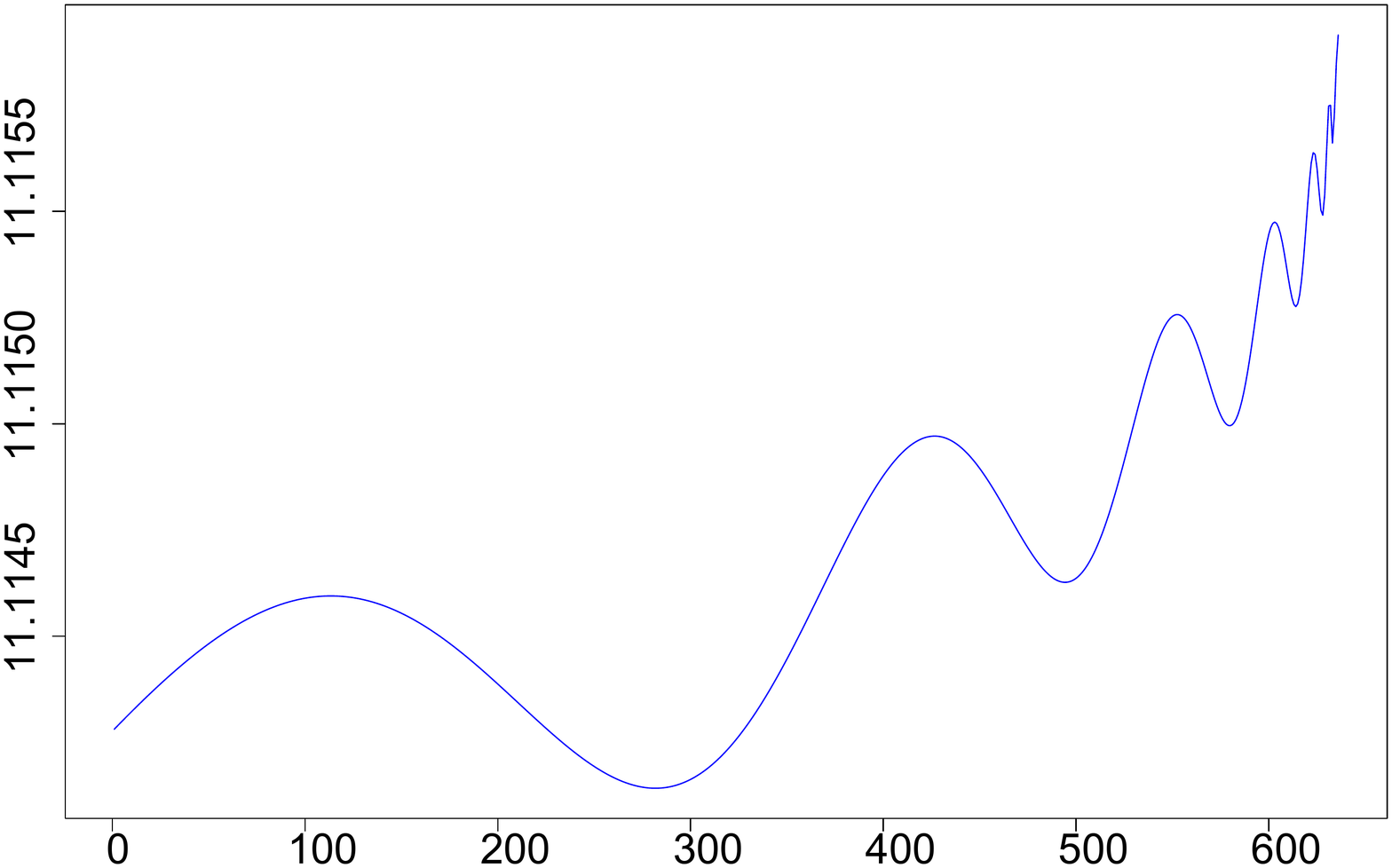}&
\includegraphics[width=0.32\textwidth]{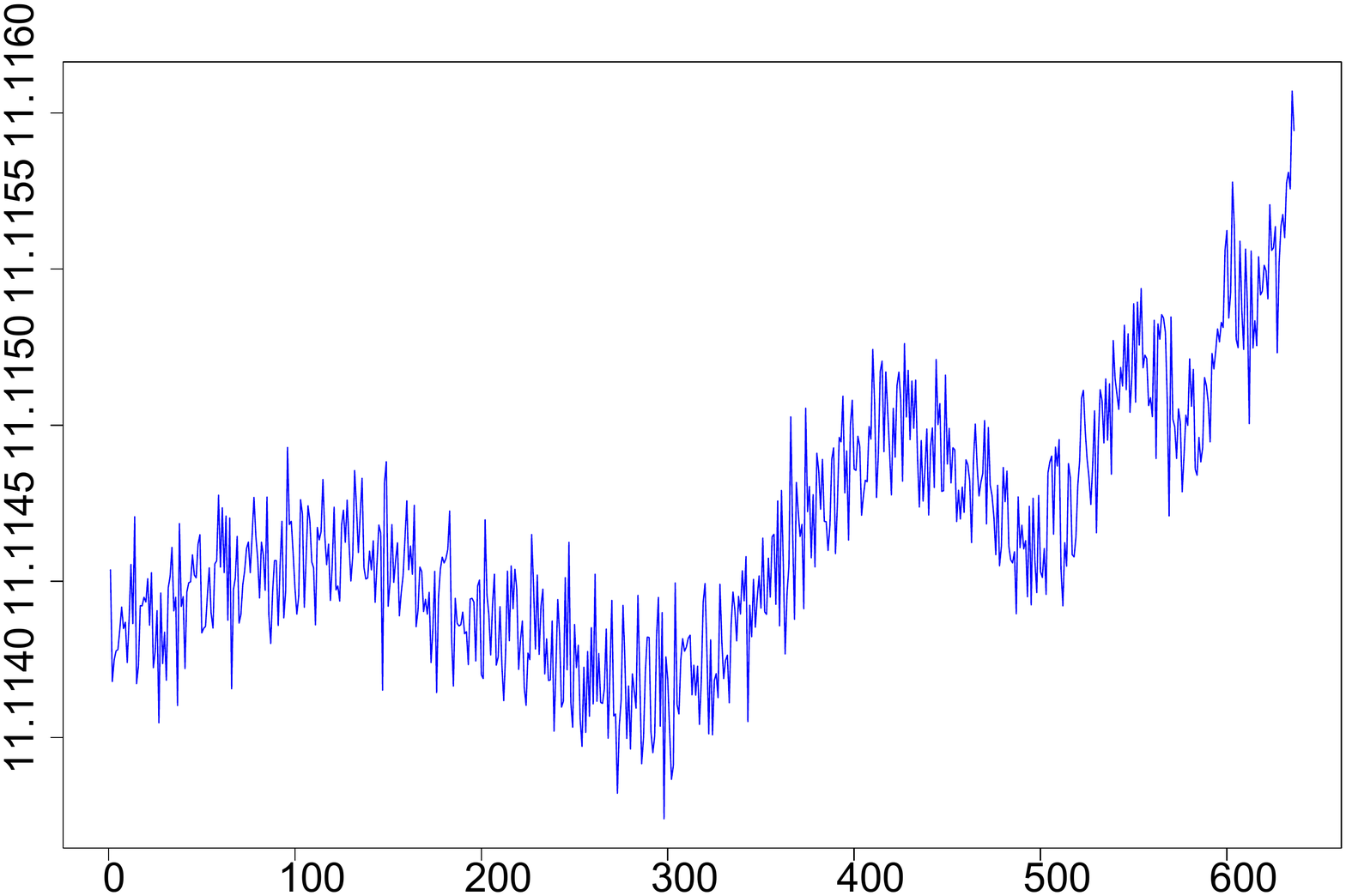}\\
\end{array}$
\caption{Synthetic LPPLS signal: without noise (left),  with noise (right). }
\label{fig:LPPLS_synthetic}
\end{figure}

The LPPLS model can be fitted to data and allows for predictions of the time of the crash \cite{yan2012detection,gerlach2018dissection}.
However, we note that the LPPLS model is fairly sensitive to the system  parameters, to where to start and end the  fit of a bubble, and to the  length of
the data window  (see \cite{bree2013prediction} and the references listed there).


\subsection{Persistent homology}
\label{sec:persistent}
In this section we summarize the main topological tools that we will use: persistent homology and persistence landscapes.
See \cite{edelsbrunner2022computational,bubenik2015statistical,bubenik2017persistence}.
Persistent homology converts a given point cloud into a geometric shape and characterizes it in terms of its homology generators.
Persistence landscapes convert the information derived from persistent homology into elements of a functional Banach space.
Through this process, a point cloud can be characterized by a single numerical value, which is the norm of the corresponding persistence landscape.

\subsubsection{Point-cloud} The input of persistent homology is a  point-cloud   $Z = \{z_0, \ldots ,z_{m-1}\}$ embedded in some Euclidean space $\R^N$.
The computation of persistent homology involves a sequence of steps.

\subsubsection{Vietoris-Rips simplicial complex.} We associate to the point cloud $Z$ a topological space, thought of as a `shape', depending on some parameter $\eps>0$,  where
$\eps$ represents the resolution at which the data is analyzed.
The Vietoris-Rips simplicial complex $R(Z, \eps)$, or, simply Rips complex, is defined  as follows:
\begin{itemize}
\item For each $k= 0,1, \ldots$, a $k$-simplex of vertices $\{z_{i_0},\ldots, z_{i_{k}}\}$ is part of the simplicial complex $R(Z,\eps)$ if and only if the mutual
distance between  any pair of its   vertices is less than~$\eps$, that is
\[d(z_{i_j},z_{i_{j'}})<\eps, \textrm { for all } z_{i_j},z_{i_{i_{j'}}}\in
\{z_{i_0},\ldots, z_{i_{k}}\}. \]
\end{itemize}
In other words, a $k$-simplex is included in $R(Z,\eps)$   whenever the vertices of that simplex are `indistinguishable from one
another' at  a  resolution   level $\eps$.

\subsubsection{Filtration of simplicial complexes.} The Rips simplicial complexes $R(Z, \eps)$ form a filtration with respect to the resolution parameter $\eps$, that is, \[R(Z,\eps)\subseteq
R(Z,\eps')\textrm{ provided } \eps<\eps'.\]

\subsubsection{Homology groups of simplicial complexes.}
At each resolution level $\eps$, the corresponding shape can be characterized in terms of its homology.
\begin{itemize}\item For each complex $R(Z, \eps)$, we   compute its $n$-dimensional homology
$H_n(R(Z,\eps))$ with coefficients in some field, e.g., $\mathbb{Z}_2$.
\end{itemize} The generators of the $0$-dimensional homology module $H_0(R(Z,\eps))$  correspond to the
connected components of $R(Z,\eps)$, the generators of the $1$-dimensional homology module  $H_1(R(Z,\eps))$ correspond to   `holes' in $R(Z,\eps)$,  the
generators of the $2$-dimensional homology module  $H_2(R(Z,\eps))$ correspond to  `voids' in $R(Z,\eps)$, etc.
For the rest of the paper we will use only  $1$-dimensional homology.

\subsubsection{Filtration of homology modules.} The filtration property of the Rips complexes induces a filtration on the corresponding homologies, that is
\[H_n(R(Z,\eps))\subseteq H_n(R(Z,\eps')) \textrm{ provided }\eps<\eps'\, \textrm { for each }n.\] These inclusions determine canonical homomorphisms
\[H_n(R(Z,\eps))\hookrightarrow H_n(R(Z,\eps')), \textrm { for }\eps<\eps'.\]

\subsubsection{Persistent homology.} For each non-zero $n$-dimensional homology generator $\alpha$  there exists a pair of  values $\eps_1<\eps_2$, such that:
\begin{itemize}
\item   $\alpha$ is in $H_n(R(Z,\eps_1))$ but is not in the image of any element in $H_n(R(Z,\eps_1-\delta))$ under the corresponding homomorphism, for
$\delta>0$;
\item the  image of $\alpha$ in $H_n(R(Z,\eps'))$ is non-zero for all $\eps_1< \eps'<\eps_2$, but the  image of $\alpha$ in $H_n(R(Z,\eps_2))$ is zero.
\end{itemize}
In this case, one says that  the generator $\alpha$ is `born' at the parameter value $b_\alpha:=\eps_1$, and `dies' at the parameter value $d_\alpha:=\eps_2$. The
pair $(b_\alpha, d_\alpha)$ represents the `birth' and `death' indices of $\alpha$. The multiplicity
$\mu_\alpha(b_\alpha,d_\alpha)$ of the point $(b_\alpha,d_\alpha)$ equals the number of classes $\alpha$ that are born at ${b_\alpha}$ and die at ${d_\alpha}$.
This multiplicity is finite since the simplicial complex is finite.

\subsubsection{Persistence diagram.} The information on  the $n$-dimensional homology generators at all resolution levels can be encoded in a \emph{persistence diagram}
$P_n$. Such a diagram consists of:
\begin{itemize}
\item  for each $n$-dimensional homology class $\alpha$ one assigns a point $p_\alpha=p_\alpha(b_\alpha,d_\alpha)\in\mathbb{R}^2$ together with its
    multiplicity $\mu_\alpha=\mu_\alpha(b_\alpha,d_\alpha)$;
\item in addition, $P_n$ contains all points on the positive diagonal  of $\mathbb{R}^2$; these points represent  all trivial homology generators that are
    born and instantly die at every level; each point on the diagonal has infinite multiplicity.
\end{itemize}
The axes of a persistence diagram are birth values on the horizontal axis and death values on the vertical axis.

\subsubsection{Persistence landscapes.} The space of persistence diagrams can be embedded into a Banach space.
One such embedding is based on \emph{persistence landscapes}, consisting of sequences of functions in the  Banach space $L^p(\N\times \R)$, with $p\geq 1$.
\begin{itemize}\item For each birth-death point $(b_\alpha,d_\alpha) \in P_n$, we first define a piecewise linear function
\begin{equation}\label{eqn:landscape_1}
f_{(b_\alpha,d_\alpha)}(x)=\left\{
                          \begin{array}{ll}
                            x-b_\alpha, & \hbox{if $x\in\left(b_\alpha,\frac{b_\alpha+d_\alpha}{2}\right]$;} \\
                            -x+d_\alpha, & \hbox{if $x\in\left(\frac{b_\alpha+d_\alpha}{2},d_\alpha\right)$;} \\
                            0, & \hbox{if $x\not\in (b_\alpha, d_\alpha)$.}
                          \end{array}
                        \right.
\end{equation}

\item To a persistence diagram $P_n$ consisting of a finite number of off-diagonal
points, we associate a sequence of functions
$\lambda_n=(\lambda_i)_{i\in\mathbb{N}}$, where
$\lambda_n(i):\mathbb{R}\to[0,+\infty]$ is
given by
\begin{equation}\label{eqn:landscape_2} \lambda_n(i)(x)=i\textrm{-max}\{f_{(b_\alpha,d_\alpha)}(x)\,|\, (b_\alpha,d_\alpha)\in P_n\}
\end{equation}
where  $i\textrm{-max}$ denotes the $i$-th largest value of a  function.
We set $\lambda_n(i)(x) = 0$ if
the $i$-th largest value does not exist. Thus, the persistence landscapes form a subset of the Banach space $L^p(\N\times \R)$.
\end{itemize}

\subsubsection{Norm of a persistence landscape.}
For each persistence landscape we can compute its norm, and so we can compare persistence landscapes by computing the mutual distances, using the metric derived from the norm.
\begin{itemize}
\item The norm of $\lambda_n\in L^p(\N\times \R)$  is given by
\begin{equation}\label{eqn:landscape_3}\|\lambda_n\|_p=\left (\sum_{i=1}^{\infty}\|\lambda_n(i)\|_p^p\right)^{1/p}.\end{equation}
Above, $\|\cdot\|_p$ denotes the $L^p$-norm, $p\geq 1$, i.e., $\|f\|_p=\left ( \int _\mathbb{R} |f|^p\right )^{1/p}$, where the integration is with respect
to  the Lebesgue measure on $\R$.
\end{itemize}

We recall from \cite{bubenik2015statistical,bubenik2017persistence} that the norms $\|\cdot\|_1$ and  $\|\cdot\|_\infty$ can be computed via direct formulas
\begin{equation}
\begin{split}
\|\lambda_n\|_1=&\frac{1}{4} \sum_{\alpha} (d_\alpha-b_\alpha)^2,\\
\|\lambda_n\|_\infty=&\frac{1}{2} \sup_{\alpha}(d_\alpha-b_\alpha).
\end{split}
\end{equation}
Throughout the paper we will only refer to the $L^1$-norms of persistence diagrams.

\subsubsection{Stability}  Persistence diagrams and persistence landscapes  are `stable' under small perturbations of a point-cloud \cite{chazal2014persistence,bubenik2015statistical}.

\subsection{Topological data analysis of time-series}
\label{sec:TDA_time_series}
As input, consider a  time series of  numerical values of length $L$:
\begin{equation}\label{eqn:X}X=\{x_0,x_1,\ldots, x_{L-1} \}.\end{equation}

\subsubsection{Time-series of time-delay coordinate vectors.}

We choose and fix an embedded dimension $N$ and a time-delay $d\geq 1$.
\begin{itemize}
\item We transform the time series $X$ into a sequence of $N$-dimensional time-delay coordinate vectors \begin{equation}\begin{split}\label{eqn:z}
    z_0&=(x_0,x_{d},\ldots,x_{(N-1) d}),\\ z_1&=(x_1,x_{1+d},\ldots,x_{1+(N-1)d}),\\  &\cdots \\z_{t}&= (x_{t},x_{t+d},\ldots,x_{t+(N-1)d}),\\ &\cdots \\
    z_{L-1-(N-1)d}&=(x_{L-1-(N-1)d},x_{L-1-(N-2)d},\ldots,x_{L-1}).\end{split}\end{equation}
\end{itemize}
 In our applications below, we will only be interested in detecting
 $D=1$-dimensional objects (loops), in which case it will be sufficient to choose $N\geq 2D+1=3$.
The theoretical considerations behind the choice of the  embedding dimension $N$  rely on Takens' embedding theorem \cite{takens1981detecting}  and its generalization by Sauer, Yorke and Casdagli \cite{tim1991embedology}.
Also, in our applications we choose the delay $d$ empirically. If we choose $d$ too small, then the
coordinates $x_{t+id}$  and $x_{t+(i+1)d}$ can be too
close to each other, and so they do not represent two `independent'
coordinates in a statistical sense. Similarly, if $d$ is too large, then $x_{t+id}$  and $x_{t+(i+1)d}$ are completely independent of each
other, and so geometric information gets lost.
Some empirical rules on choosing a suitable  delay $d$ are discussed in \cite{abarbanel1993analysis}.

\subsubsection{Sliding windows.}
To detect qualitative changes along a time series, we apply a sliding window 
and assess how its features change  along the sliding window. In our case:
\begin{itemize}
\item We apply  a sliding window of size $w$ to the sequence \eqref{eqn:z}, for $w$ sufficiently large with $N\ll w\ll L$,  obtaining a   time-varying
    point-cloud embedded in  $\mathbb{R}^N$:
\begin{equation}\label{eqn:Z}
Z^t=\{z_t,z_{t+1},\ldots, z_{t+w-1}\}, \textrm { for } t\in\{0,\ldots,K-1\},\end{equation}
\end{itemize}
where we denote $K-1=L-1-(N-1)d-(w-1)$.


\subsubsection{Time series of norms of persistent landscapes.}
Fix $n\geq 1$ and $p\geq 1$.
\begin{itemize}
\item For each point-cloud $Z^t$ we compute the norm of the persistent landscape $\|\lambda ^t_n\|_p$ , thus obtaining a time-series of norms
\[ \|\lambda ^t_n\|_p,  \textrm { for } t\in\{0,\ldots, K-1\}.\]
\end{itemize}

\section{TDA applied to oscillatory time-series}
\label{sec:TDA_quasi}
The main question is  to understand what features of a time-series yield $1$-dimensional `holes' in the persistence diagram and
result in `peaks' of the norms of the persistence landscapes.

Time-series generated by periodic and quasi-periodic signals are some basic type of signals that result in  `holes' in the persistence diagram.
We will also consider oscillatory time series whose frequency changes over time.

\subsection{Periodic time-series}
First, consider time-series generated by a periodic signal of the form
\[X(t)=A_1\sin(\omega_1 t+\phi_1)+\sigma G(t) \]
where $A_1, \omega_1,\phi_1$ are the parameters of the deterministic signal, $\sigma$ is the noise intensity, and $G(t)$ is Gaussian noise with zero mean
and unit variance.
When $\sigma=0$, the embedding of $X(t)$ in $\R^N$ with $N\geq 2$   fills densely a topological circle, which is a homeomorphic copy of $\mathbb{T}^1$.
It has a single hole, which is $1$-dimensional, thus the $1$-dimensional homology is $H_1(\mathbb{T}^1)=\mathbb{Z}_2$ whose generator is a $1$-dimensional loop. When we add noise with intensity $\sigma$ small, we obtain a `noisy' circle.

To apply the TDA procedure, we first discretize the  time $t$, i.e., we let $t_{i+1}=t_i+\Delta t$, with a step size of $\Delta t>0$ small.
Then  we choose the window size $w$ and the delay $d$.
It is important that the size of the sliding window $w$ is larger than the period $T_1=2\pi/\omega_1$ of the signal, and that the delay $d$ is less than $2\pi/ (\omega_1\Delta T )$ but not too small. Otherwise,  if $T_1>w$, then the  embedding of a window does not yield a closed curve.
If $d$ is too small, the embedding of the window may yield a point cloud with a narrow `hole'.
Indeed,  if $\Delta t$ is small and $d$ is small then the components of  $z_{t}= (x_{t},x_{t+d},\ldots,x_{t+(N-1)d})$ are very close to one another, and
thus the points $z_t$ are all located within a small neighborhood of the diagonal set $\{\textbf{x}=(x_1,\ldots, x_N)\in\R^N\,:\,x_1=\ldots=x_N\}$.
In such a case, the persistence of the homology generator for the topological circle may be relatively small, and hence the corresponding norm of the persistence
landscape  may be dominated by the effect of the noise. Some precise formulations on how to choose the window size $w$ and the delay $d$ can be found in \cite{perea2015sliding}. Related works on applying TDA to periodic signals can be found in \cite{khasawneh2018topological,dlotko2019cyclicality}.

\begin{figure}
\centering
$\begin{array}{cccc}
\includegraphics[width=0.22\textwidth]{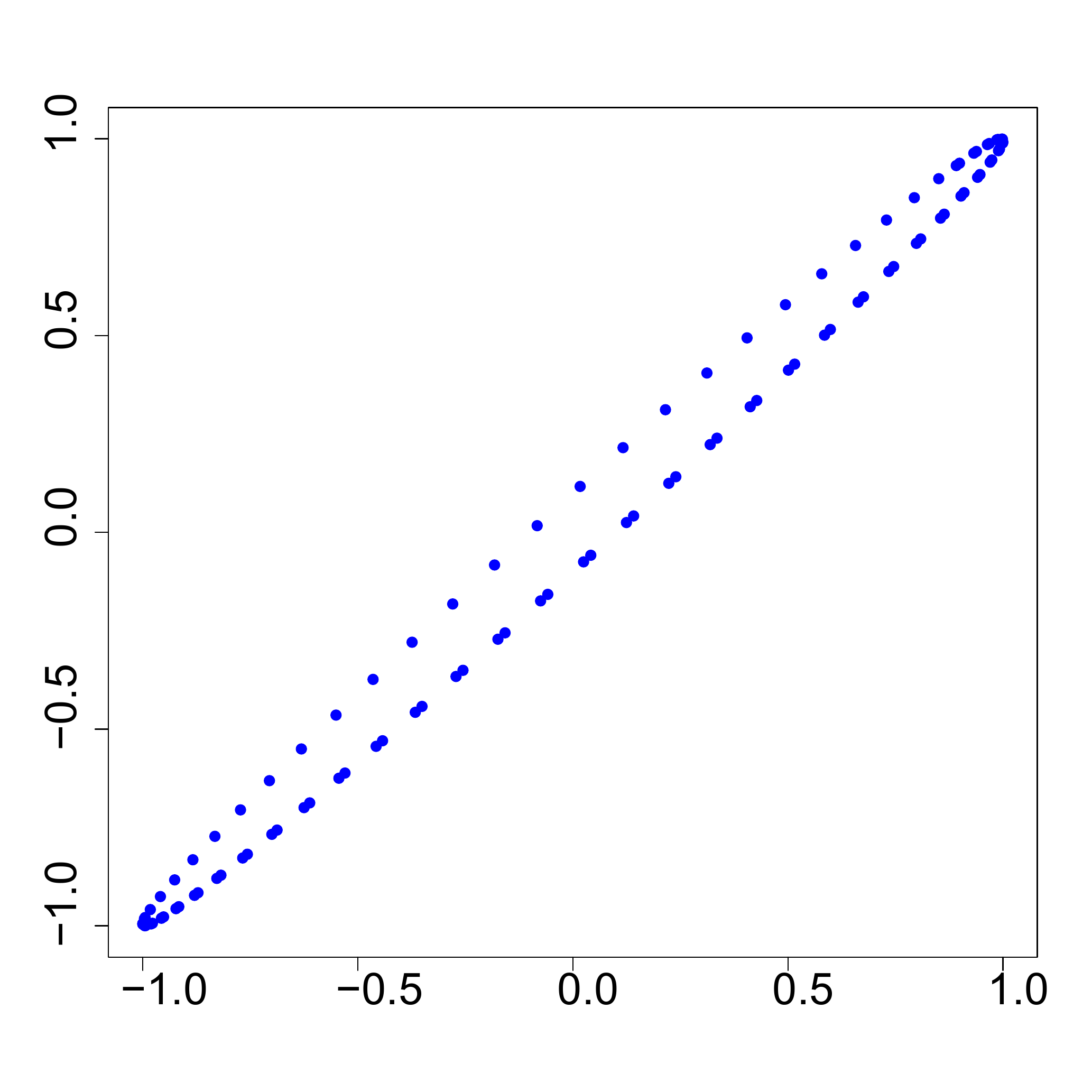}&
\includegraphics[width=0.25\textwidth]{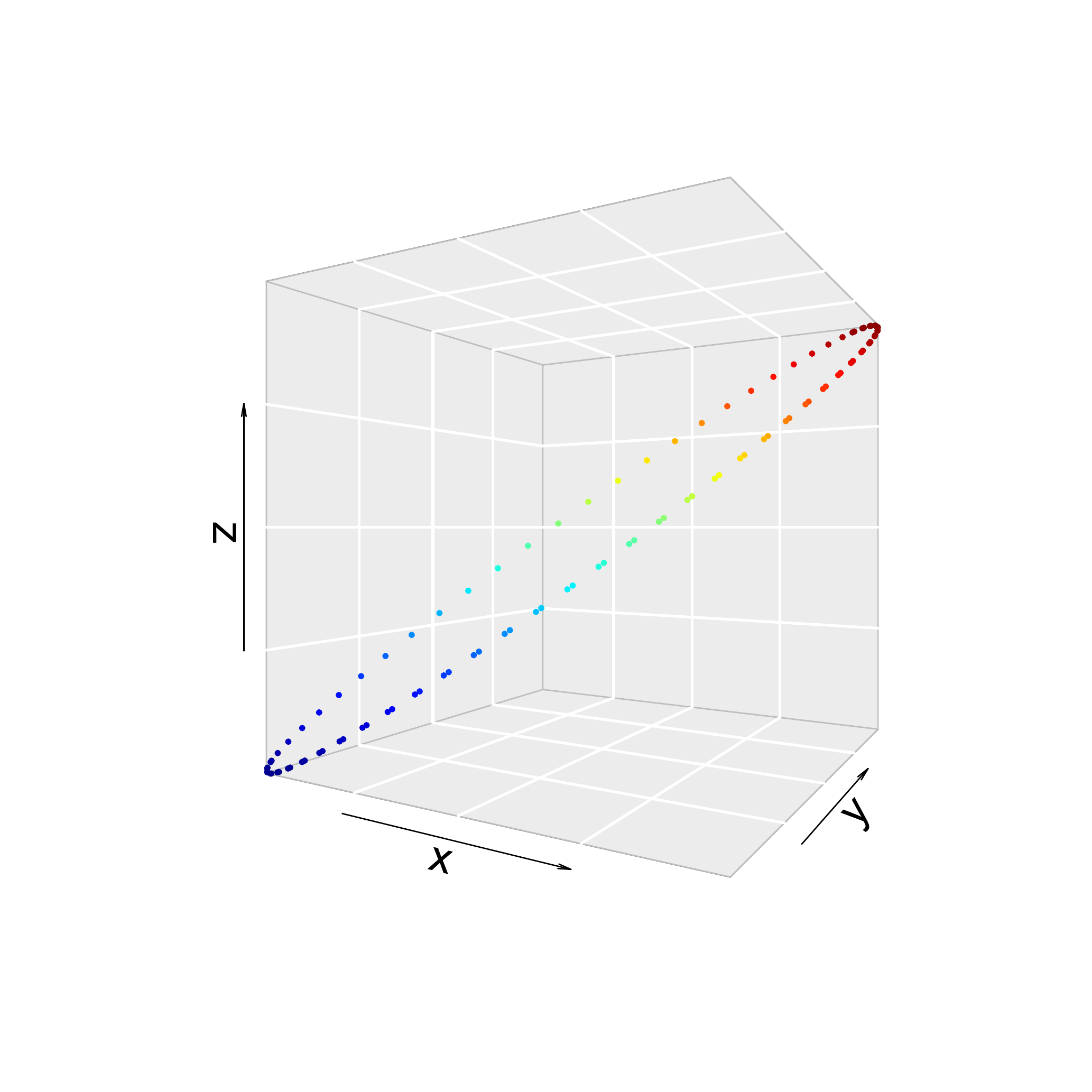}&
\includegraphics[width=0.22\textwidth]{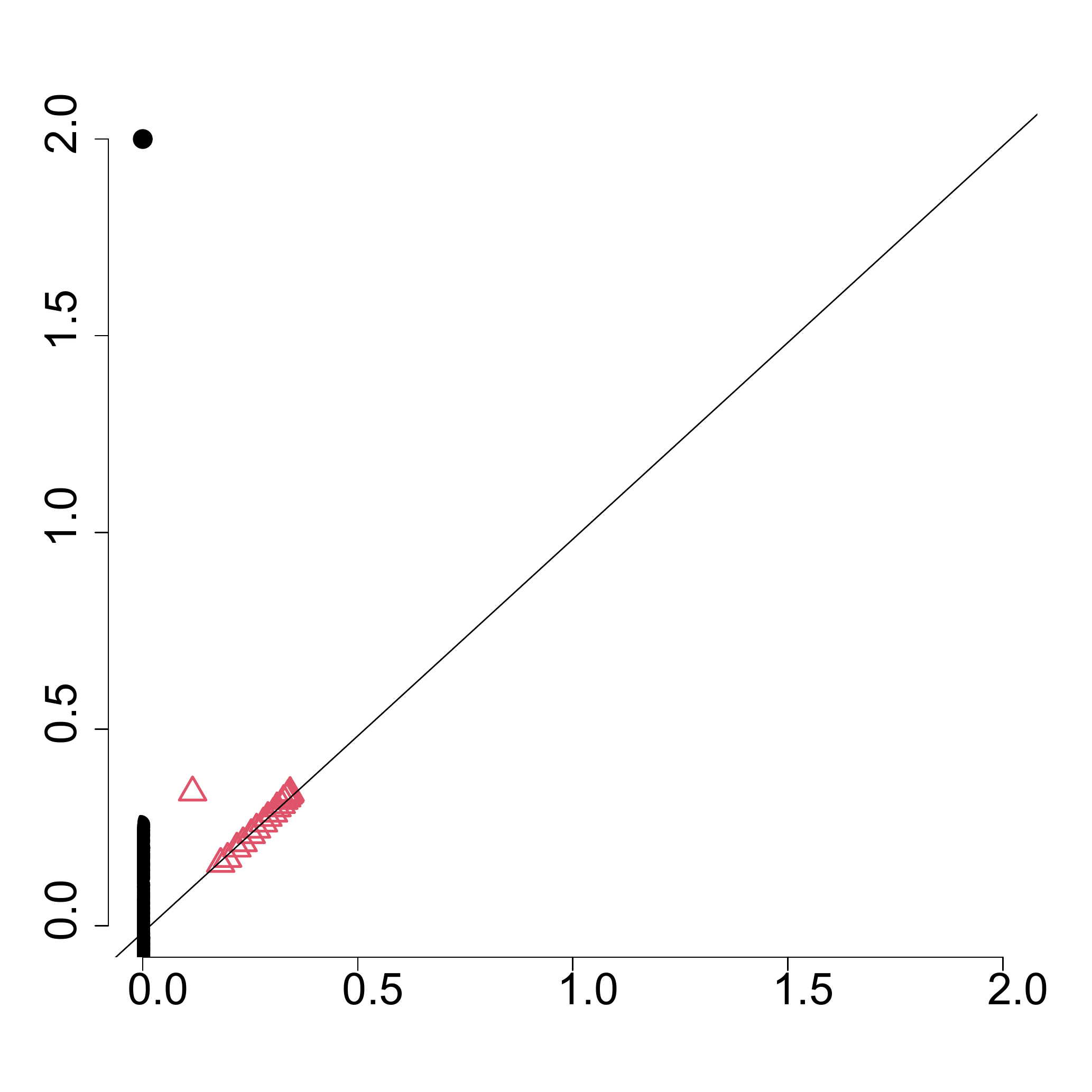}&
\includegraphics[width=0.24\textwidth]{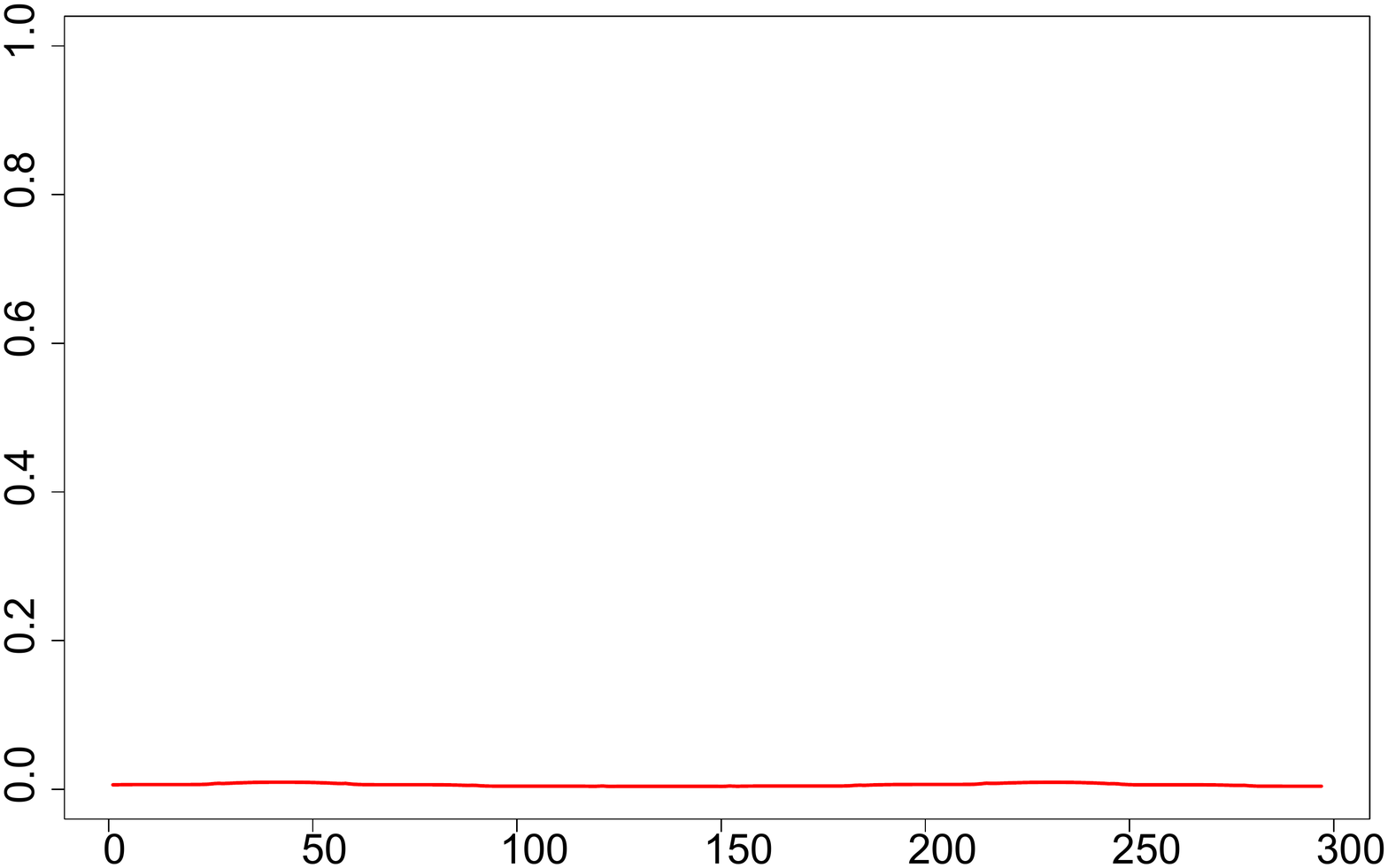}\\
\includegraphics[width=0.22\textwidth]{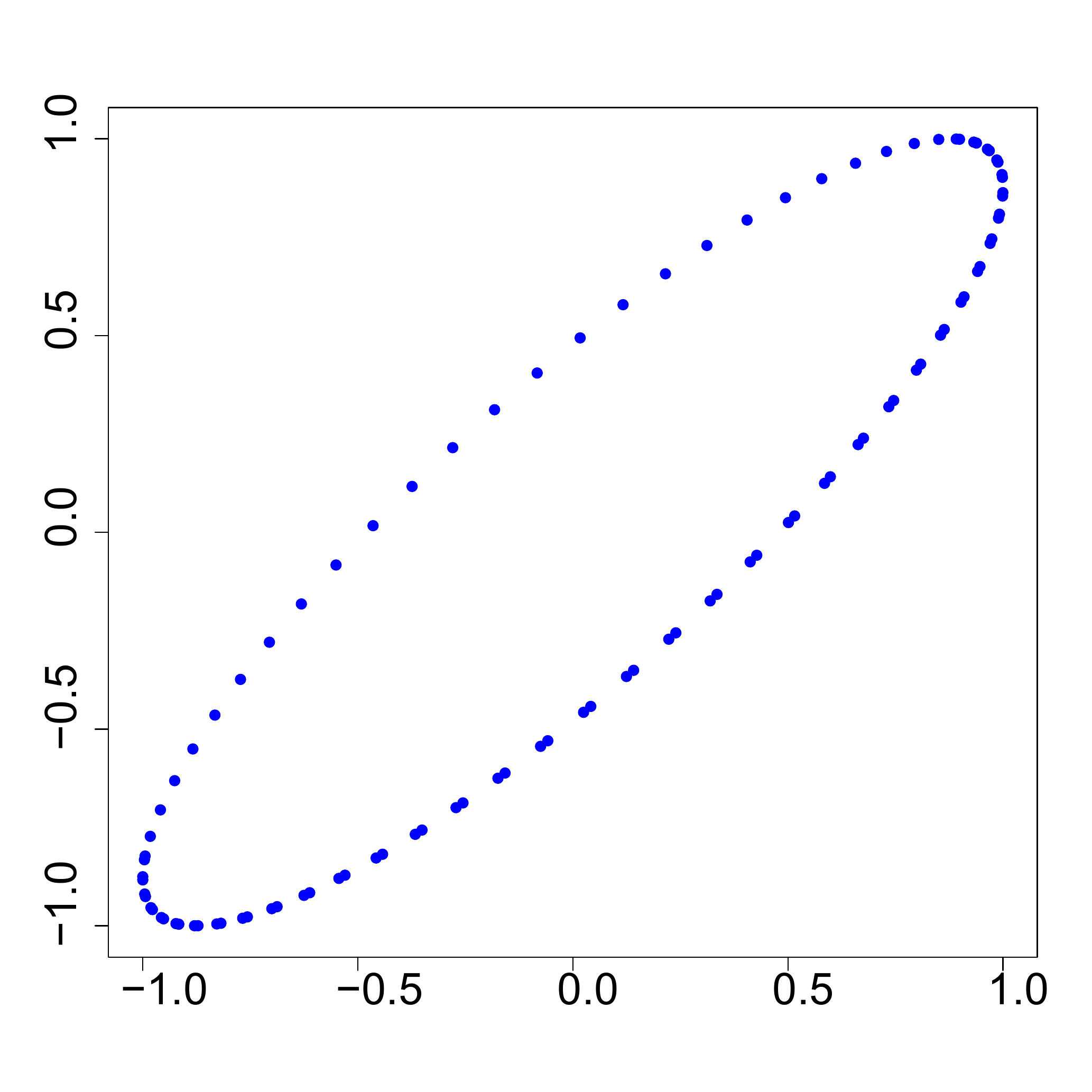}&
\includegraphics[width=0.25\textwidth]{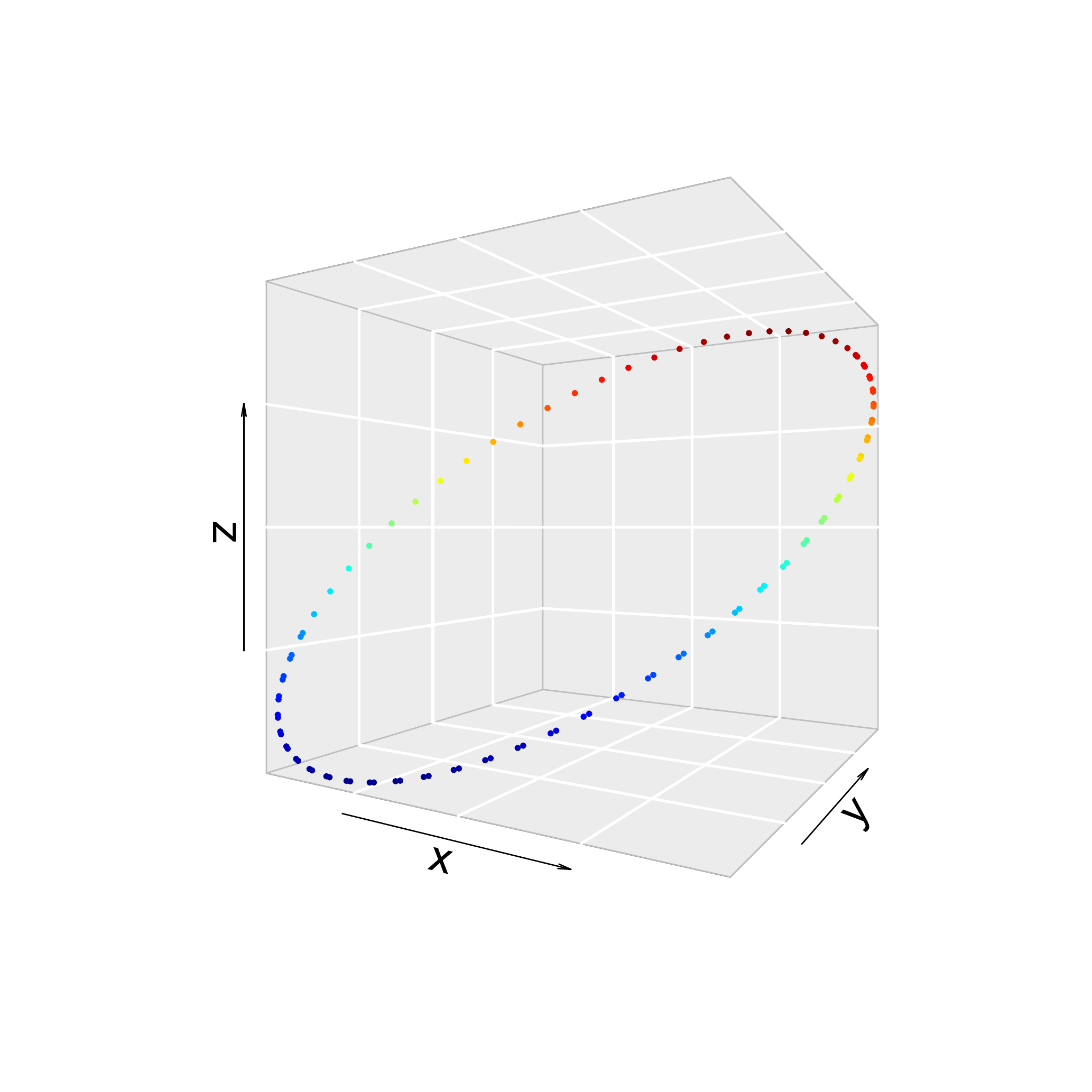}&
\includegraphics[width=0.22\textwidth]{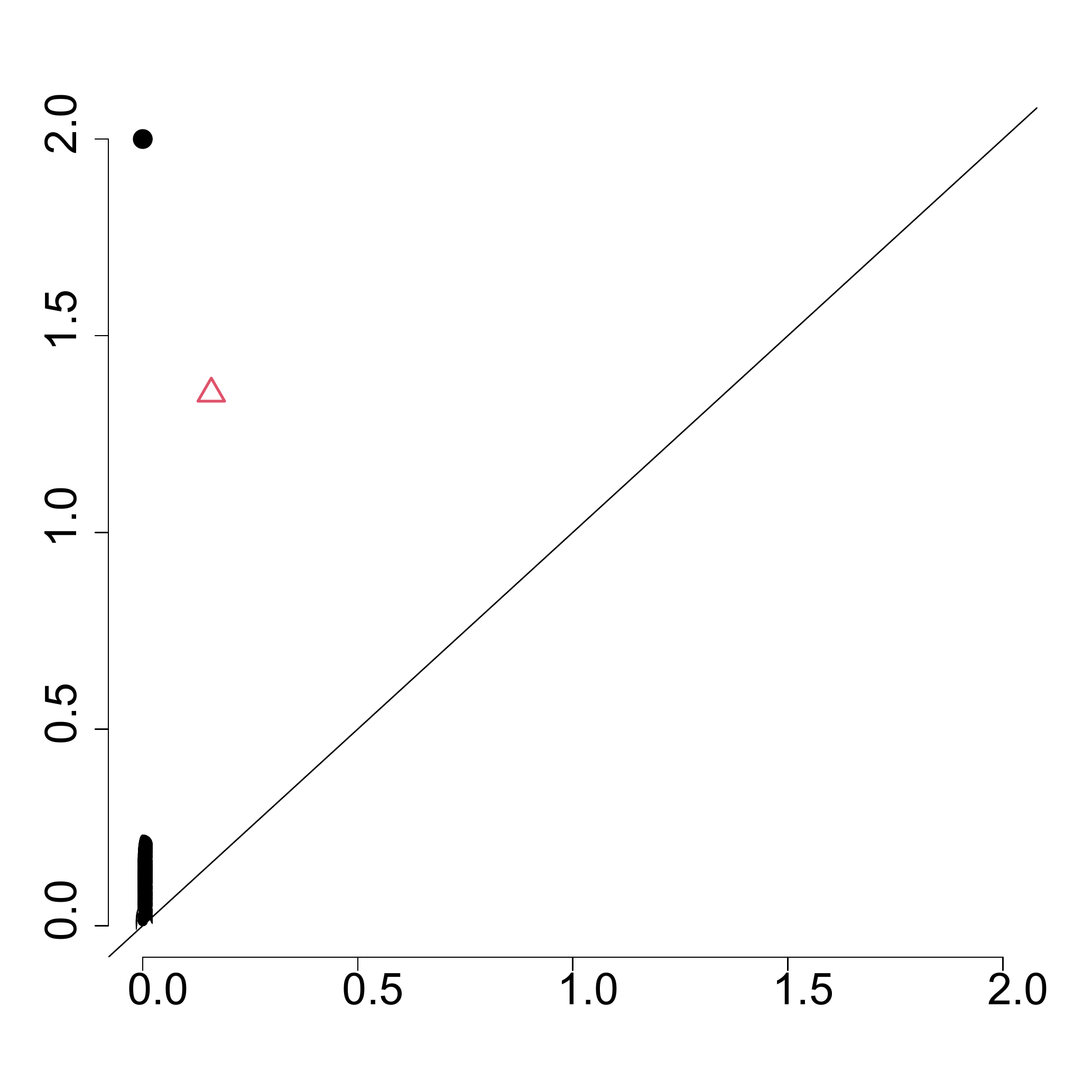}&
\includegraphics[width=0.24\textwidth]{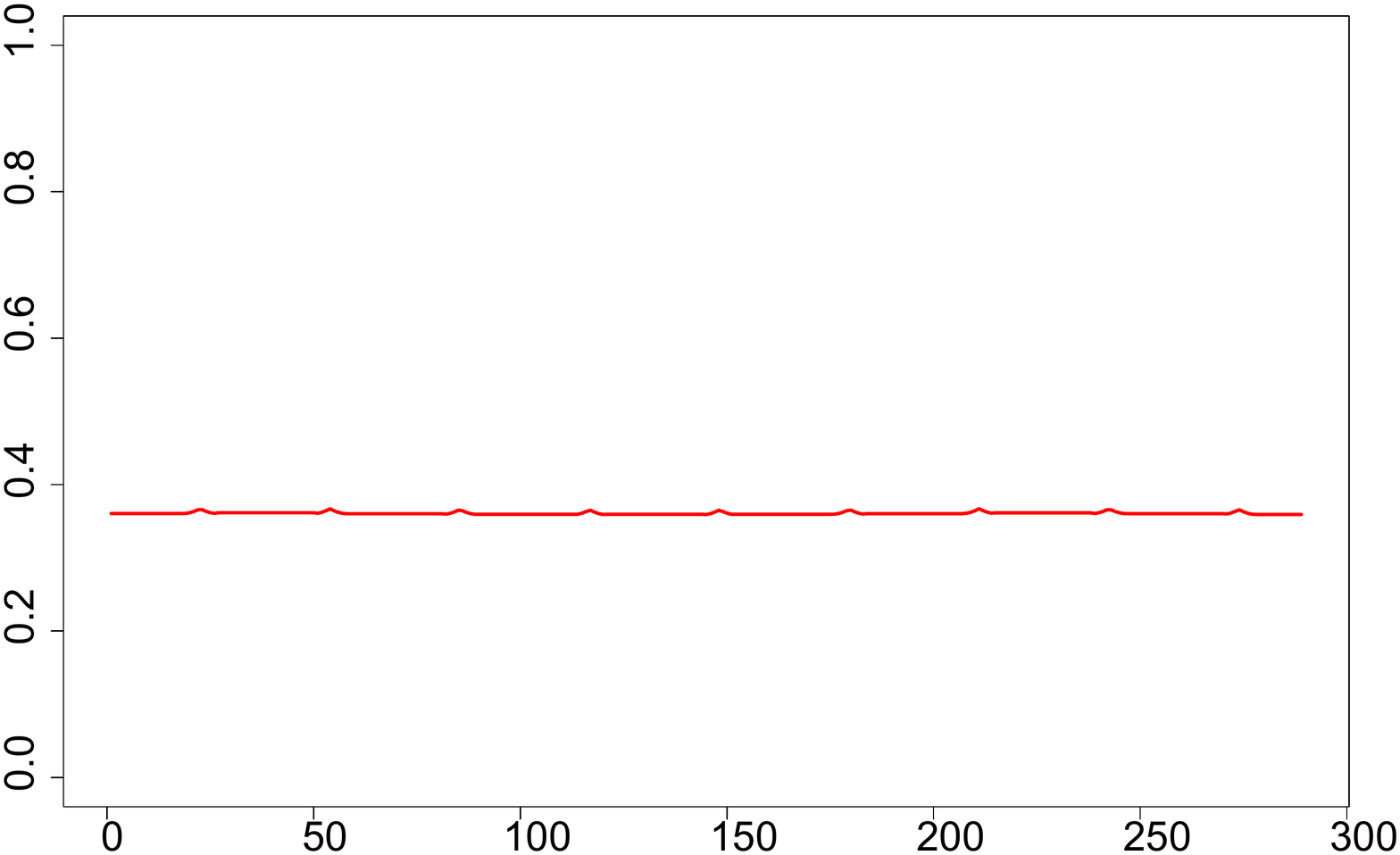}\\
\end{array}$
\caption{Periodic signal without noise reconstructed in 2D and 3D, persistence diagram and norms of persistence landscapes:  $d=1$ (top), and  $d=5$ (bottom)}
\label{fig:circle}
\end{figure}

An example of a periodic time-series without noise is shown in Fig. \ref{fig:circle}. 
We note that delay $d=5$ yields a  stronger signal than $d=1$ in both the persistence diagram  and the norm  of the persistence landscape.

In either case, the norms of the persistence landscapes stay nearly constant, since the reconstructed set in $\R^N$ -- a topological circle -- is the same for
each sliding window. The deviations from a constant value are due to the discretization of the signal.

\begin{figure}
\centering
$\begin{array}{cccc}
\includegraphics[width=0.22\textwidth]{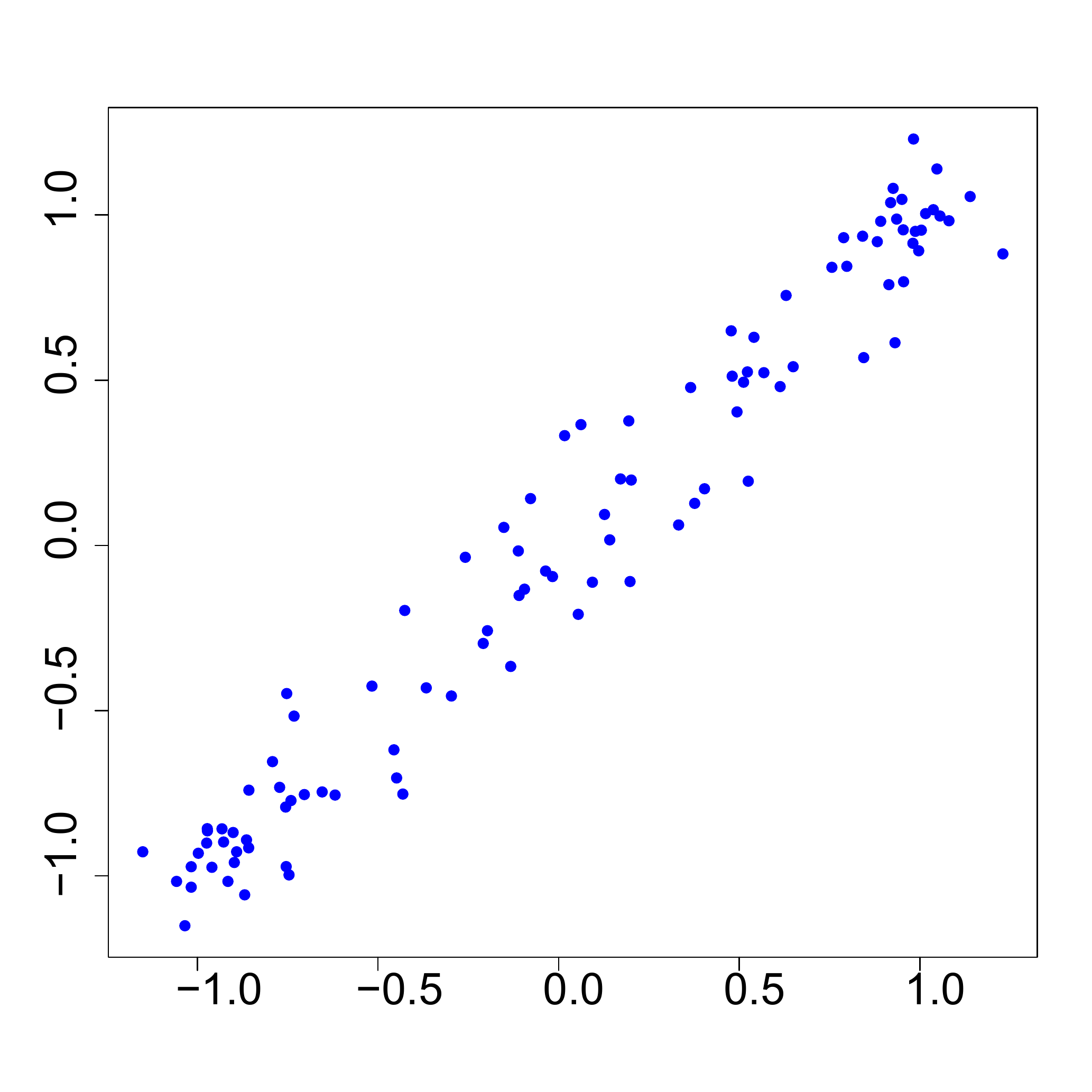}&
\includegraphics[width=0.25\textwidth]{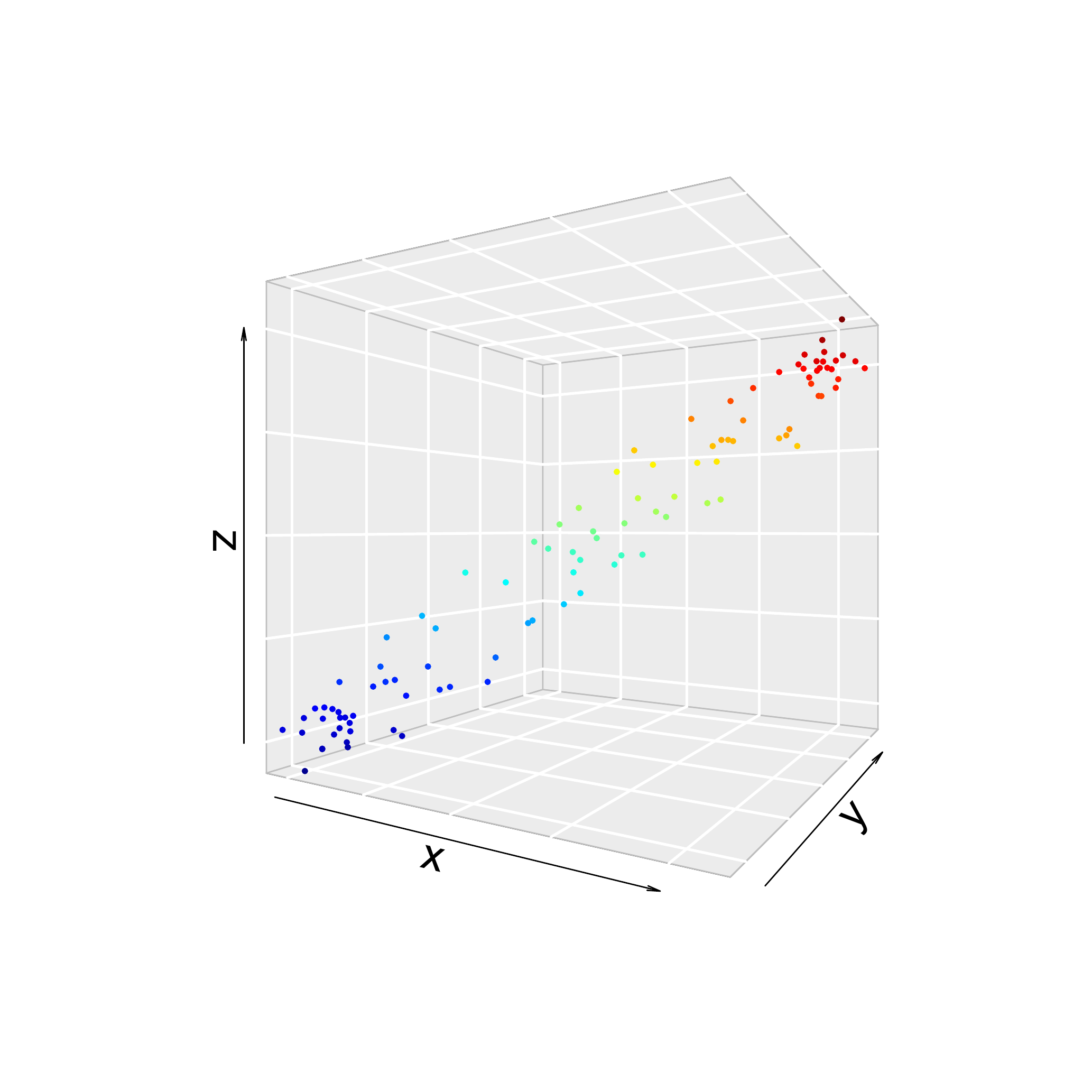}&
\includegraphics[width=0.22\textwidth]{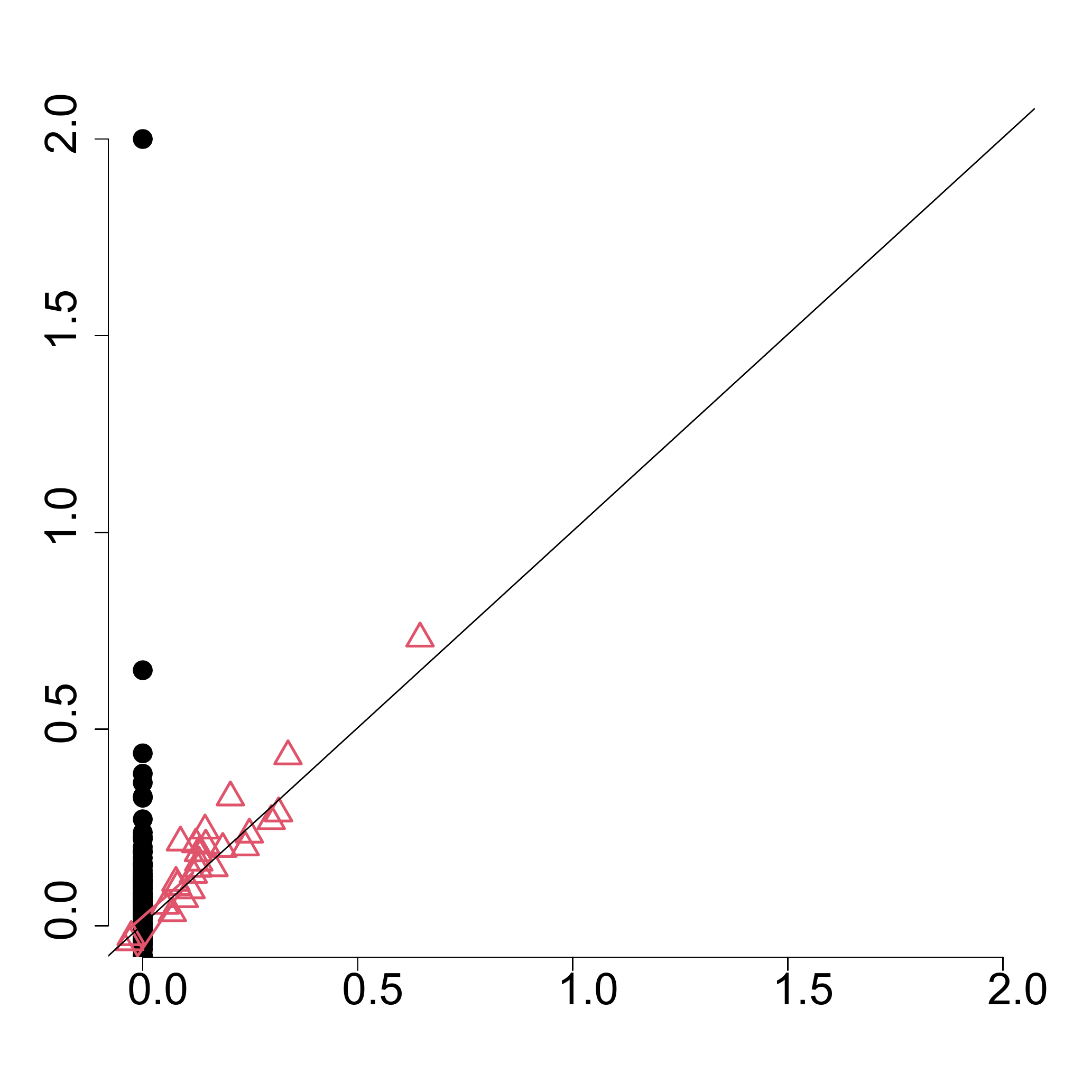}&
\includegraphics[width=0.22\textwidth]{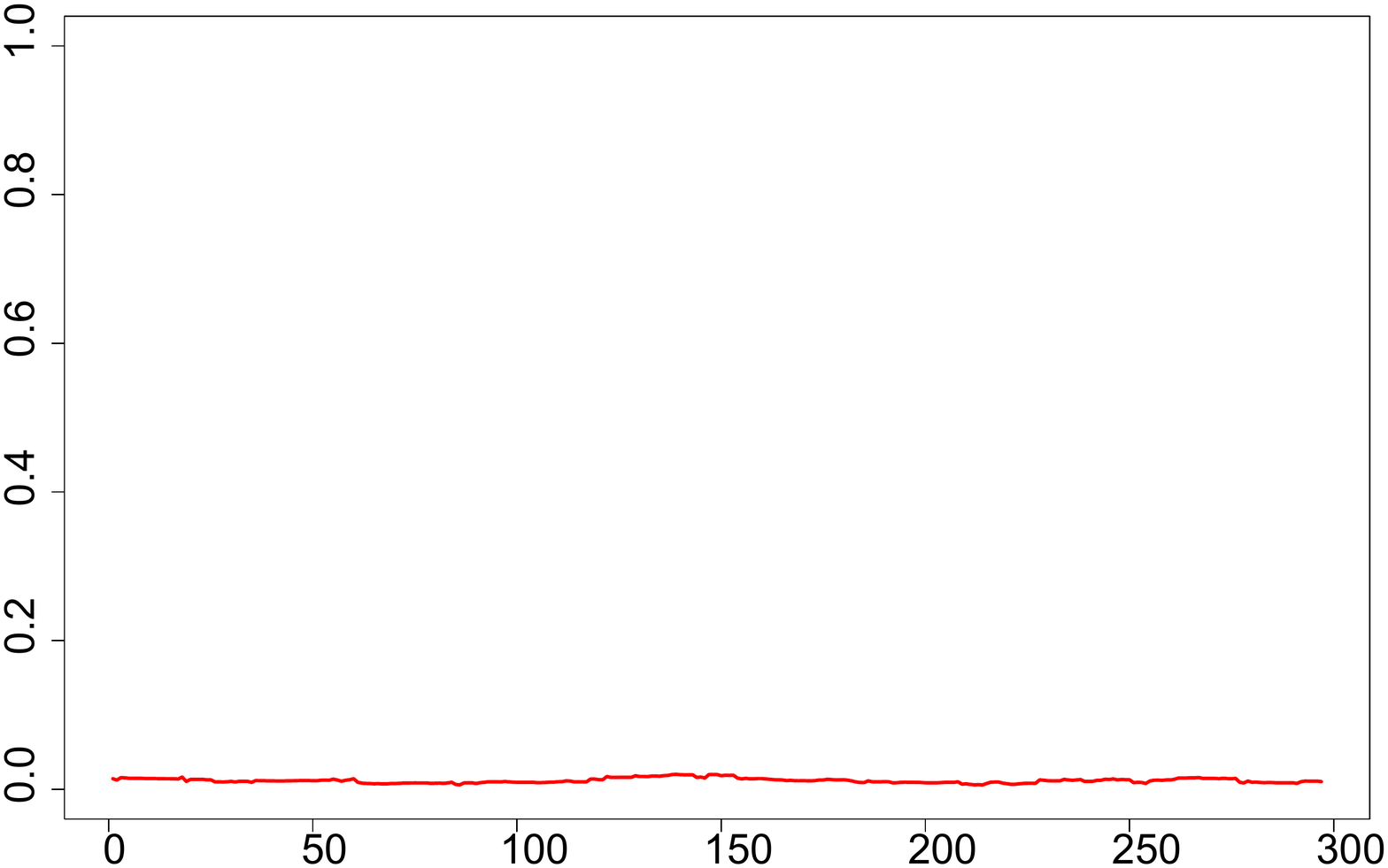}\\
\includegraphics[width=0.22\textwidth]{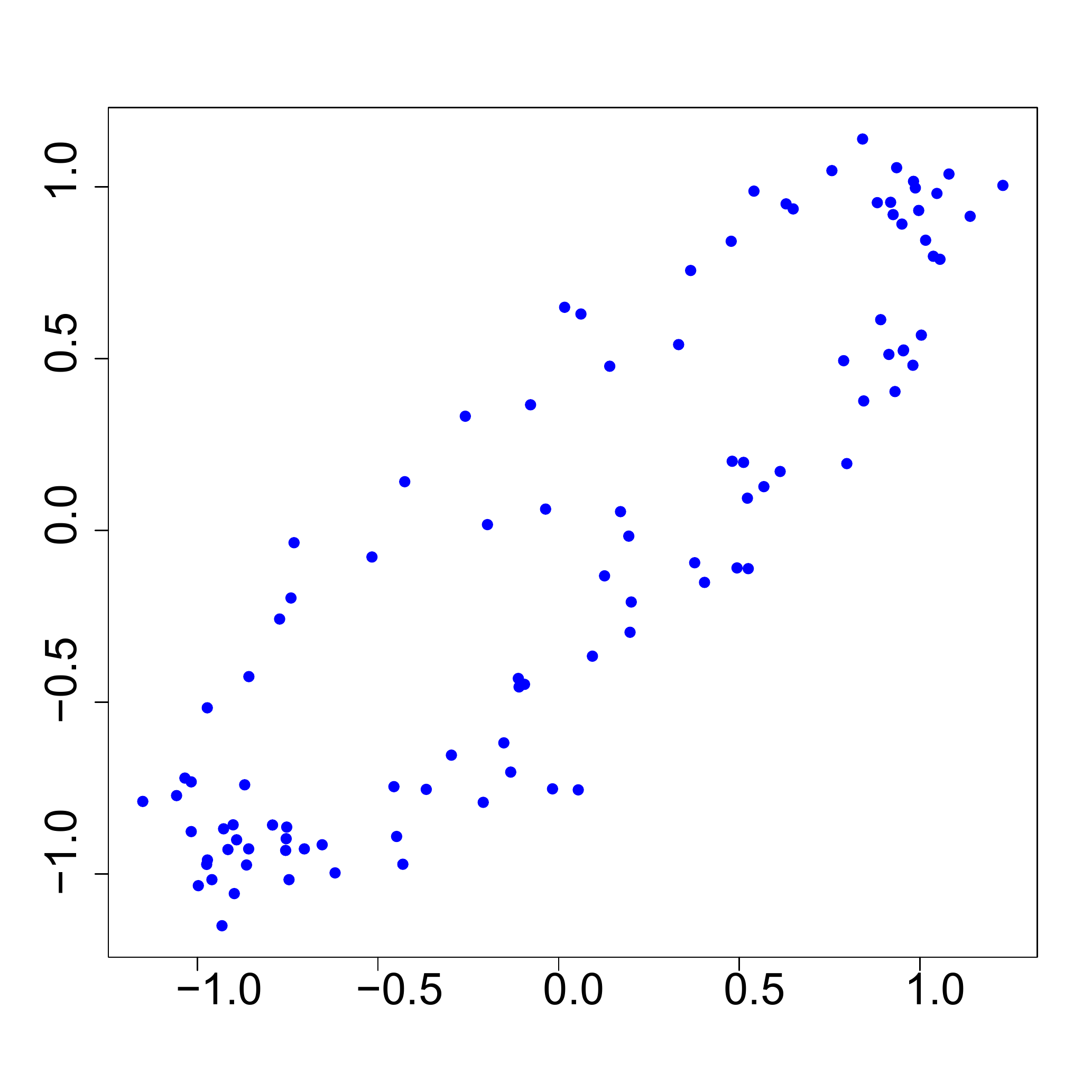}&
\includegraphics[width=0.25\textwidth]{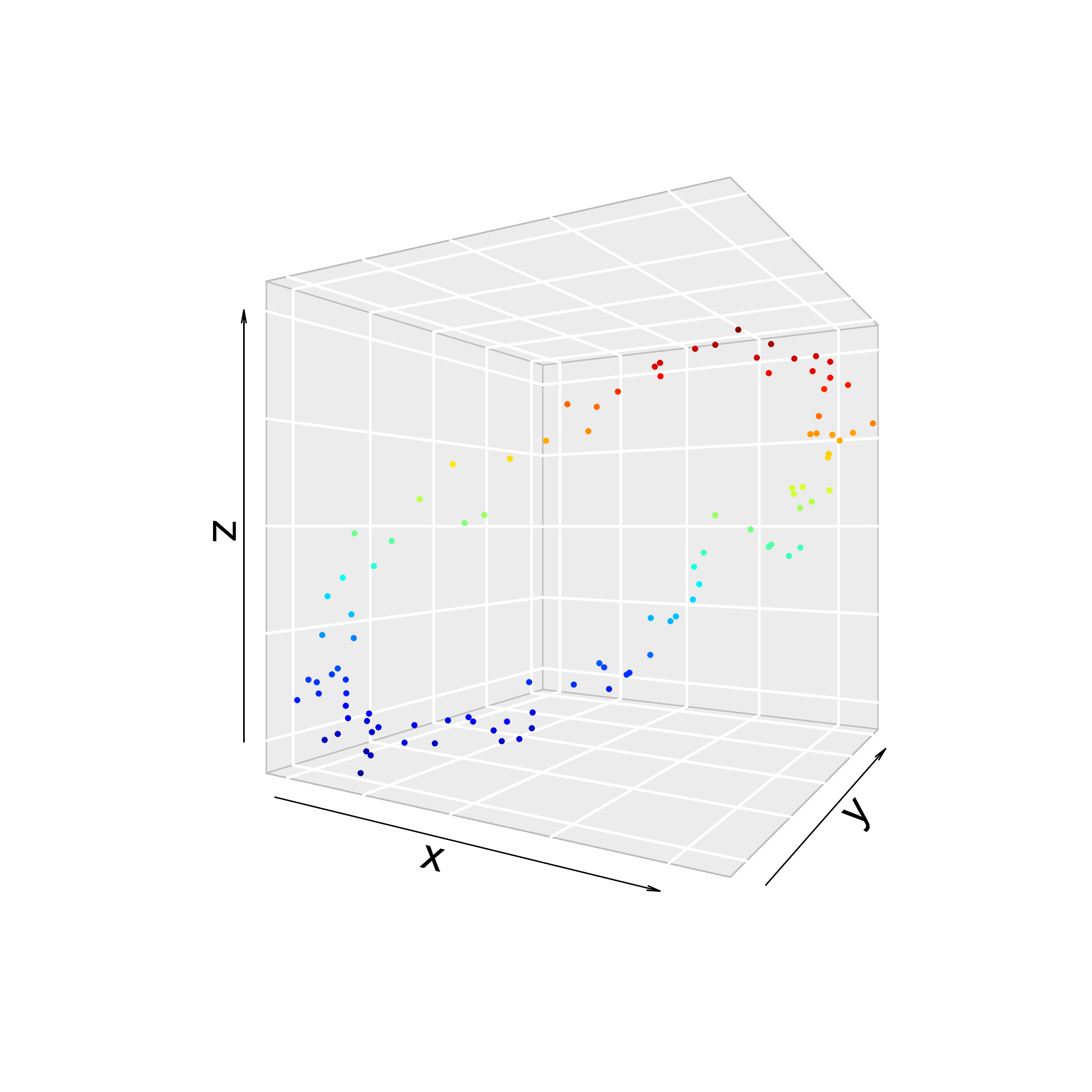}&
\includegraphics[width=0.22\textwidth]{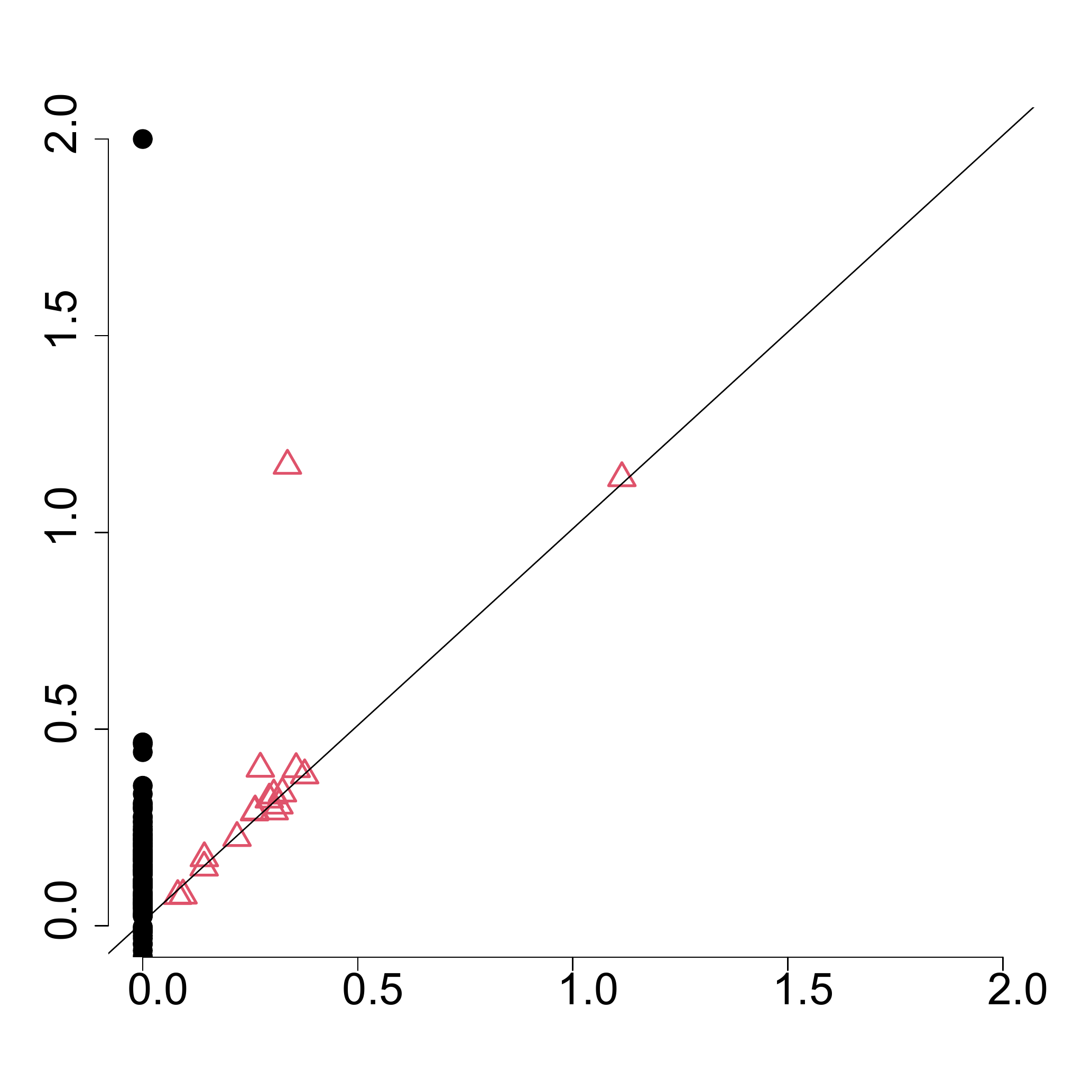}&
\includegraphics[width=0.24\textwidth]{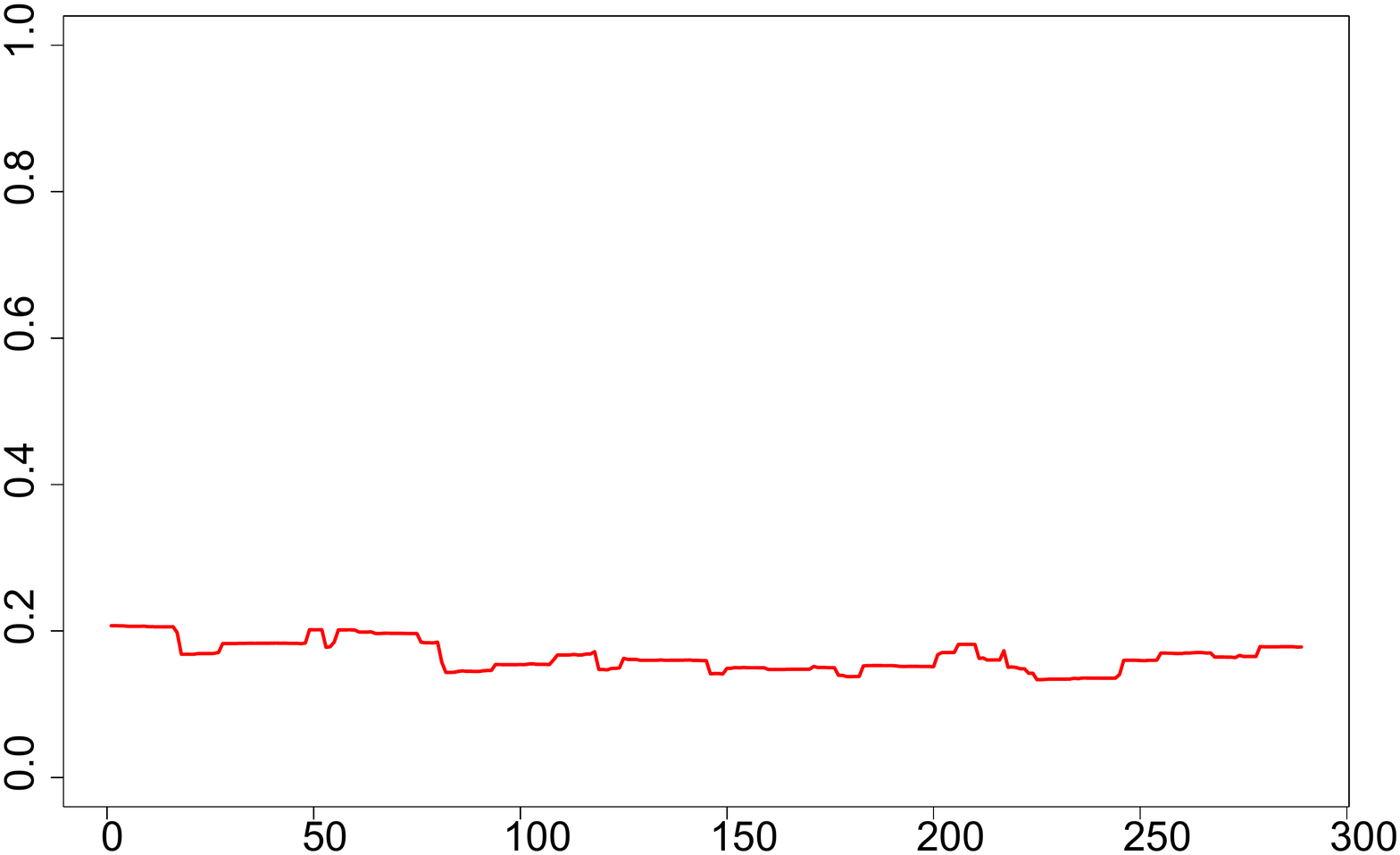}
\end{array}$
\caption{Periodic signal with noise reconstructed in 2D and 3D, persistence diagram and norms of persistence landscapes:  $d=1$ (top), and  $d=5$ (bottom)}
\label{fig:circle_noisy}
\end{figure}

An example with a periodic time-series with noise is shown in Fig. \ref{fig:circle_noisy}; in this case we add noise with $\sigma=0.1$.
We note that delay $d=5$ still  yields a  stronger signal than $d=1$.
We also notice that adding the noise weakens the TDA signal from the no-noise case. This is because  the noise   narrows the  `hole' in the point cloud.

\subsection{Quasi-periodic time-series}
\begin{figure}
\centering
$\begin{array}{cccc}
\includegraphics[width=0.22\textwidth]{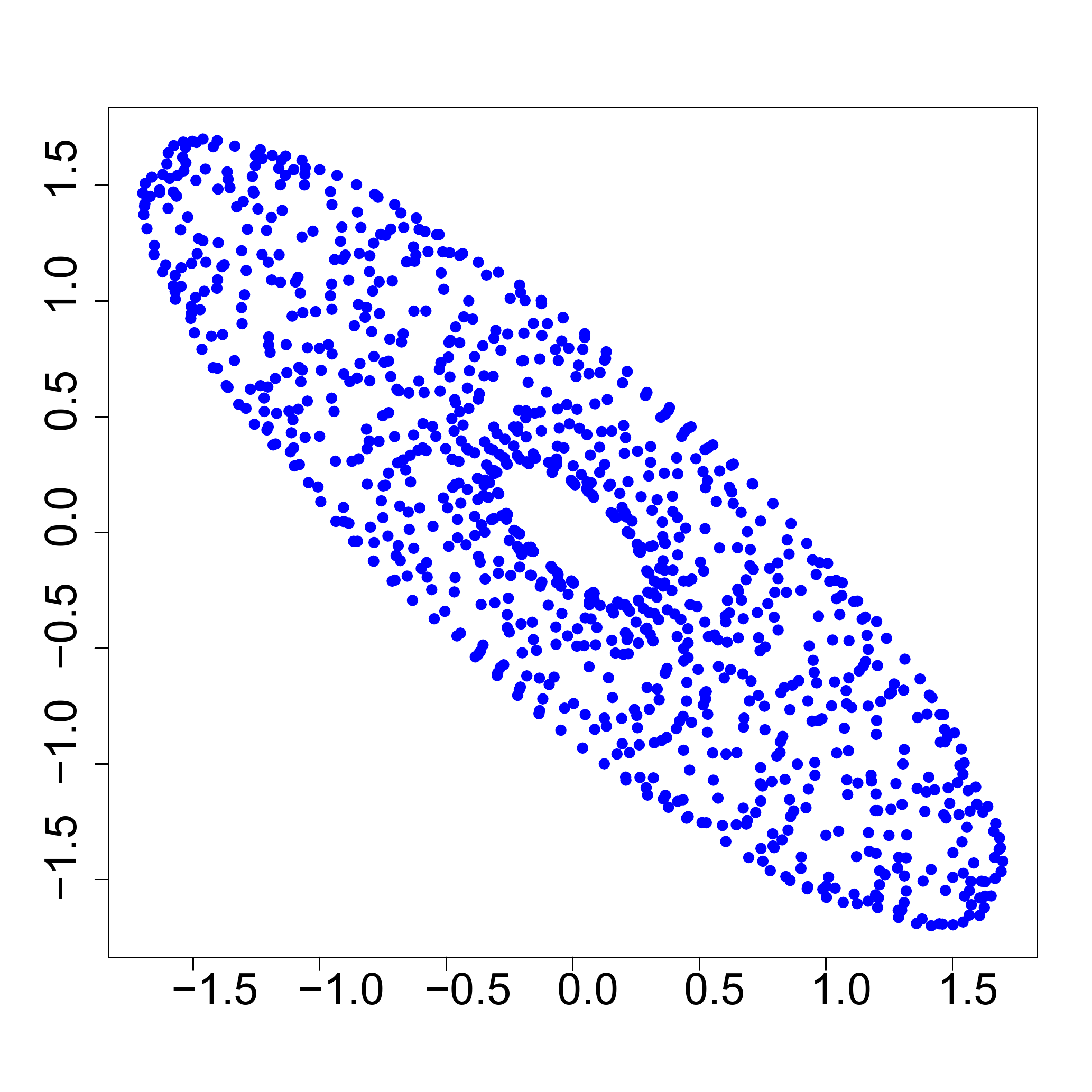}&
\includegraphics[width=0.25\textwidth]{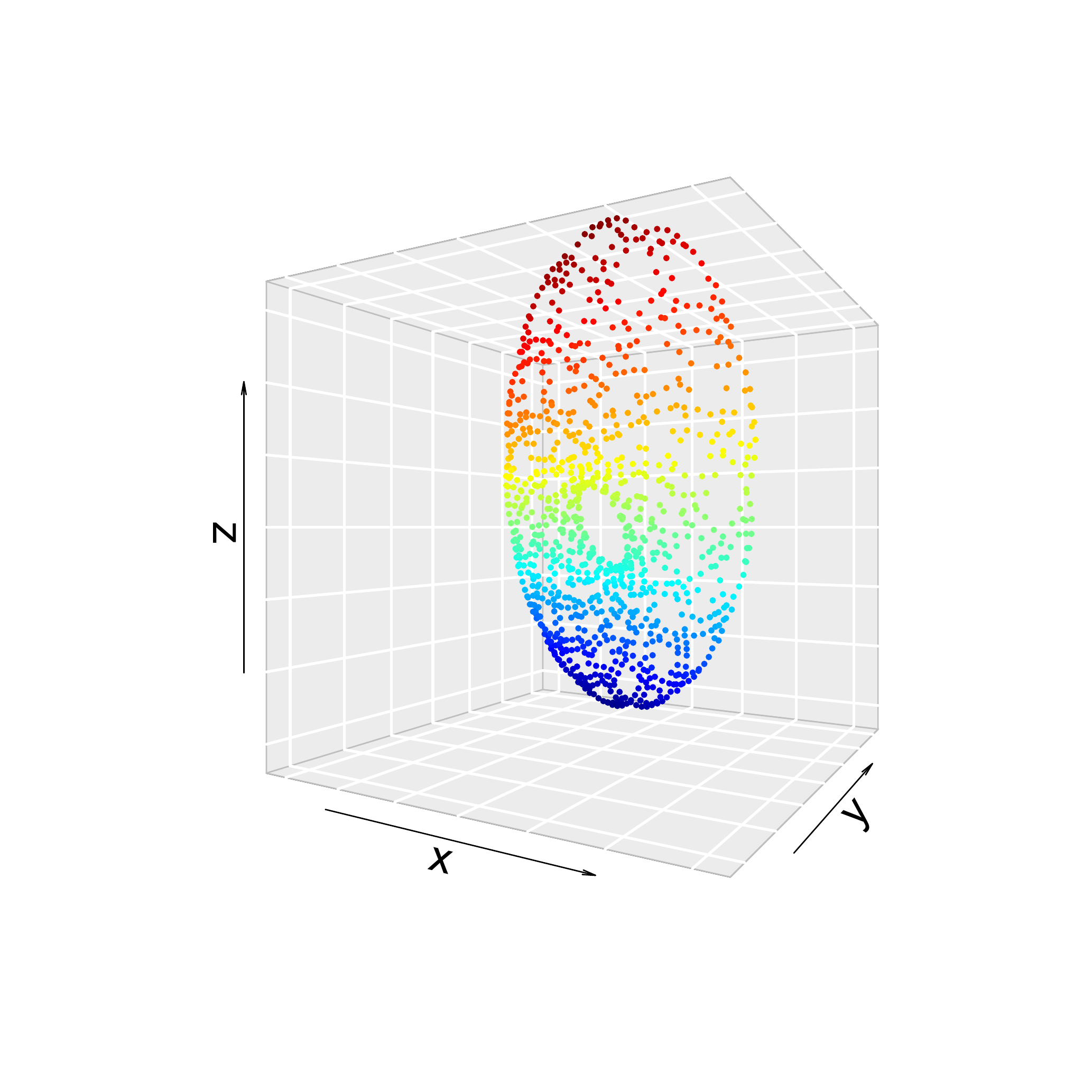}&
\includegraphics[width=0.22\textwidth]{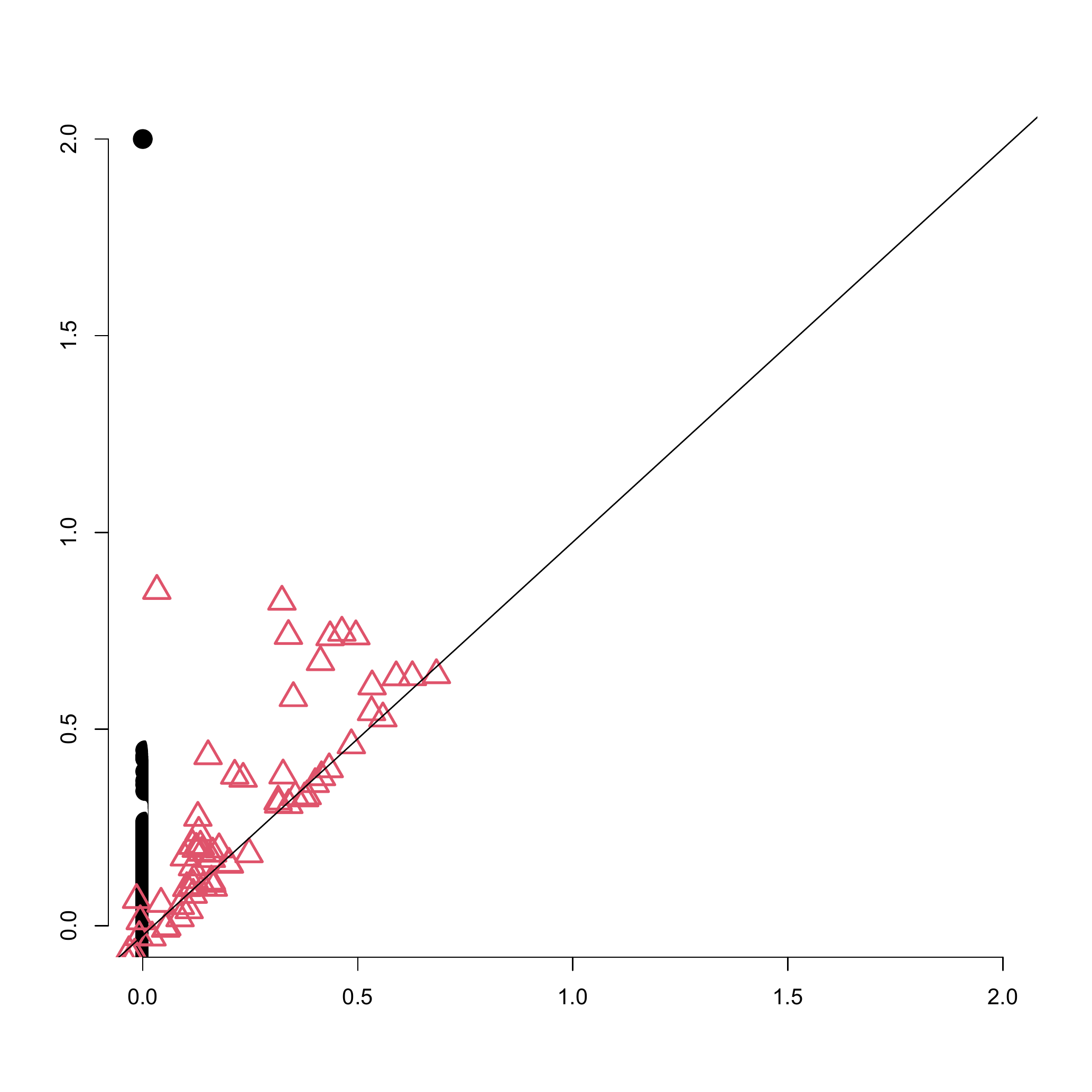}&
\includegraphics[width=0.24\textwidth]{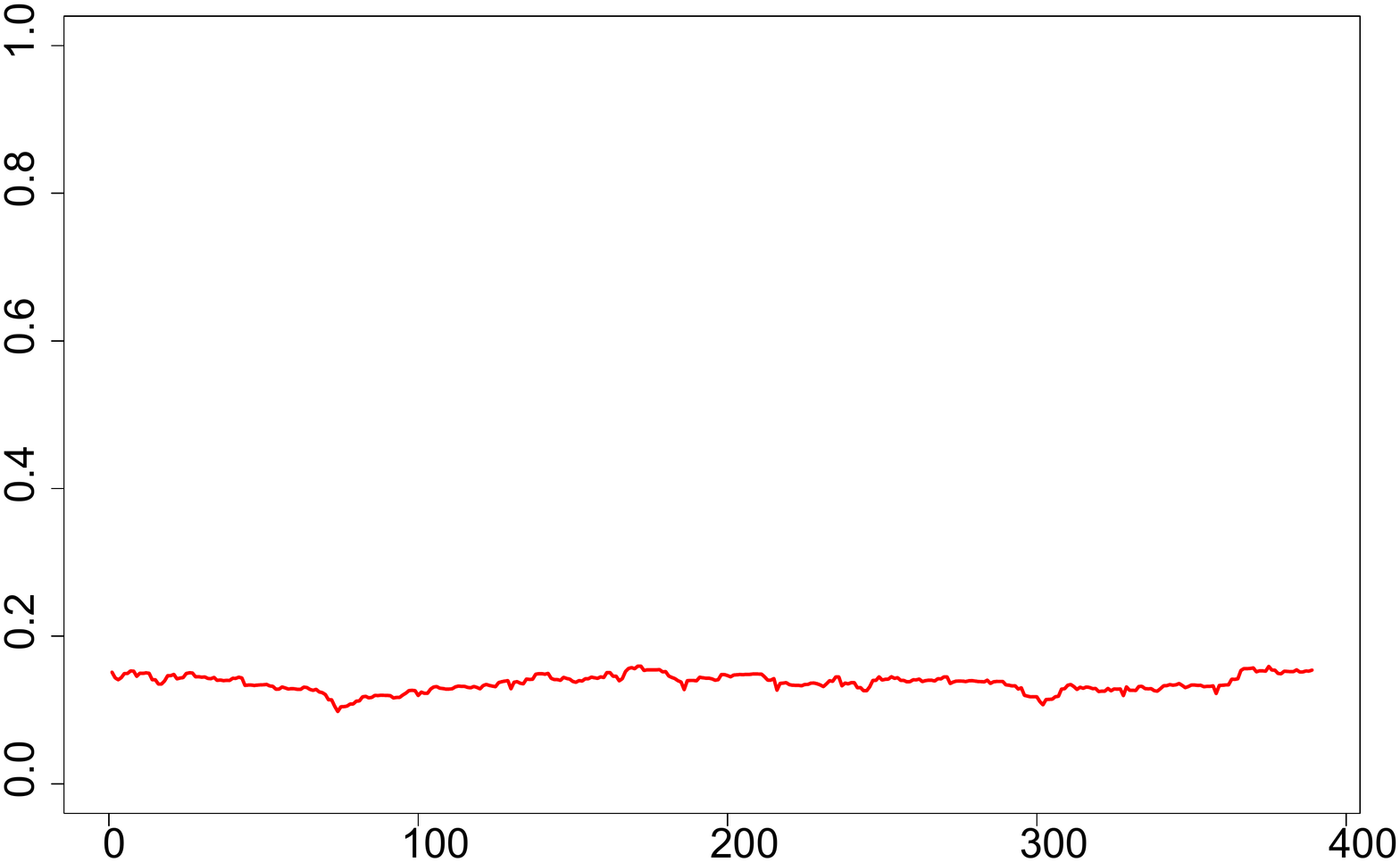}\\
\end{array}$
\caption{Quasi-periodic signal without noise reconstructed in 2D and 3D, persistence diagram and norms of persistence landscapes:  $d=1$ (top), and  $d=5$ (bottom)}
\label{fig:torus}
\end{figure}

Second, consider time-series generated by a quasi-periodic signal with two incommensurate frequencies
\[X(t)=A_1\sin(\omega_1 t+\phi_1)+A_2\sin(\omega_2 t+\phi_2)+\sigma G(t) \]
where $A_1, \omega_1,\phi_1,A_2, \omega_2,\phi_2$ are the parameters of the deterministic signal, and $\sigma$, $G(t)$
are as before; we assume that $\omega_1/\omega_2$ is an irrational number.
When $\sigma=0$, for suitable window size $w$ and delay $d$,  the embedding of $X(t)$ in $\R^N$ with $N\geq 3$   yields a topological $2$-dimensional torus $\mathbb{T}^2$. This has two $1$-dimensional homology generators, since $H_1(\mathbb{T}^2)=\mathbb{Z}^2_2$. In addition, there is a single $2$-dimensional homology generator, i.e.,  $H_2(\mathbb{T}^2)=\mathbb{Z}$. See Fig.~\ref{fig:torus}.

The norms of the persistence landscapes stay nearly constant, since the reconstructed set in $\R^N$ -- a topological torus -- is the same  for each sliding
window.
The deviations from a constant value are due to the discretization of the signal.

Similarly, we can consider general quasi-periodic signals with $m\geq 1$ incommensurate frequencies
\[X(t)=\sum_{i=1}^{m} A_i\sin(\omega_i t+\phi_i)+\sigma G(t). \]
There are specific conditions on how to choose the windows size $w$, the delay $d$ and the embedding dimension $N$; see \cite{perea2015sliding,perea2016persistent}. The reconstructed set is an $m$-dimensional torus $\mathbb{T}^m$, whose homology is given by
$H_k(\mathbb{T}^m)=\mathbb{Z}_2^{m \choose k}$.


\subsection{Oscillatory time-series with changing frequency}
Another model that produces `holes' in point-cloud is an oscillatory signal of the form
\[X(t)=A_1\sin(\omega_1 \ln(t_c-t)+\phi_1)+\sigma G(t),\]
where $t_c$ represents a critical time and $t\in [0,t_c)$.
Note that when $t\to  t_c$ the frequency of the oscillations $\omega_1 \ln(t_c-t)\to\infty$ at a logarithmic rate, while the amplitude of the oscillations stays constant.
This type of signal is related to the LPPLS model \eqref{eqn:LPPLS_expected}.

There is a strong dependence of the TDA signal on the range $[0, t_c)$ of the time-series, on the parameter $\omega_1$, and on  size of the sliding window $w$. Note that at any moment of time $t$ the approximate period of the signal is $T=\frac{2\pi}{\omega_1 \ln(t_c-t)}$.
Each full oscillation in the time-series produces  a hole in point-cloud.  As $t$ increases, more holes appear in point-cloud for the corresponding sliding window.
When $w$ is larger than the initial period   $T=\frac{2\pi}{\omega_1 \ln(t_c)}$, then the TDA signal will capture the first oscillation  in the time-series as well as the subsequent ones. However, when $w<\frac{2\pi}{\omega_1 \ln(t_c)}$, the TDA signal will only capture oscillations  in the time-series whose period
$\frac{2\pi}{\omega_1 \ln(t_c-t)}<w$.

\begin{figure}
\centering
$\begin{array}{ccc}
\includegraphics[width=0.32\textwidth]{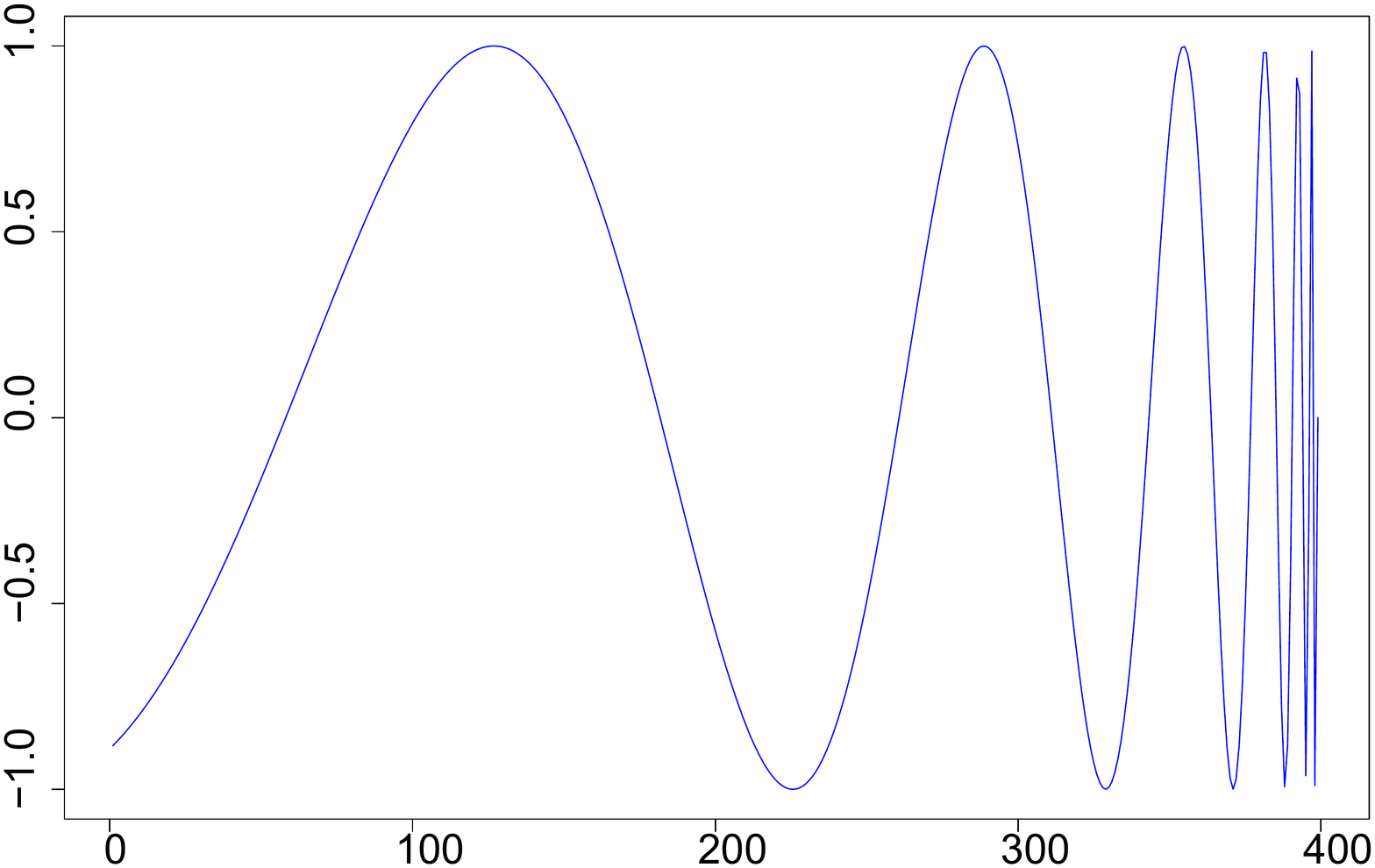}&
\includegraphics[width=0.32\textwidth]{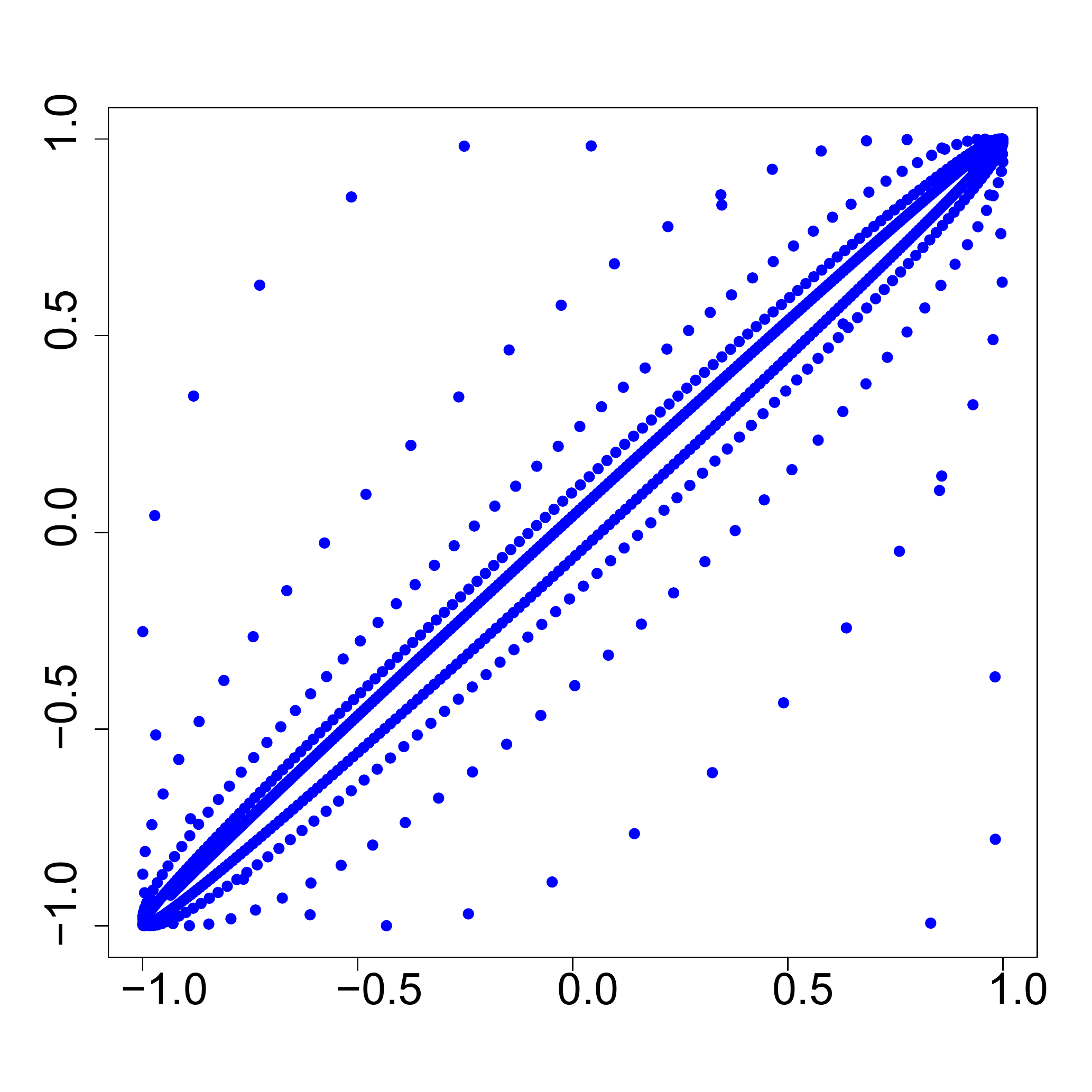}&
\includegraphics[width=0.32\textwidth]{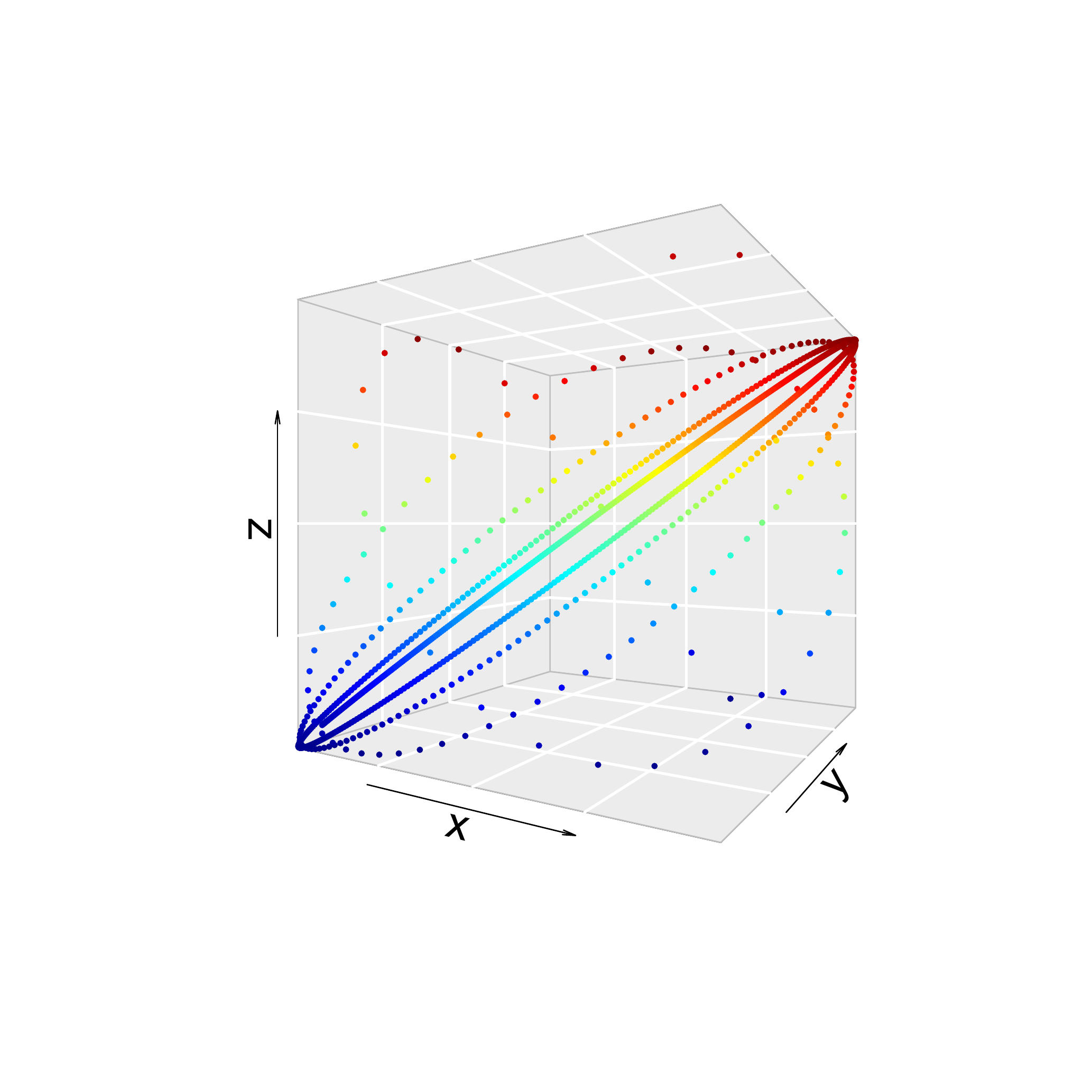}\\
\includegraphics[width=0.32\textwidth]{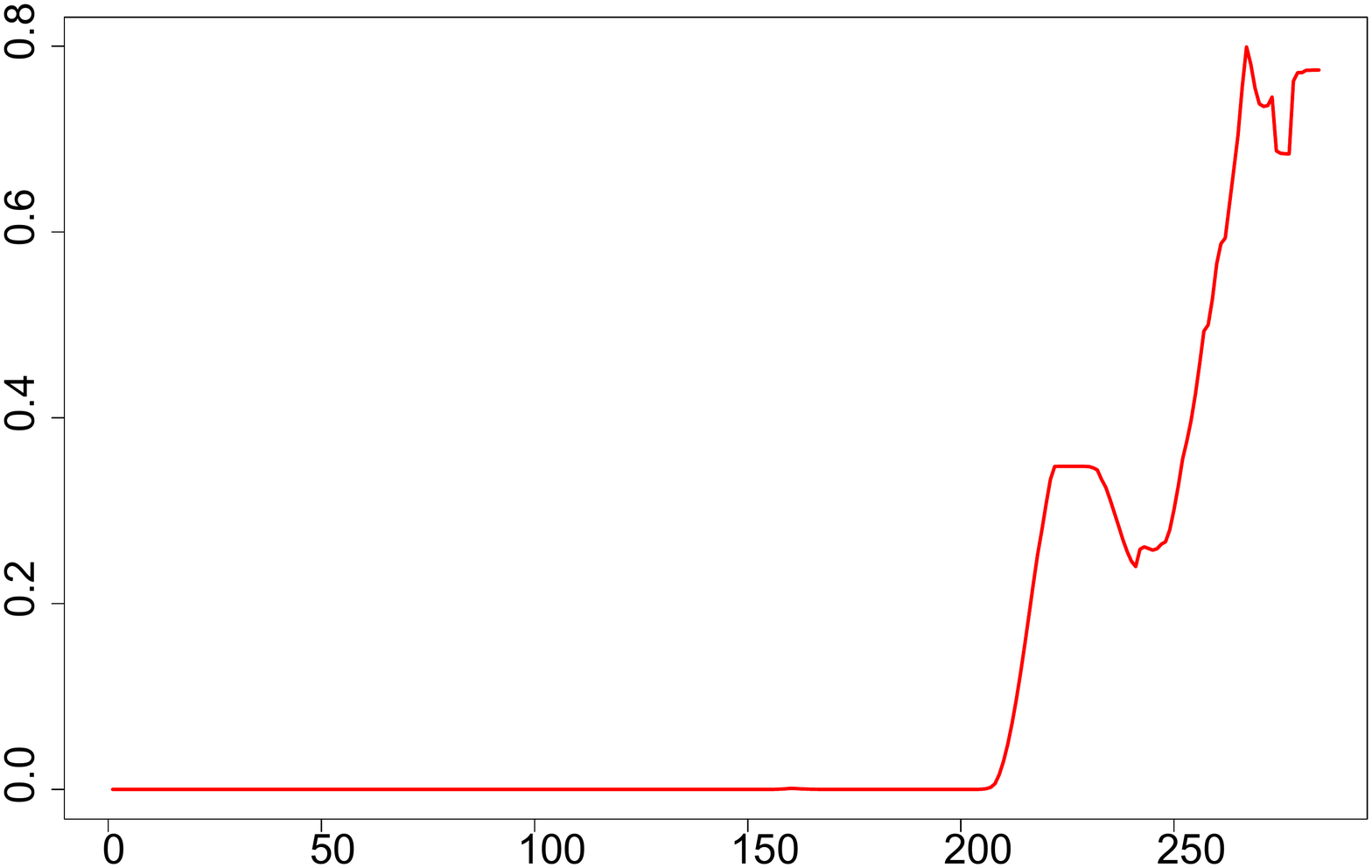}&
\includegraphics[width=0.32\textwidth]{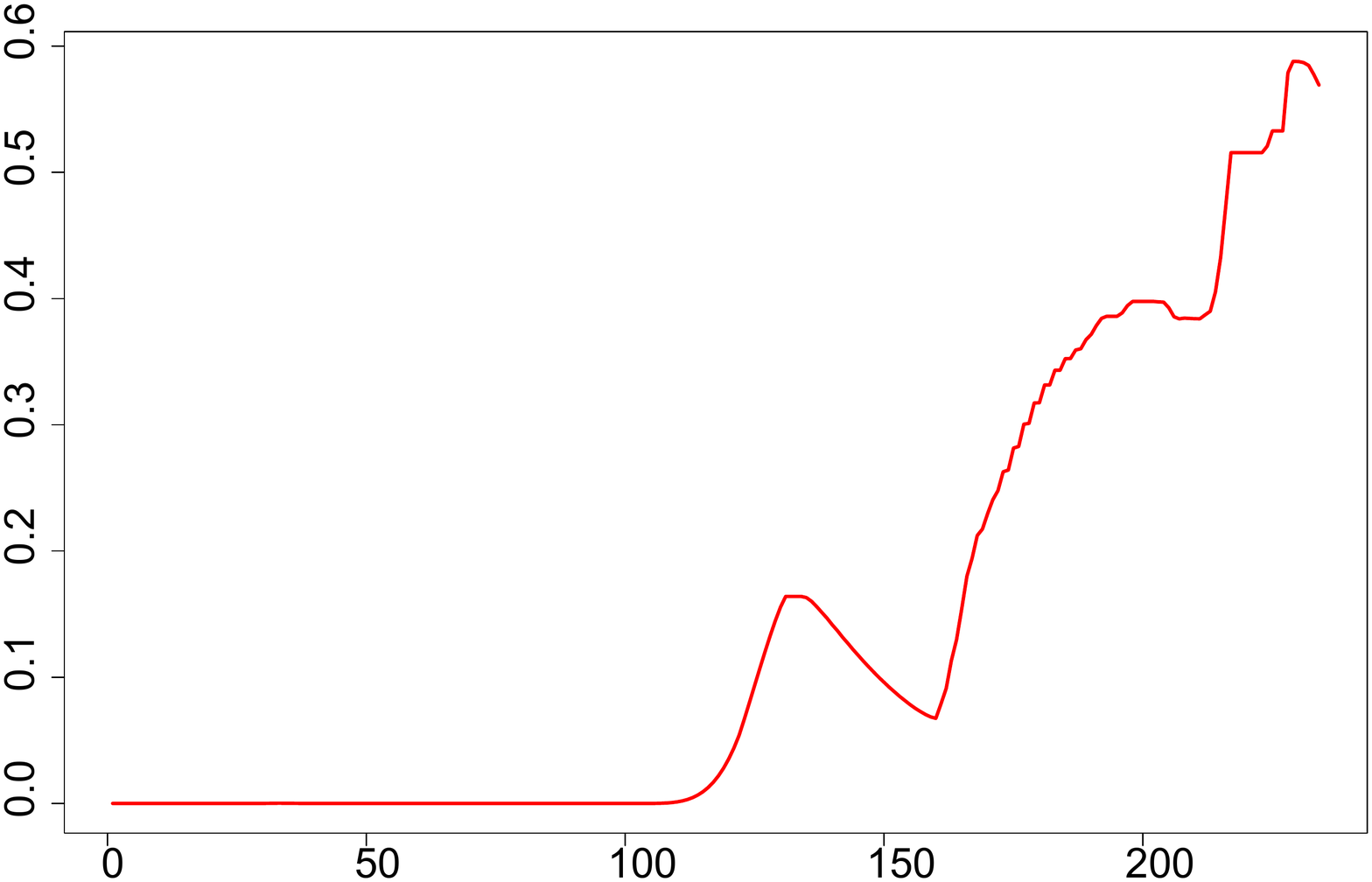}&
\includegraphics[width=0.32\textwidth]{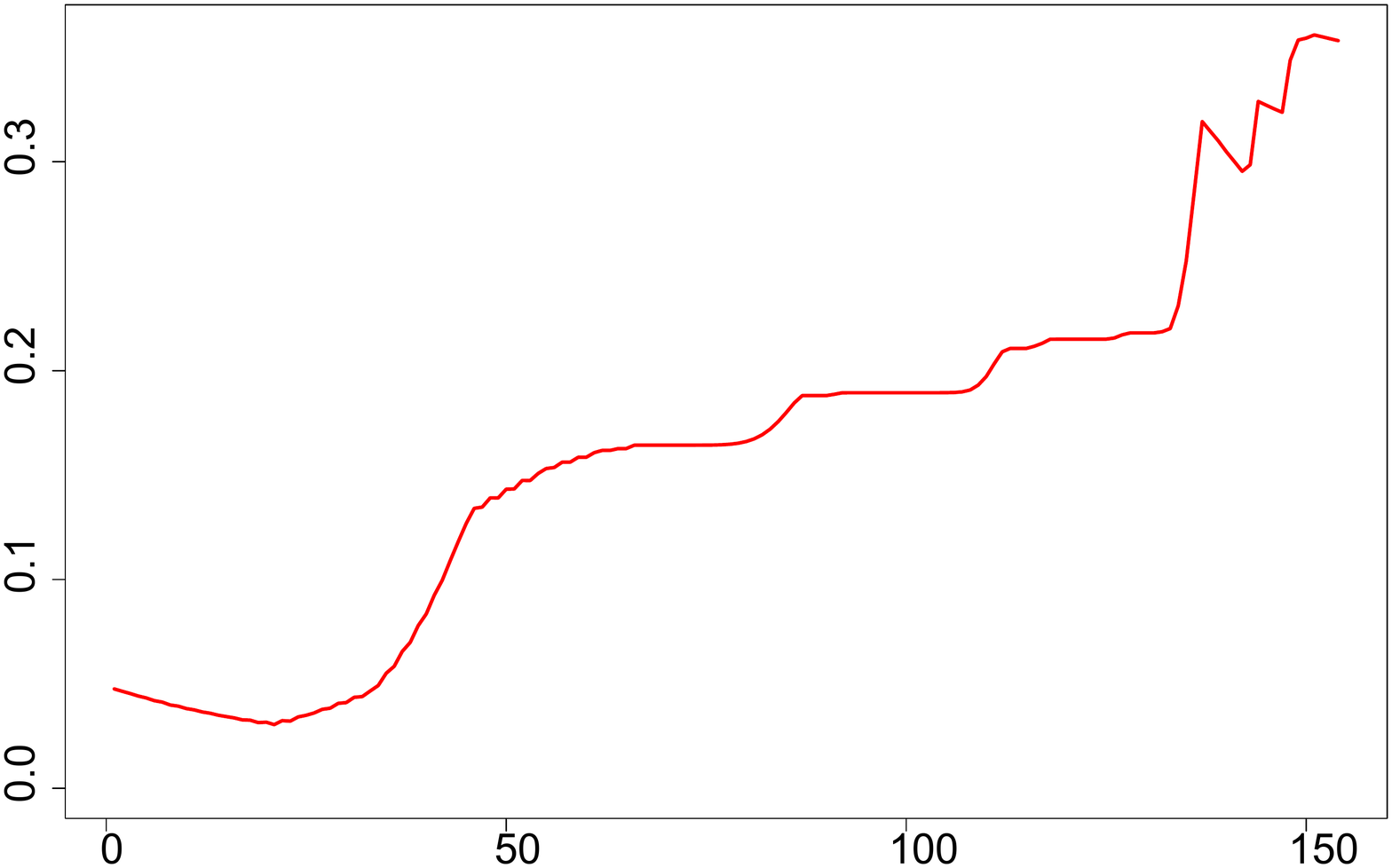}\\
\end{array}$
\caption{Oscillatory time-series with changing frequency and its  reconstructions in 2D and 3D (top), norms of persistence landscapes for $w=100,150,230$ (bottom). }
\label{fig:oscillatory}
\end{figure}

In Fig.~\ref{fig:oscillatory} we show an oscillatory signal as above with no  noise,
and its TDA signal for various sizes of the sliding window.

\section{TDA applied to the LPPLS model}
We recall that for the LPPLS model, the price of the asset oscillates with increasing frequency and decreasing amplitude about a super-exponential trend line.
When we apply the TDA procedure with a fixed window size we  observe the following:
\begin{itemize}
\item At the beginning of the time-series, the behavior is nearly periodic, and the TDA output is similar to that of  a periodic signal, producing a single,
    persistent  $1$-loop dimensional loop in the point cloud;
\item  As the time increases, the time-series exhibits more than one frequency,  and  TDA output produces
     a few  persistent $1$-dimensional loops in the point cloud;
\item  As the time approaches the critical time, the time-series exhibits oscillations of growing frequencies, and the  TDA output yields more and more  persistent  $1$-dimensional loops, which decay in size. In addition, the super-exponential trend
makes the points in the point-cloud more spread out, so the growing-in-size  `gaps' in the point-cloud contribute with additional   $1$-dimensional loops.
Consequently, the norms of the persistent landscapes spike in the proximity of the critical time.
\end{itemize}

To capture such behavior, the size of the sliding window $w$ should be suitably chosen; this depends on the length of the data set as well as on the parameters of the LPPLS model. For instance, using a larger  window size $w$ will result in   the larger loops in the earlier part of the time series to carry more weight towards the TDA signal, which will result in a spike earlier in the range as opposed to the end of the range. On the other hand,  a smaller window size $w$ will not capture the larger loops but only the smaller ones in the latter part of the time series,  which will result in a spike towards the end of the range.


In practice, one needs to explore a range of window size $w$ and observe the presence of a spike in the TDA signal that appears earlier in the range for large $w$, and shifts towards the end of the range for smaller $w$.

As an example, we consider a noisy LPPLS model for a positive bubble, with the following parameters
\begin{equation}\label {eqn:LPPLS_parameters_true}
\begin{split}
t_c= 637,\, m=3.003e-01,\, \omega= 6.889e+00,\\
A=      1.111e+01,\,
B=     -2.937e-04\\
C=   5.515e-05,
C_1=     4.372e-05,\,
C_2=   -3.362e-05.
\end{split}
\end{equation}

These parameters  corresponds to fitting the LPPLS model to the Bitcoin price between the dates 2021/03/22/--2021/04/16, sampled hourly.

We use the following TDA sliding-window size $w$, delay $d$, and embedding dimension $N$:
\begin{equation}
w= 72,\, d=5,\, N=4.
\label{eqn:TDA_parameters_1}
\end{equation}

\begin{figure}
\centering
$\begin{array}{ccc}
\includegraphics[width=0.32\textwidth]{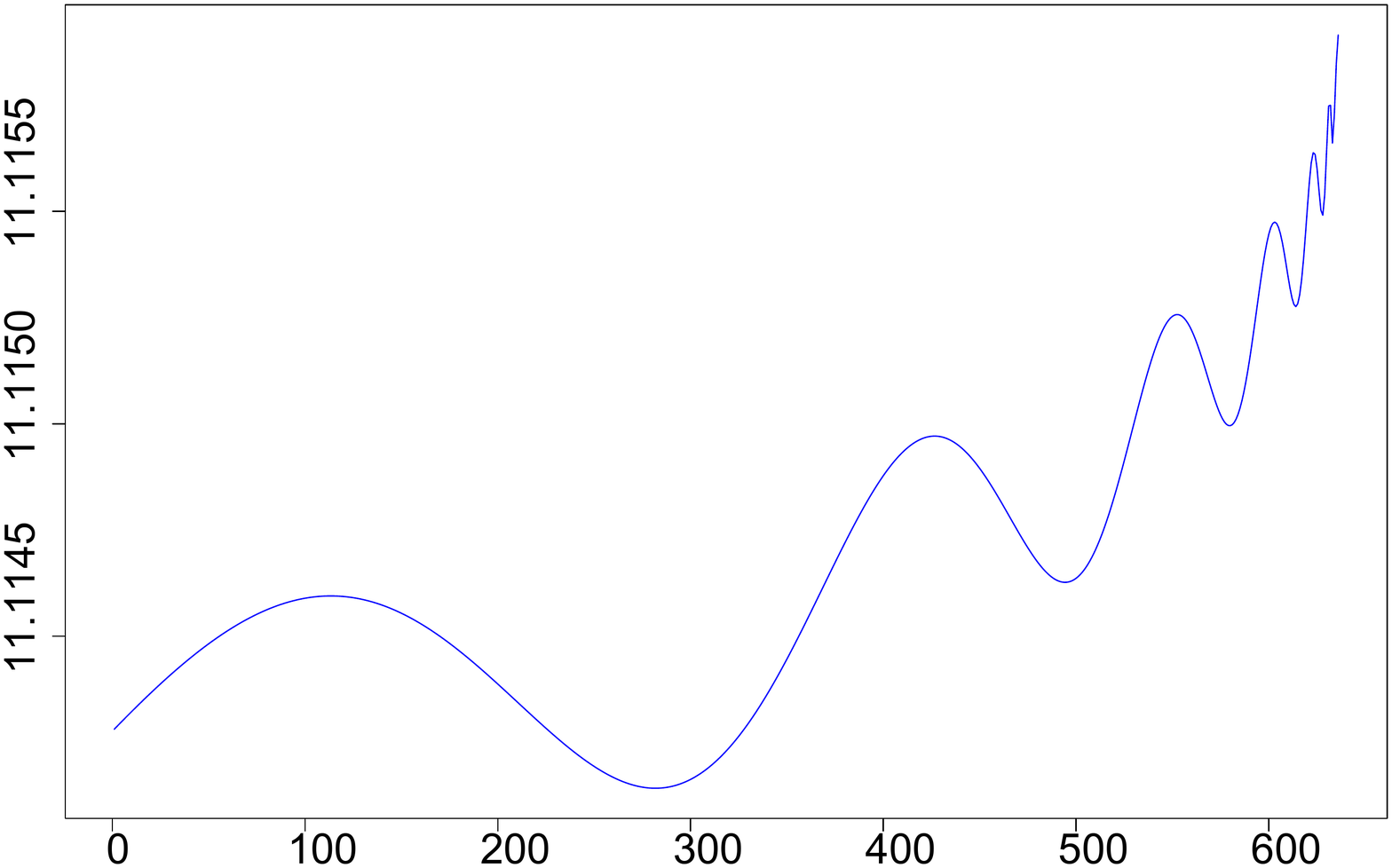}&
\includegraphics[width=0.32\textwidth]{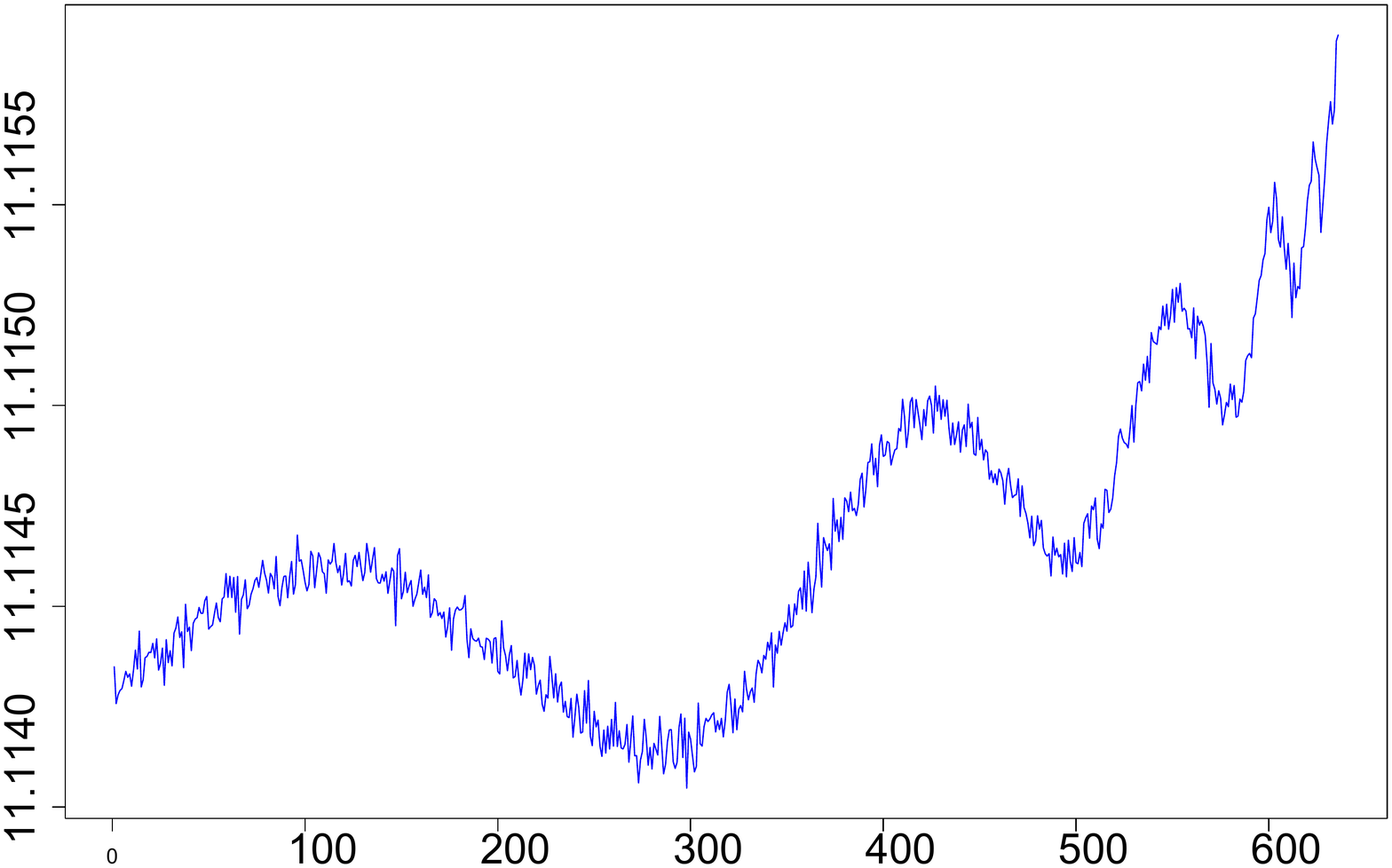}&
\includegraphics[width=0.32\textwidth]{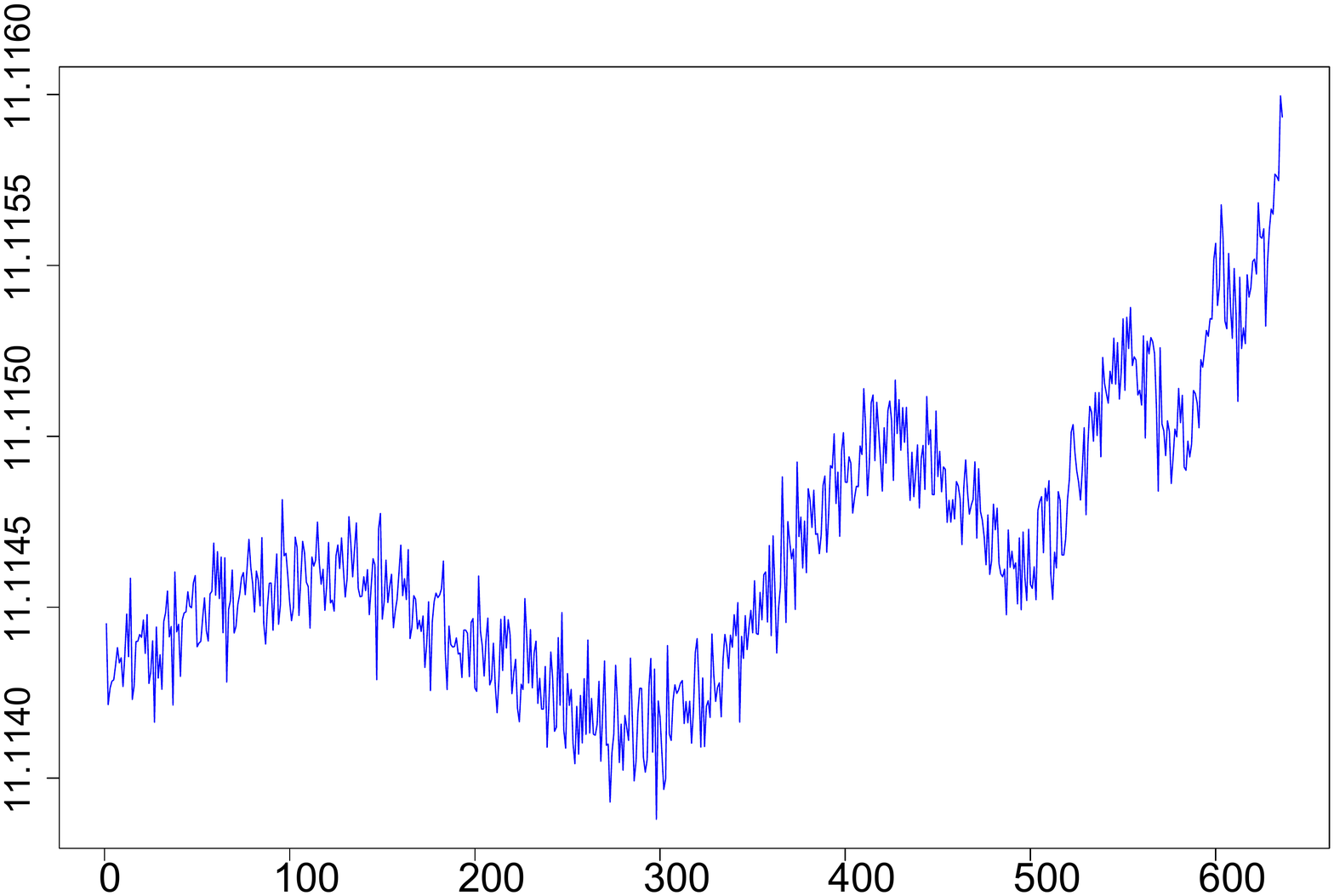}\\
\includegraphics[width=0.32\textwidth]{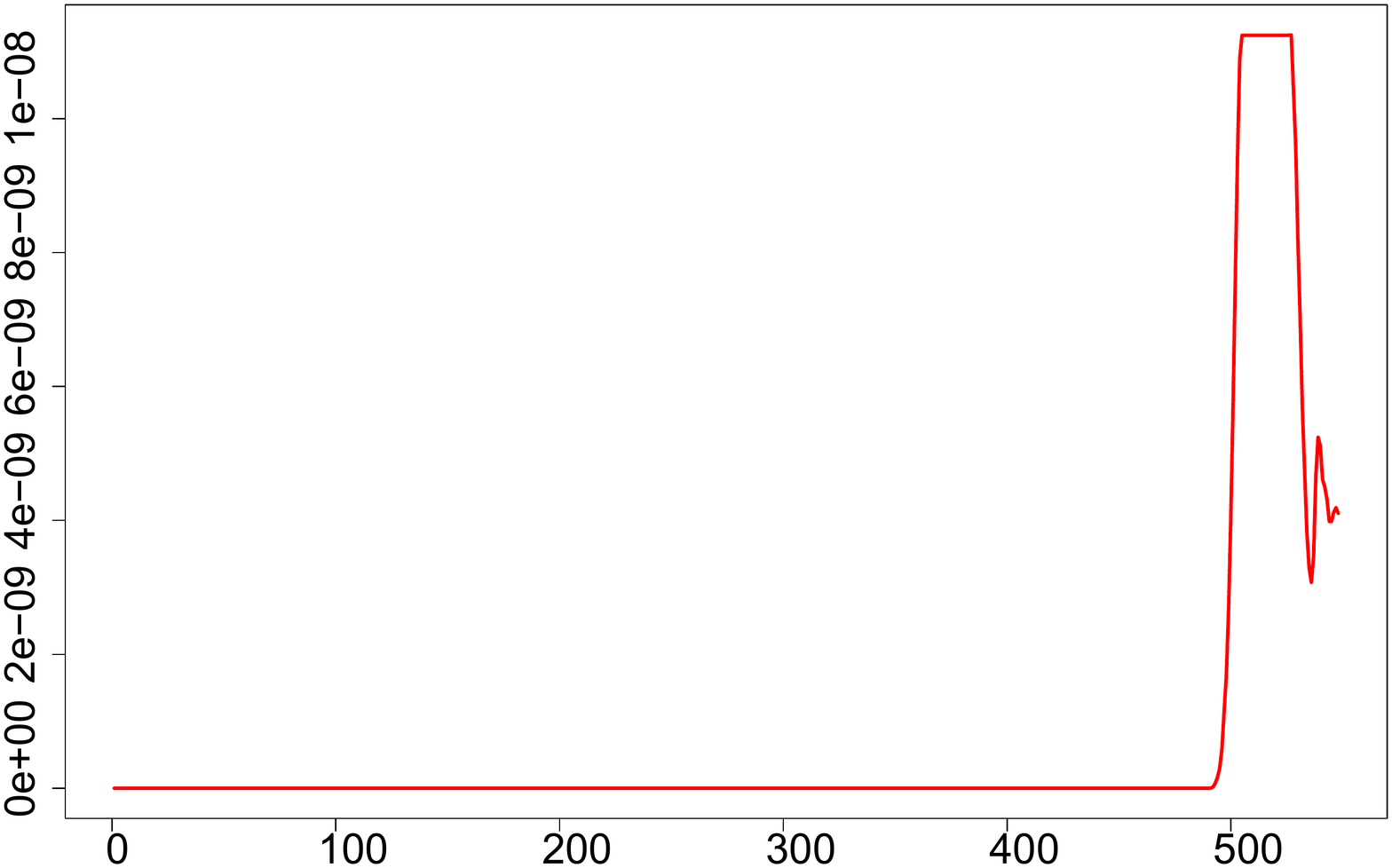}&
\includegraphics[width=0.32\textwidth]{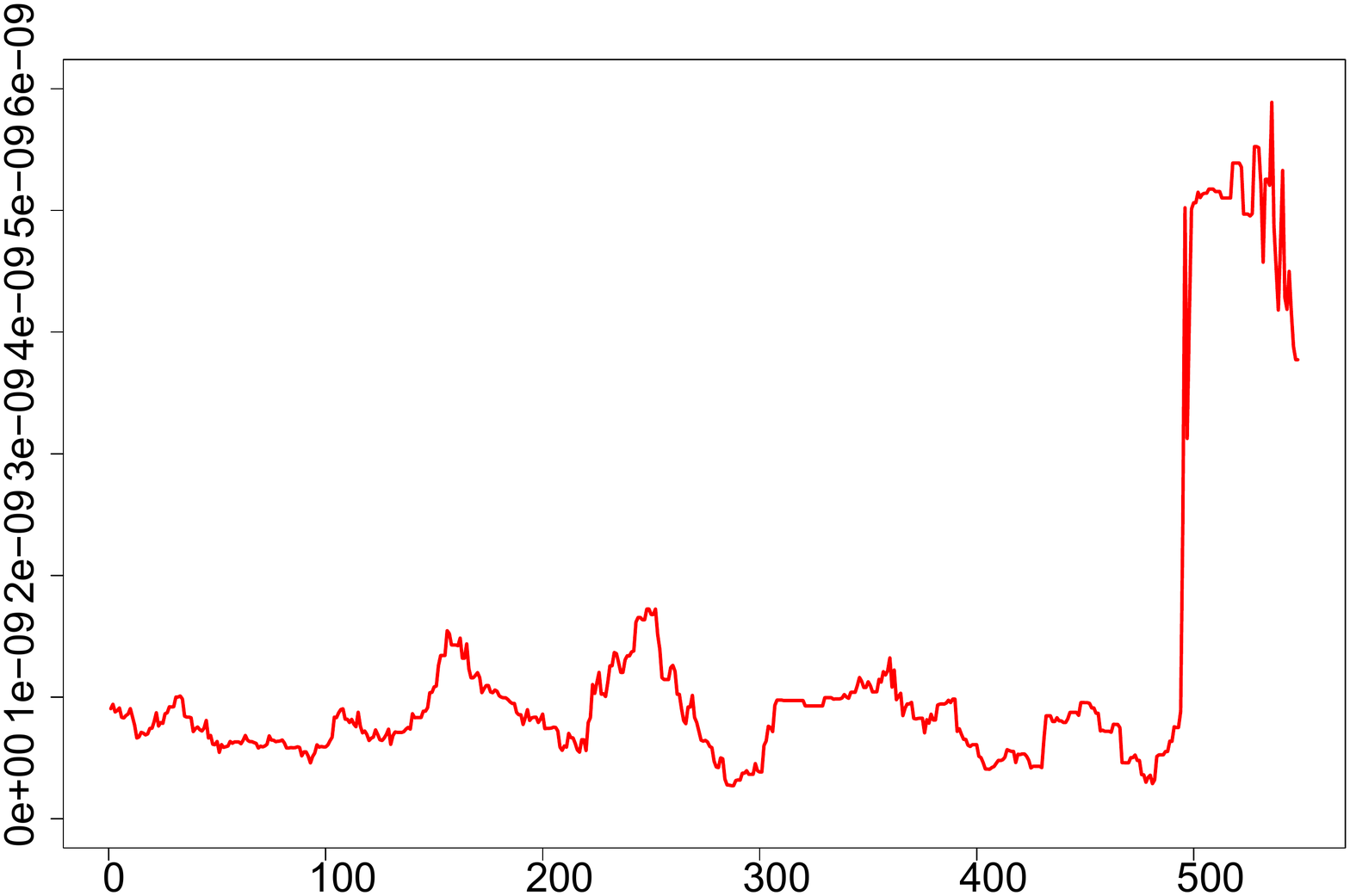}&
\includegraphics[width=0.32\textwidth]{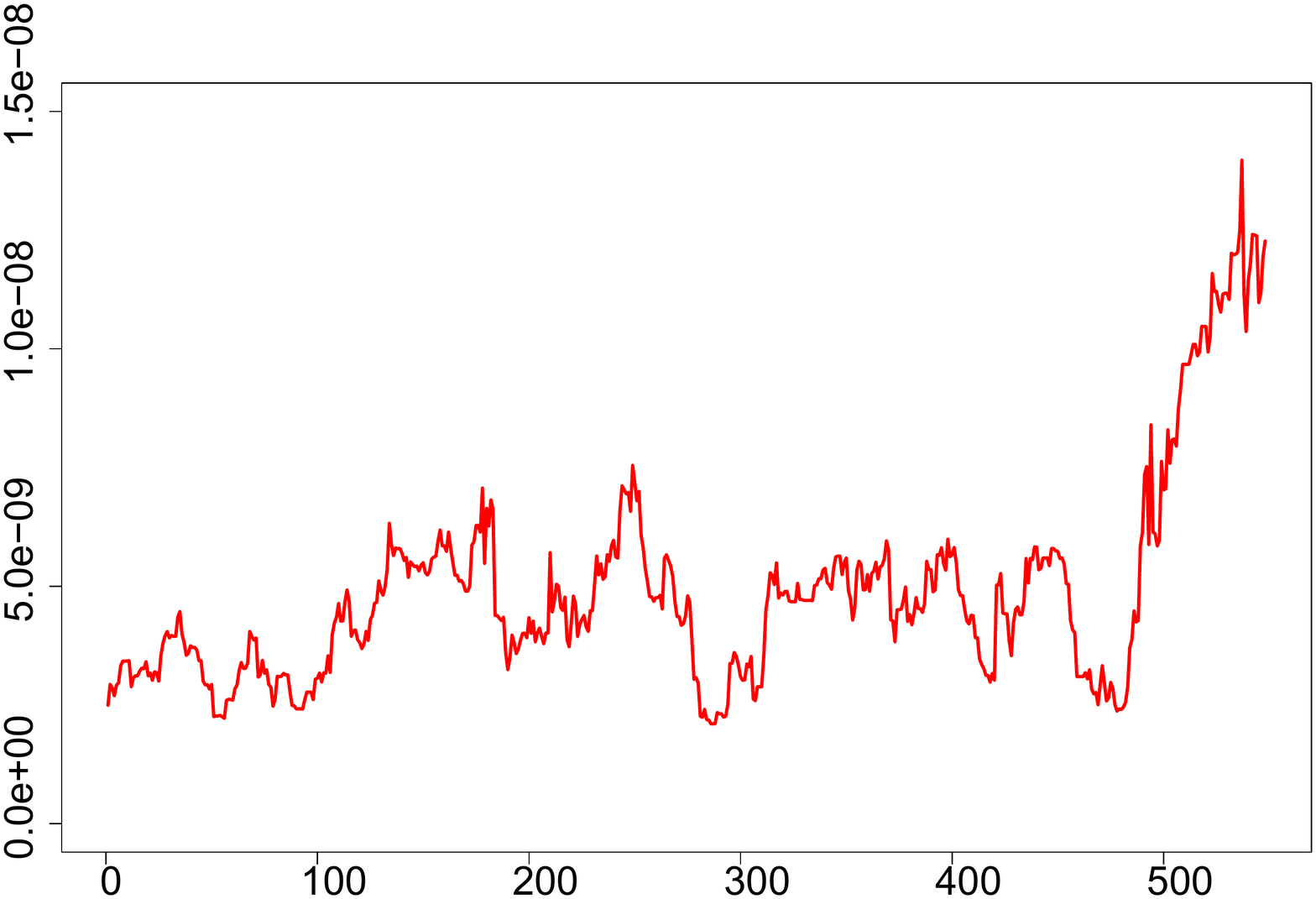}
\end{array}$
\caption{LPPLS model (top) and norm of persistence landscapes (bottom) for a positive bubble and increasing noise intensity}
\label{fig:LPPLS_model_3}\end{figure}

The TDA output for various levels of added noise is shown in Fig.~\ref{fig:LPPLS_model_3}.
We observe a peak in the TDA signal prior to the critical time, and note that the growing trend towards the peak appears earlier when the noise intensity is larger.

\subsection{Dependence of the TDA signal on parameters}
\label{sec:dependence}
We consider synthetic time series generated by \eqref{eqn:LPPLS_expected}, of $200$ data points and with $t_c=200$.
We start with the following model parameters
$m=    0.3$, $\omega=    6.7$, $A=    11$, $B=    -0.0003$, $C_1=   0.000044$, $C_2=   -0.000034$,
and with a sliding widow  of size $w=48$.
These parameters are similar to the ones in \eqref{eqn:LPPLS_parameters_true}, except for the length  of the time series, which is chosen shorter for computational efficiency.
In the experiments below we explore the changes in the TDA signal when we change the parameters $w,\omega,m,B,C_1,C_2$,
one at a time.
A  more detailed explorations would be needed to study the interdependence among these parameters, which appears to be quite intricate.

\subsubsection{Dependence on sliding window size $w$.}
\begin{figure}
\centering
$\begin{array}{cc}
\includegraphics[width=0.32\textwidth]{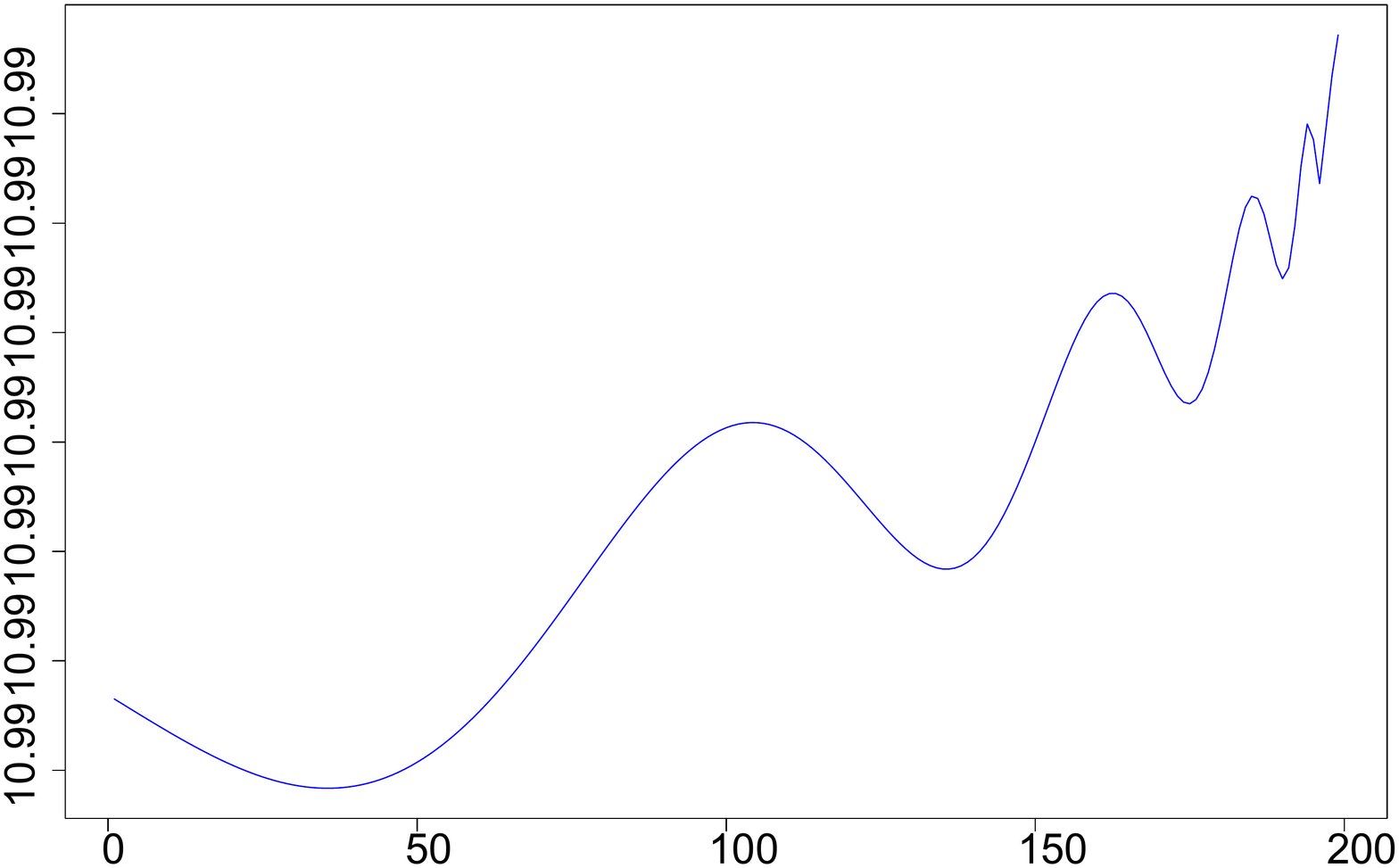} &\includegraphics[width=0.32\textwidth]{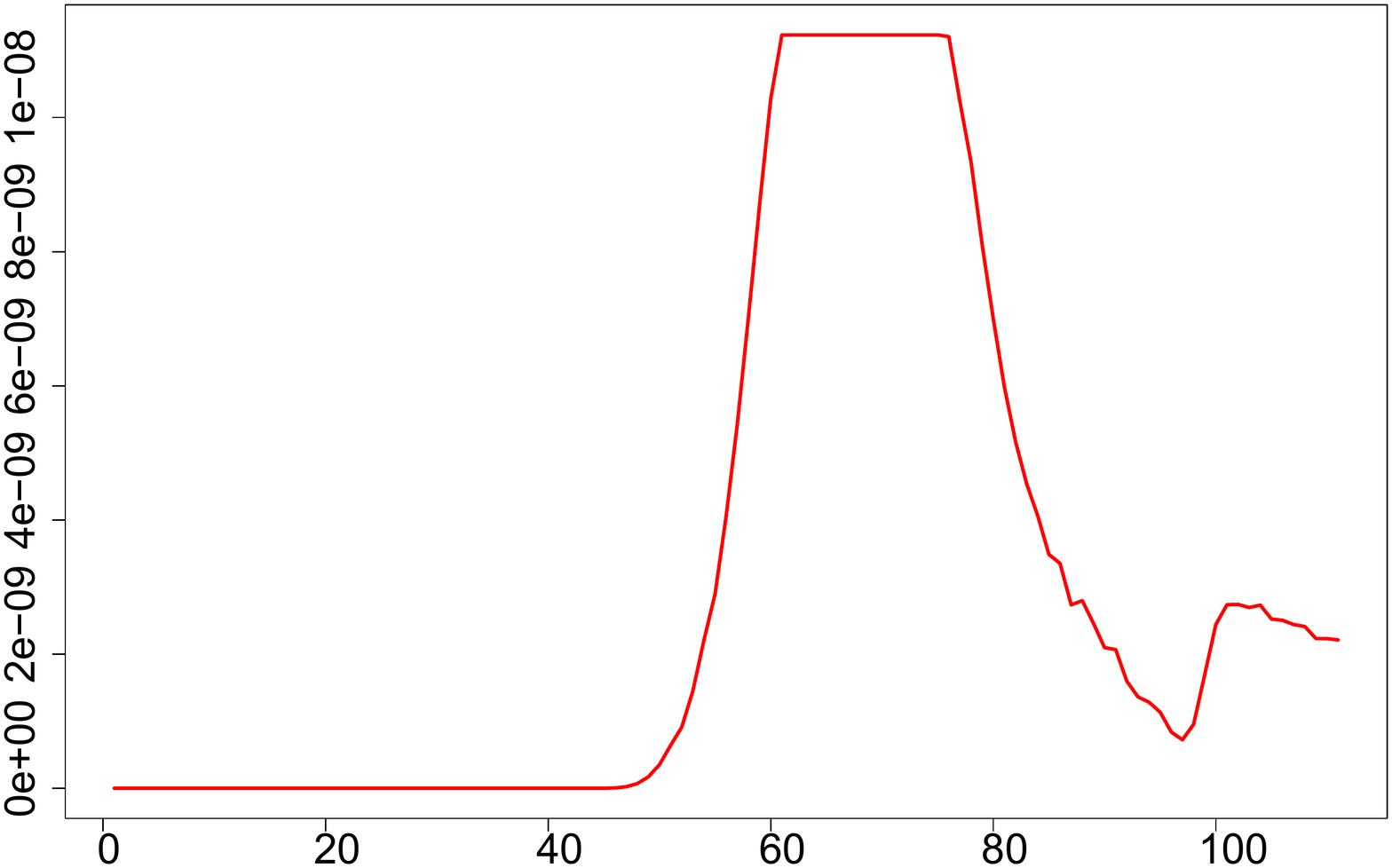}\\
\includegraphics[width=0.32\textwidth]{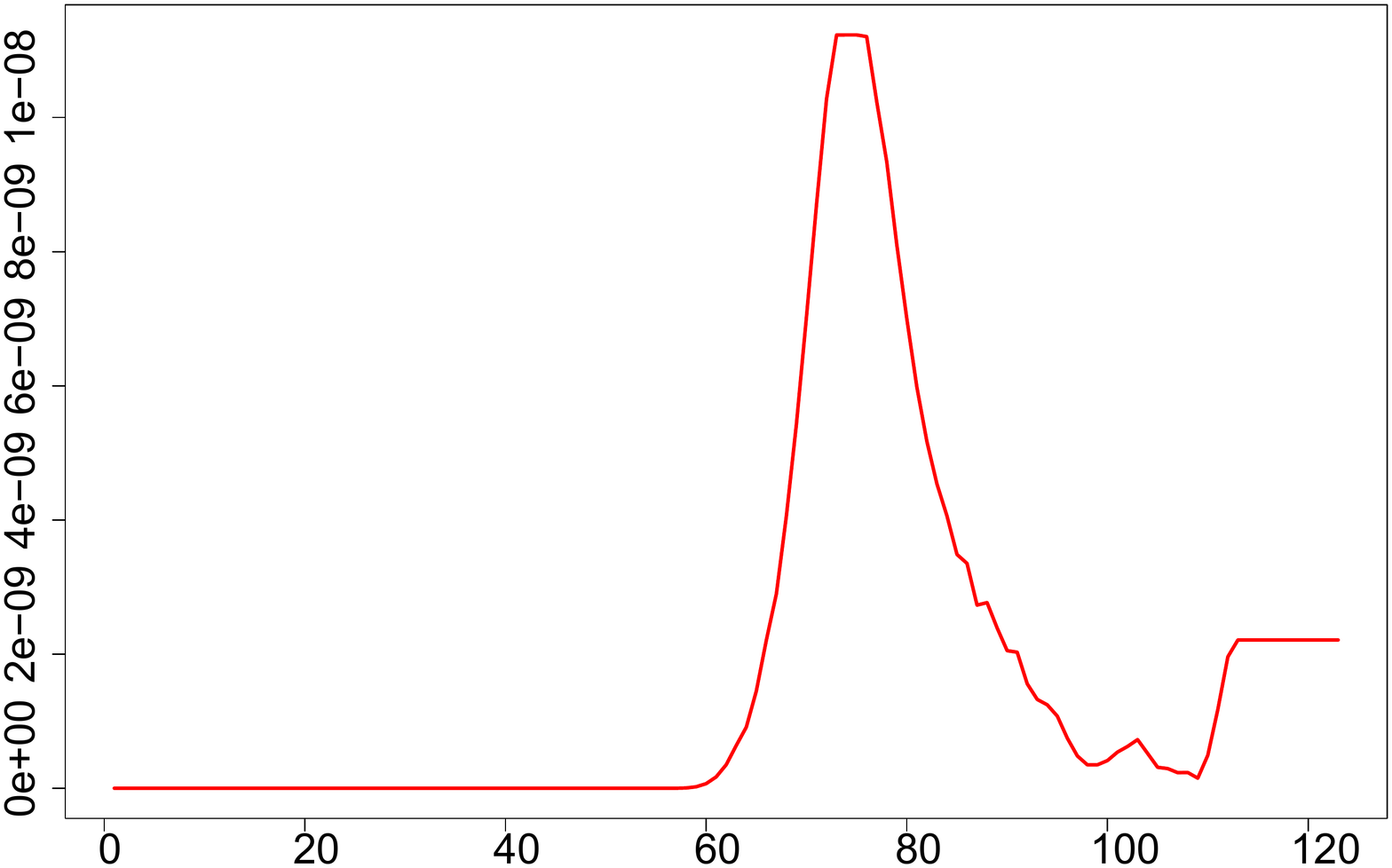}&\includegraphics[width=0.32\textwidth]{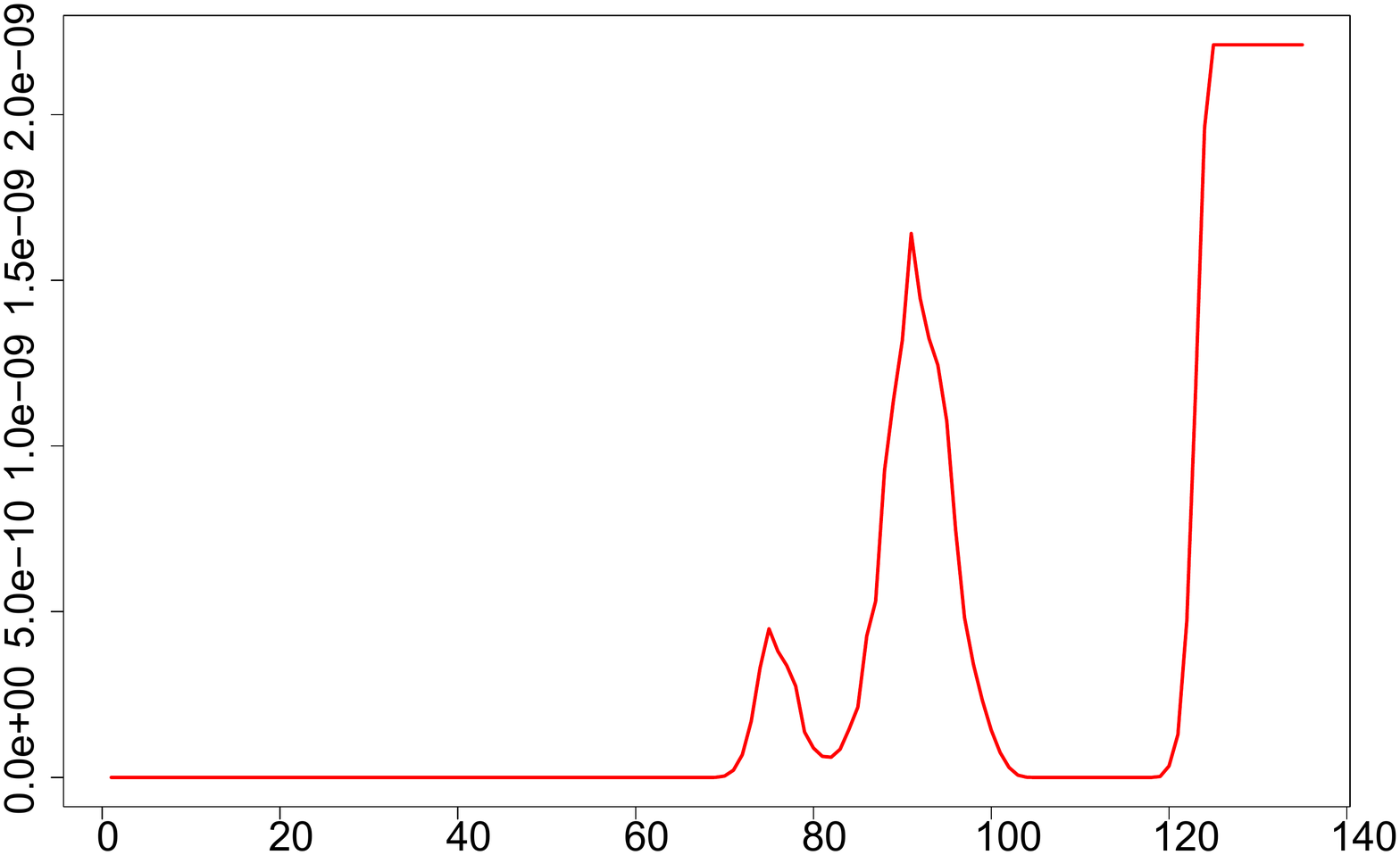}
\end{array}$
\caption{Synthetic LPPLS signal  and corresponding TDA signal for window sizes $w= 72, 60, 48$. }
\label{fig:LPPLS_model synthetic_1}
\end{figure}

The experiments displayed in Fig.~\ref{fig:LPPLS_model synthetic_1} show that as the window size is decreased, the peak of the TDA signal gets shifted towards the end of the range of the time series.
This is because  windows of larger size $w$ capture  more of the larger loops in the point-cloud, while windows of smaller  size capture more of the smaller loops in the point-cloud.
This is consistent with the remarks in Section \ref{sec:TDA_quasi}.

\subsubsection{Dependence on frequency $\omega$}
\begin{figure}
\centering
$\begin{array}{ccc}
\includegraphics[width=0.32\textwidth]{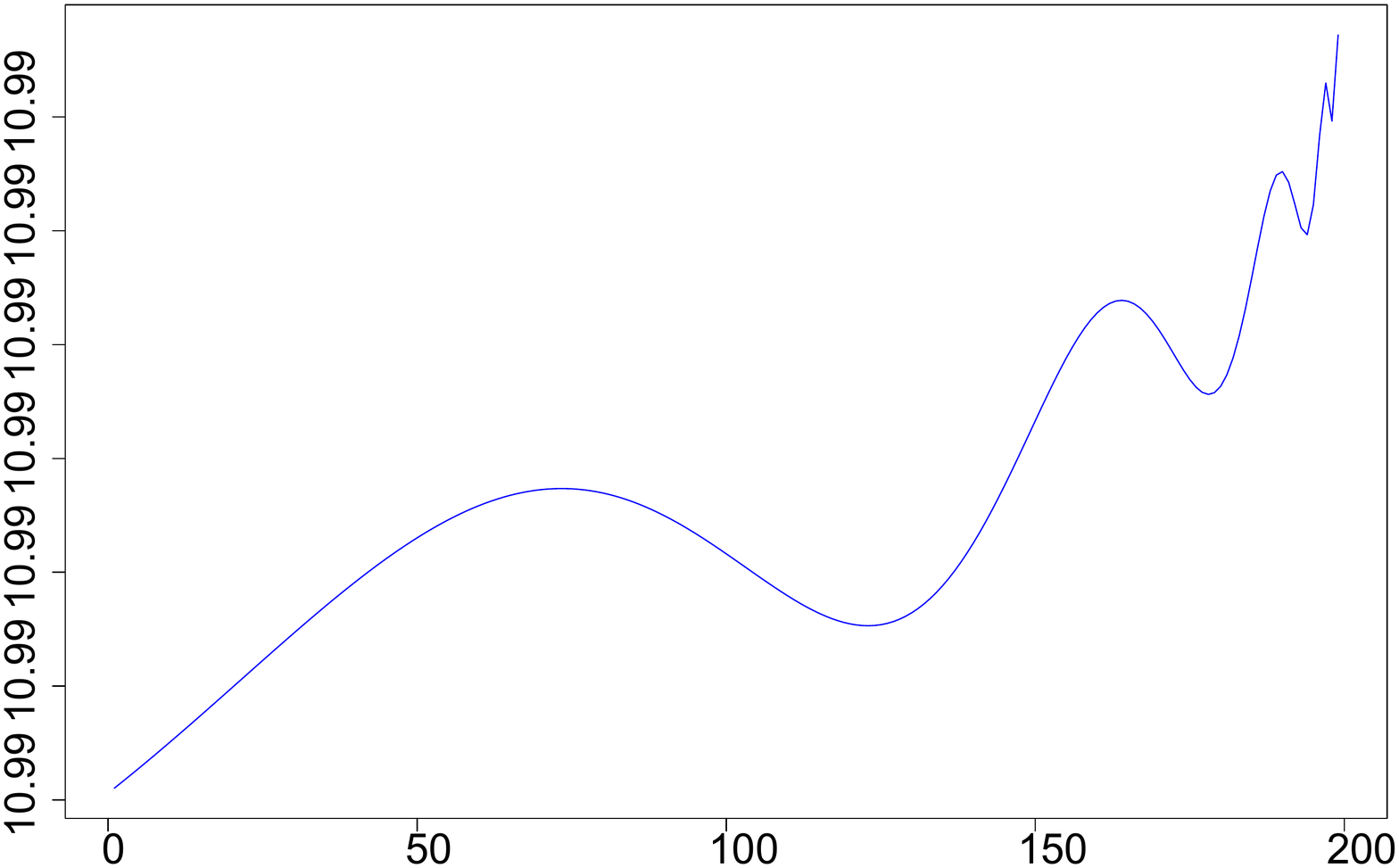}&
\includegraphics[width=0.32\textwidth]{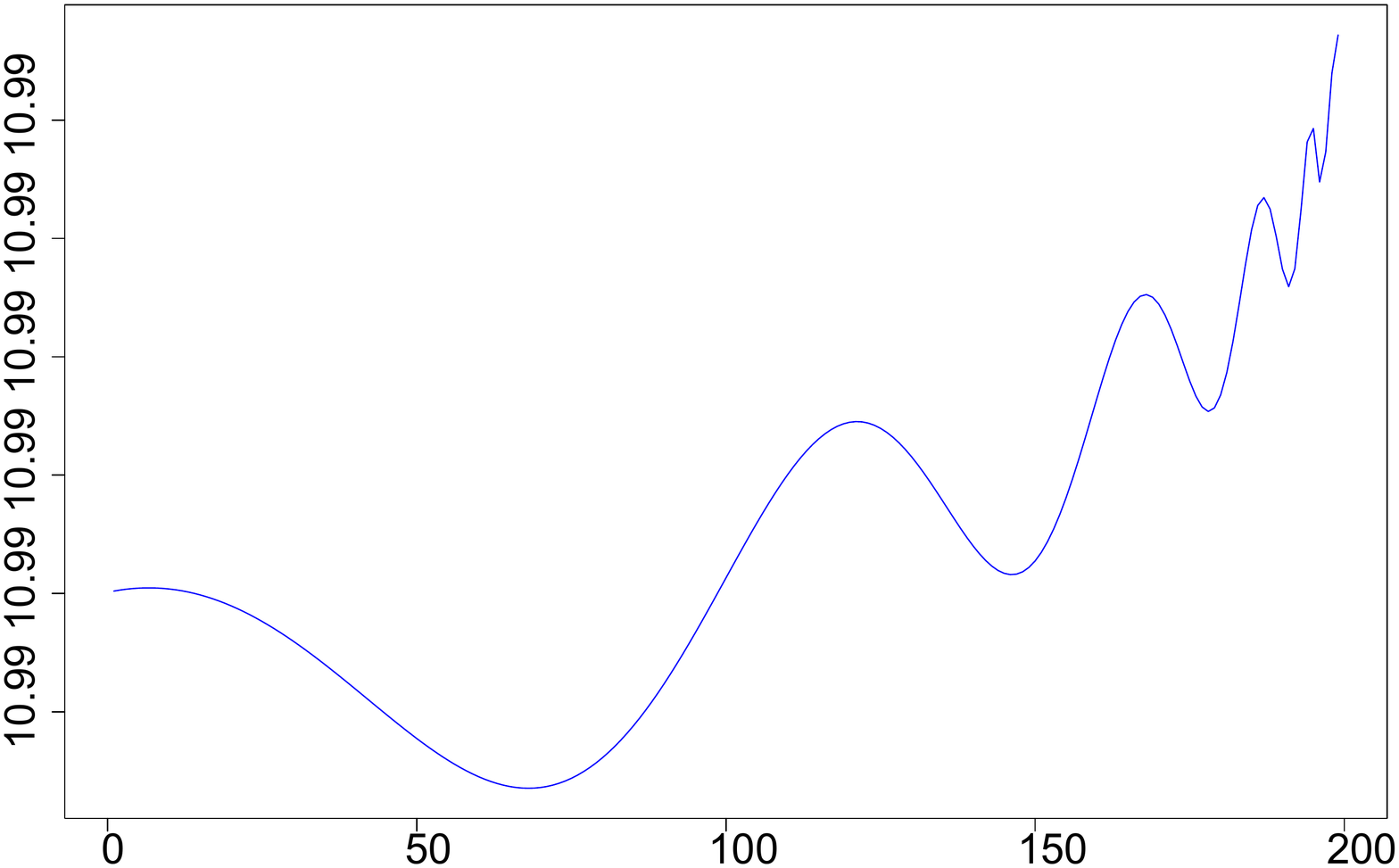}&
\includegraphics[width=0.32\textwidth]{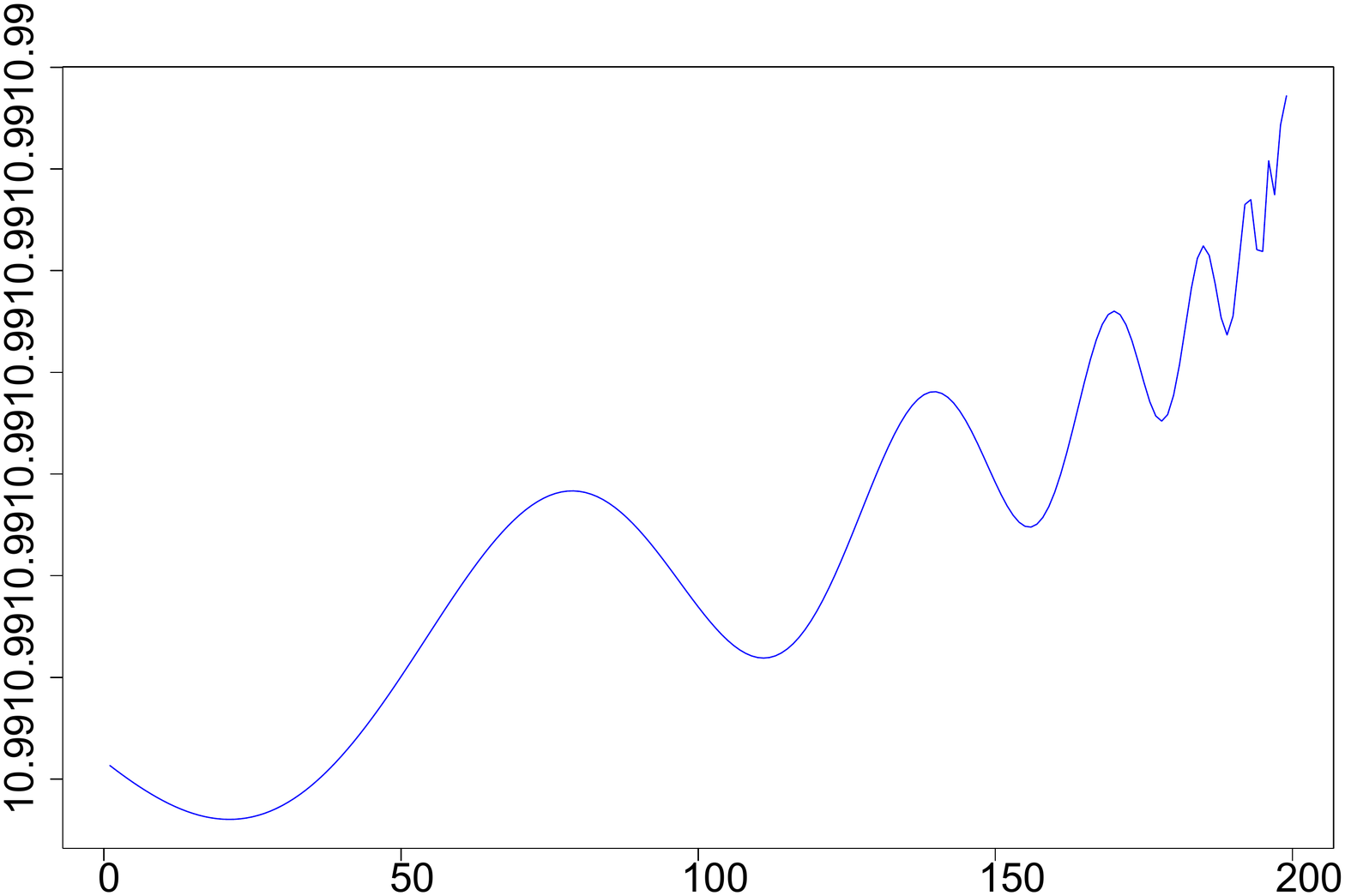}\\
\includegraphics[width=0.32\textwidth]{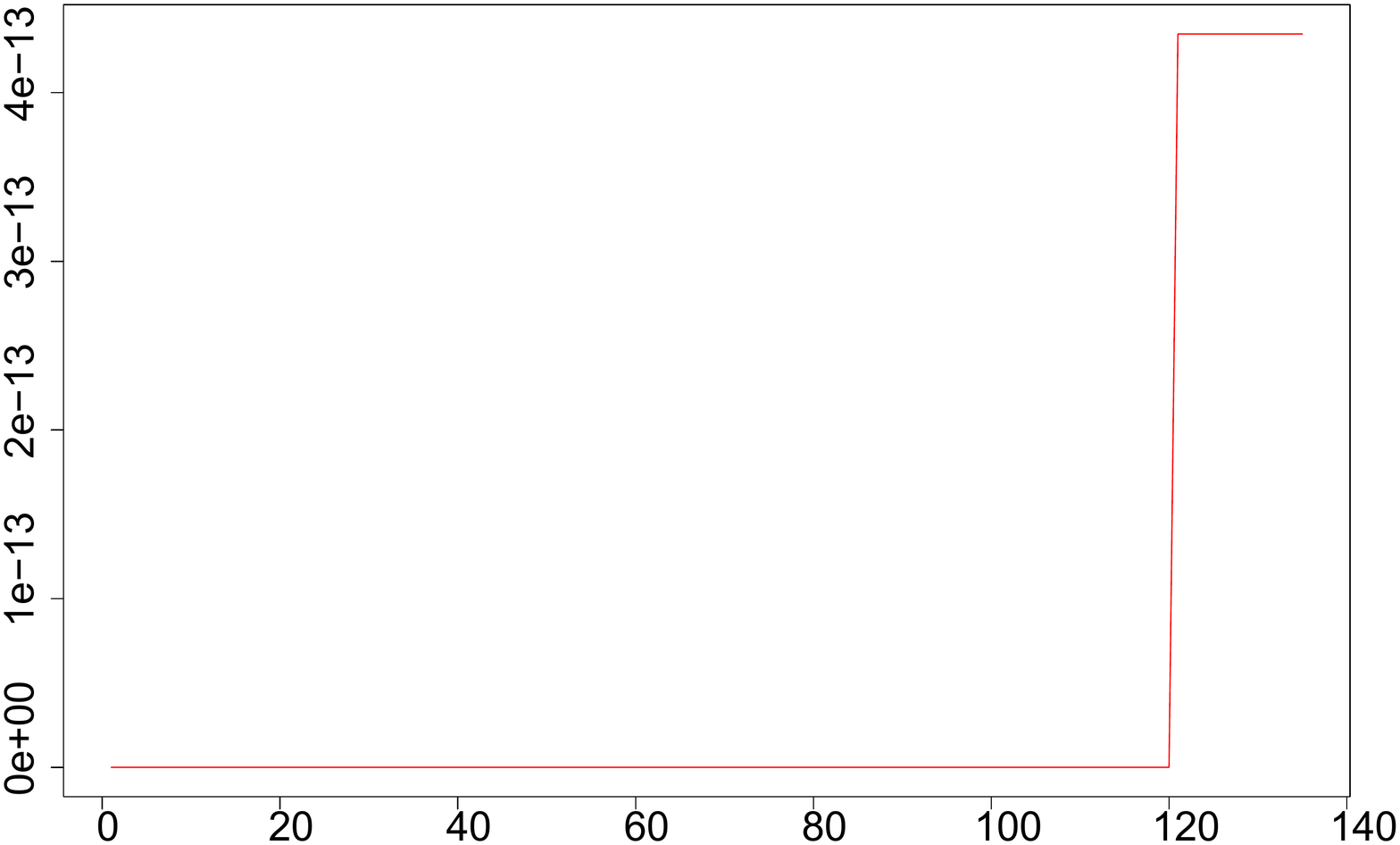}&
\includegraphics[width=0.32\textwidth]{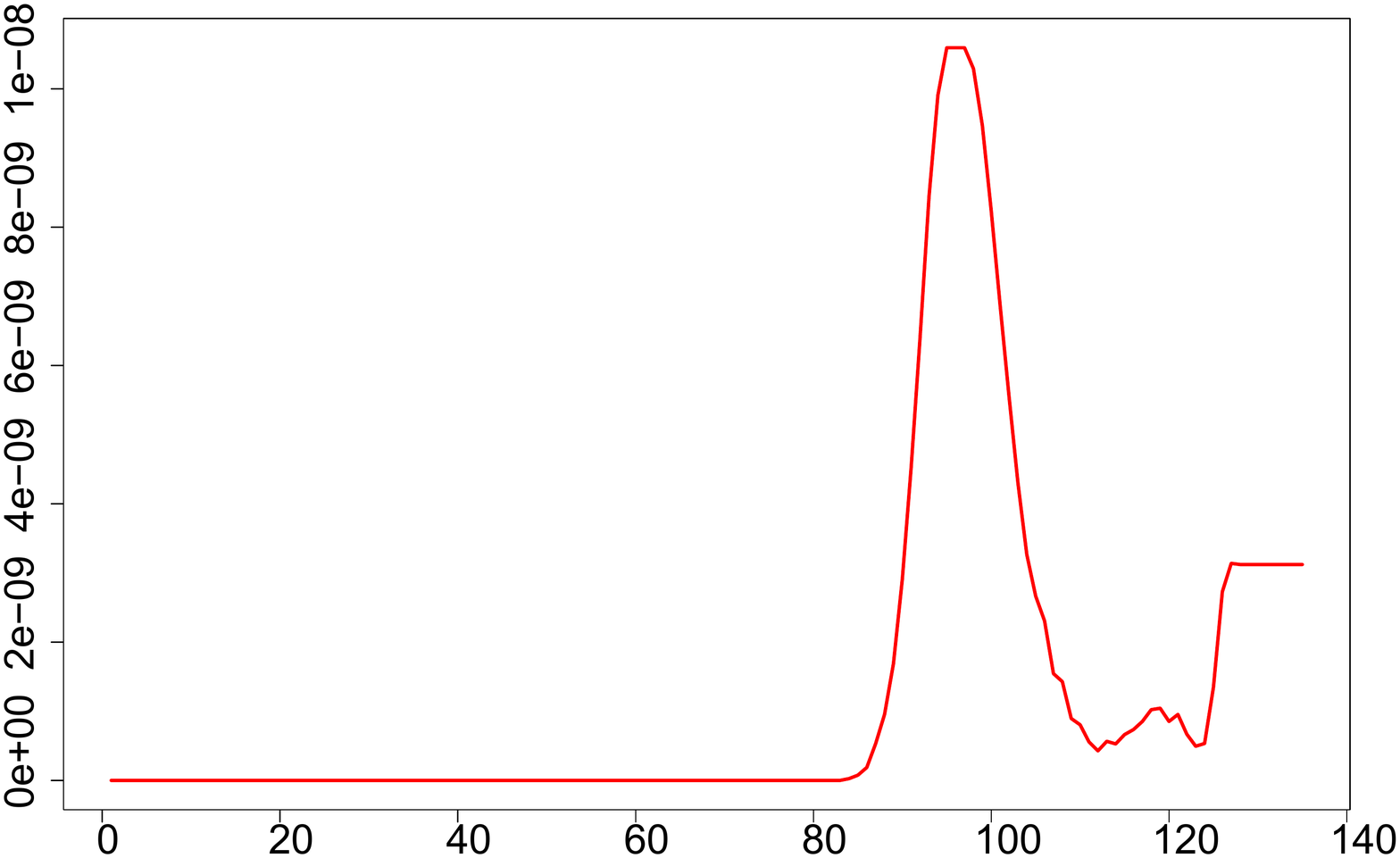}&
\includegraphics[width=0.32\textwidth]{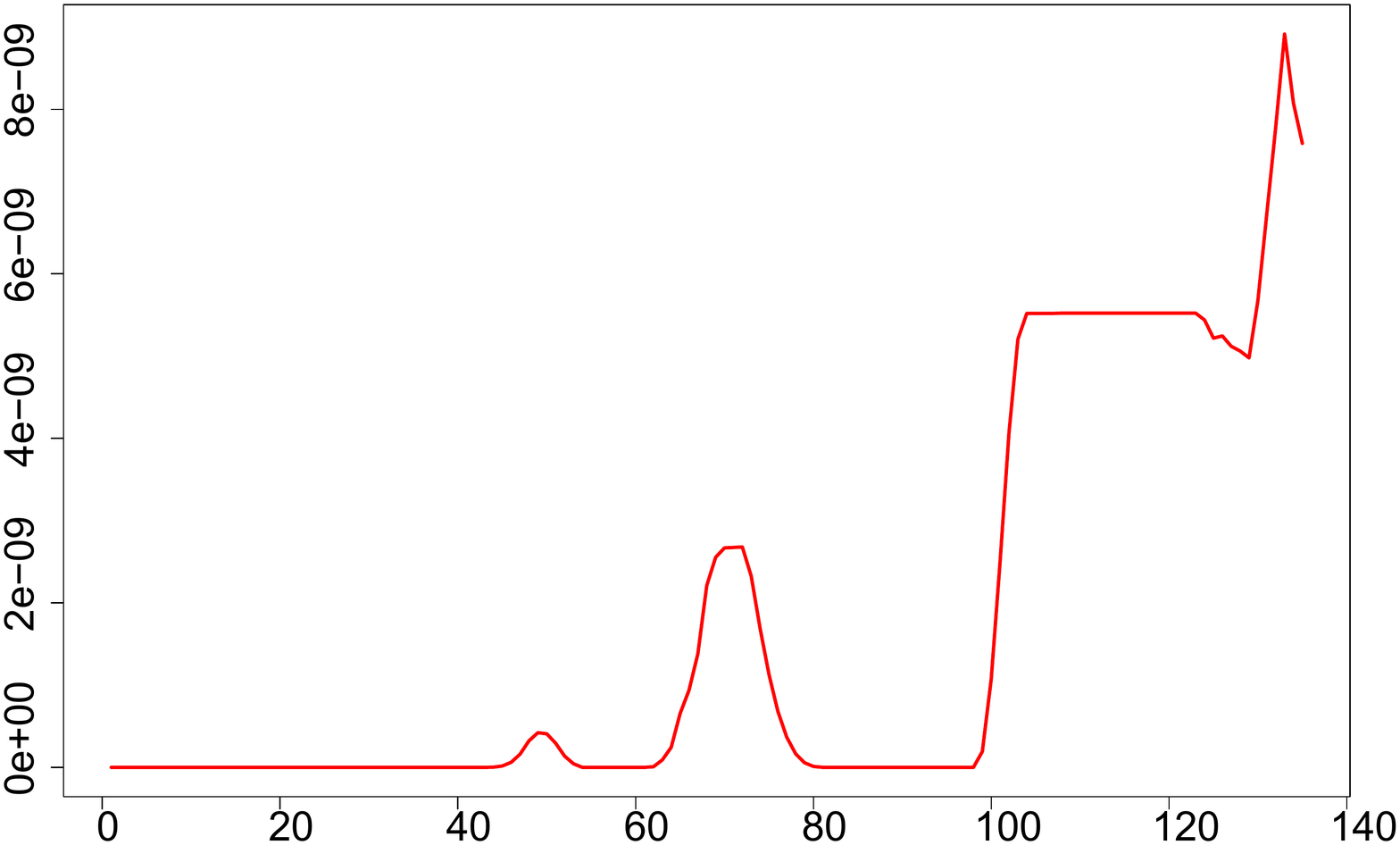}
\end{array}$
\caption{Synthetic LPPLS signal (top) and corresponding TDA signal (bottom) for $\omega= 5, 7, 9$. }
\label{fig:LPPLS_model synthetic_2}
\end{figure}

The experiments displayed in Fig. \ref{fig:LPPLS_model synthetic_2}  show that as the frequency $\omega$ is increased, the TDA signal increases in strength.

\subsubsection{Dependence on non-linearity parameter $m$}
\begin{figure}
\centering
$\begin{array}{ccc}
\includegraphics[width=0.32\textwidth]{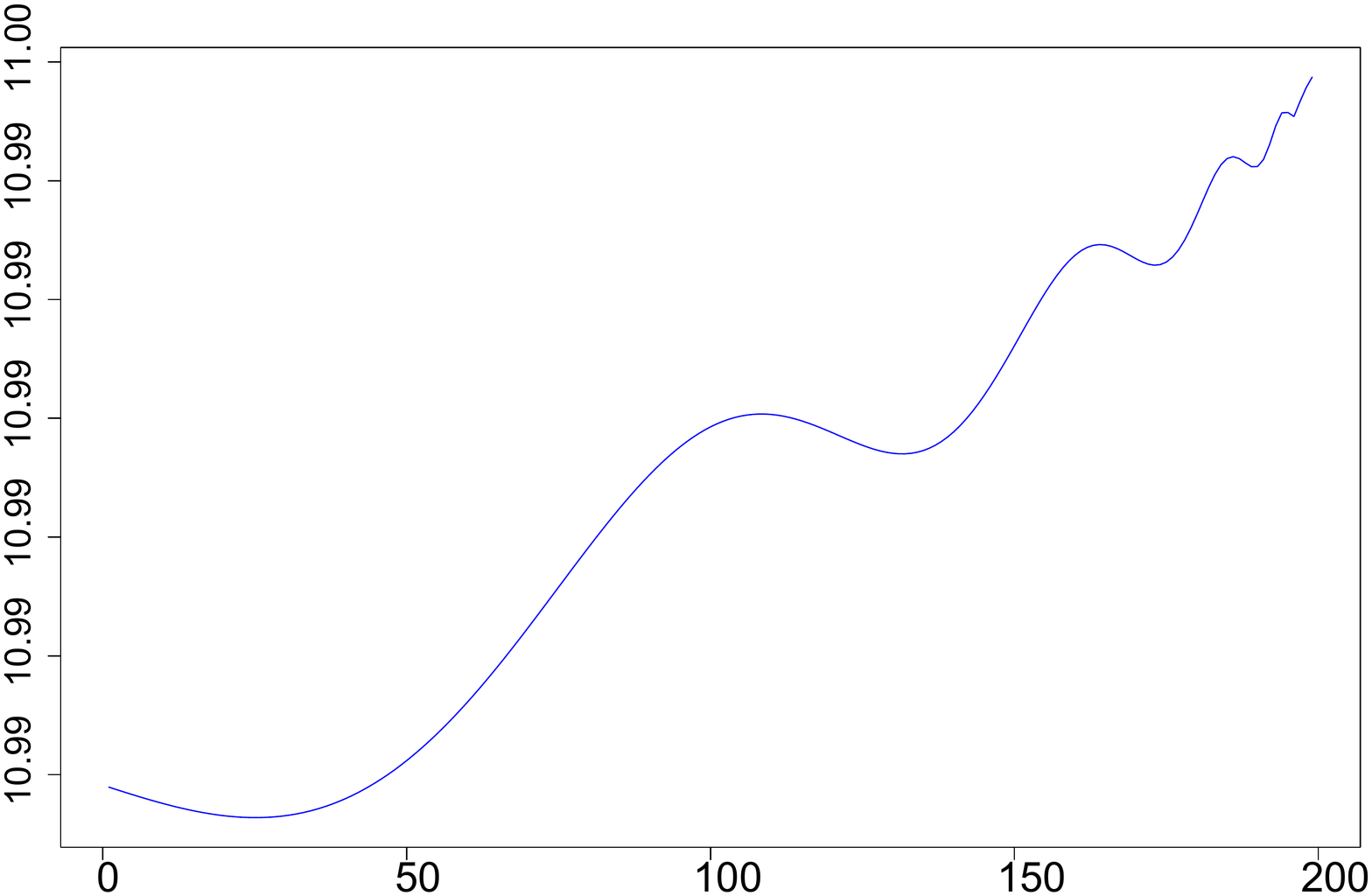}&
\includegraphics[width=0.32\textwidth]{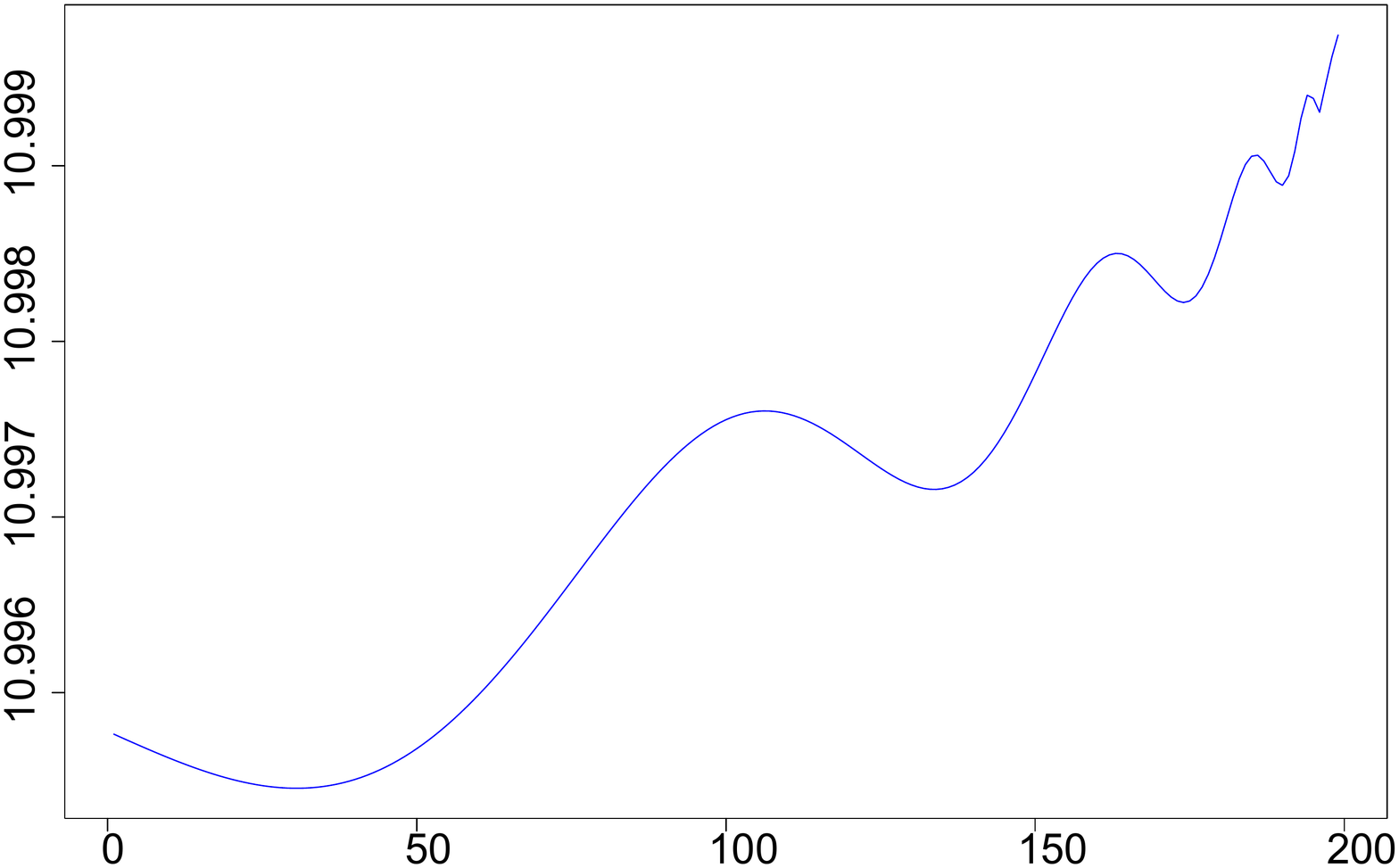}&
\includegraphics[width=0.32\textwidth]{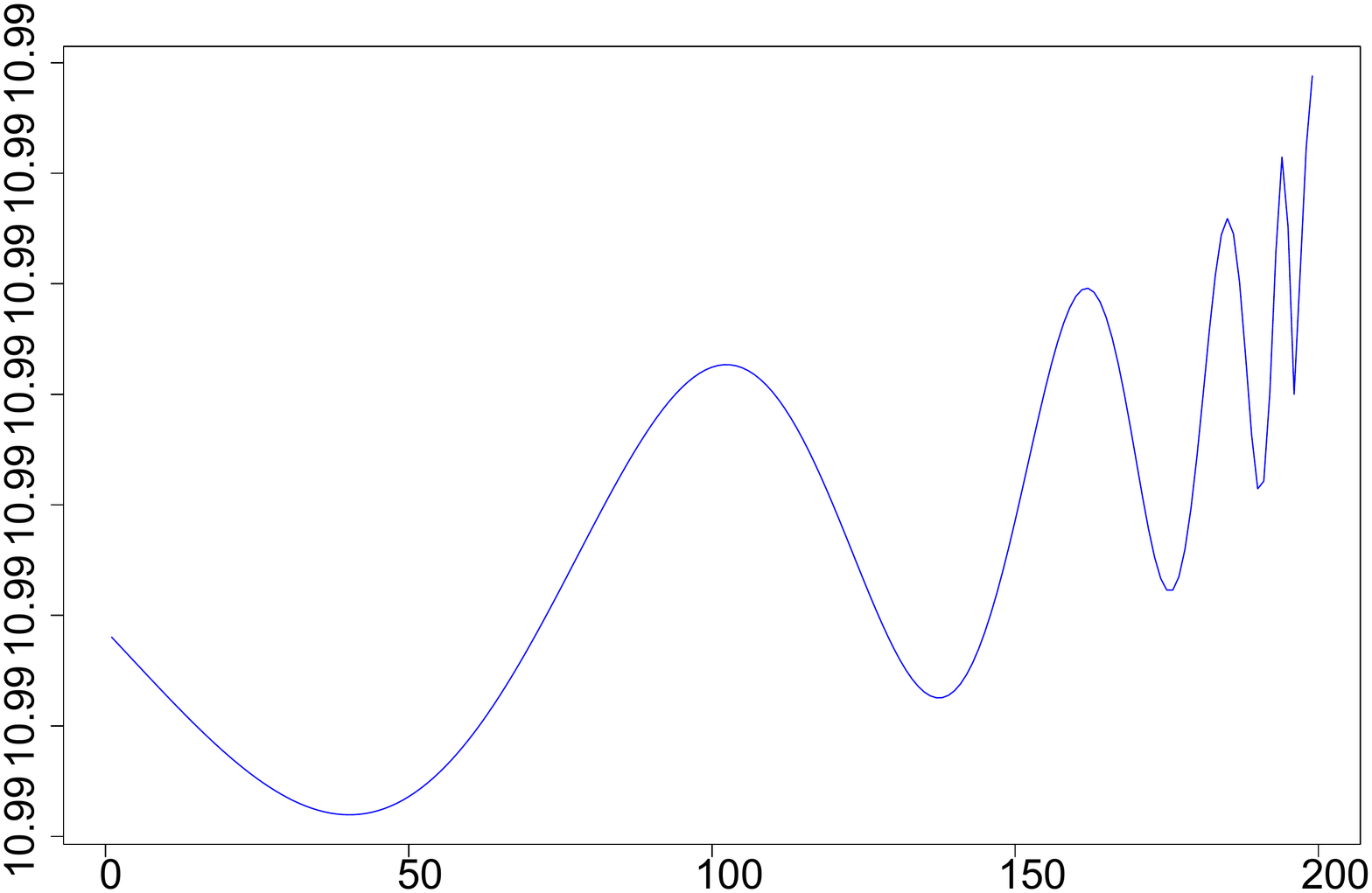}\\
\includegraphics[width=0.32\textwidth]{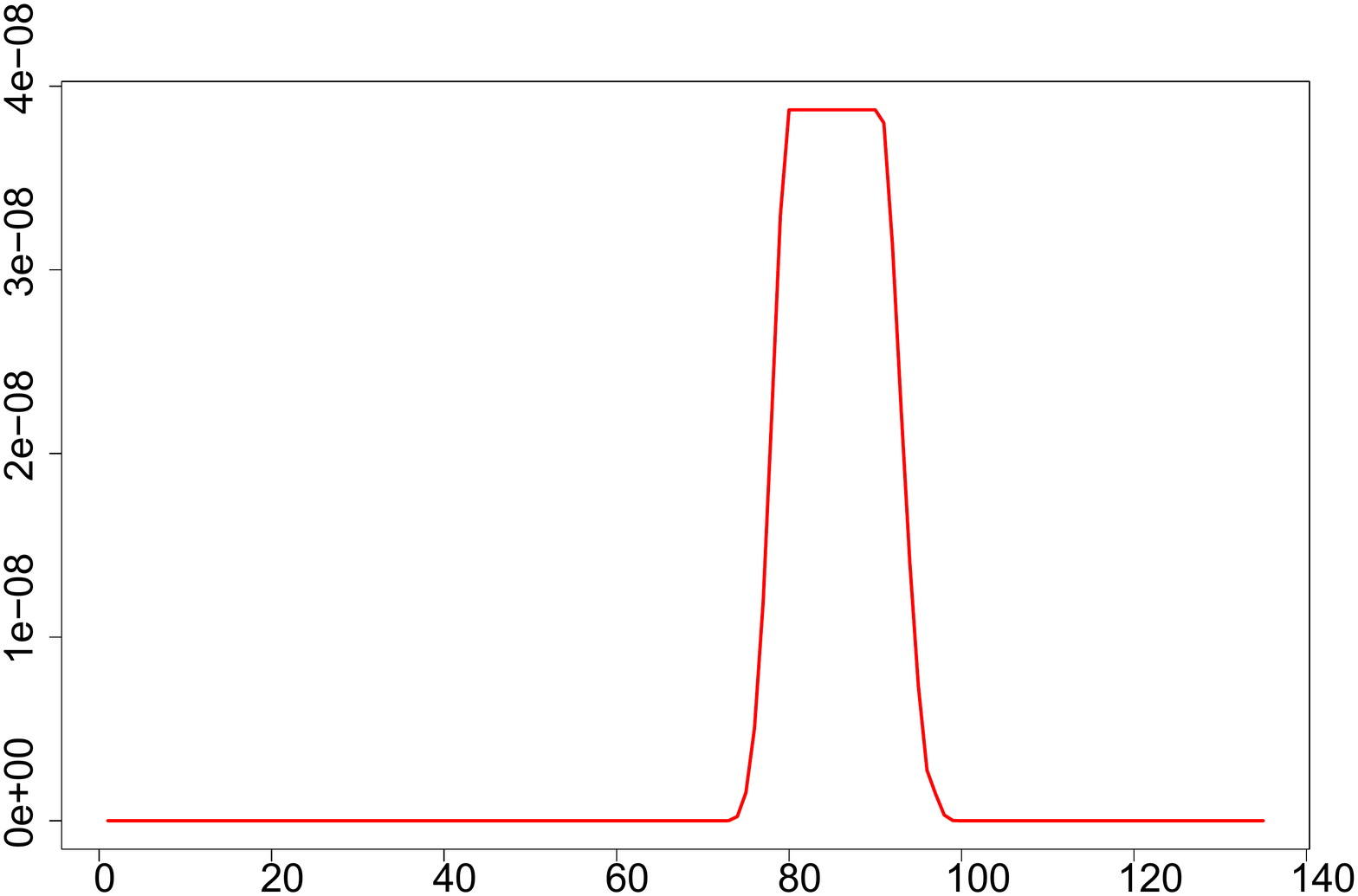}&
\includegraphics[width=0.32\textwidth]{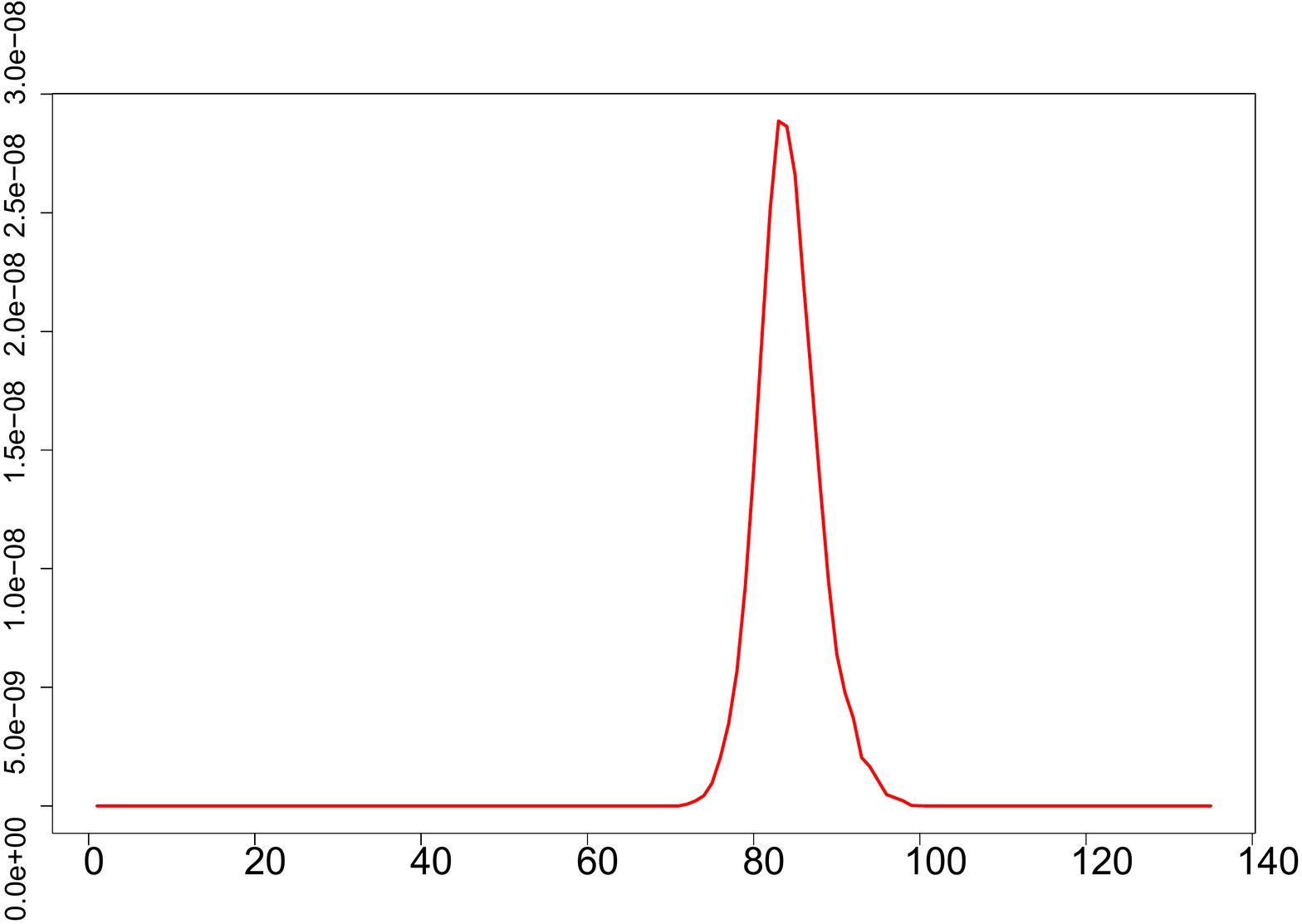}&
\includegraphics[width=0.32\textwidth]{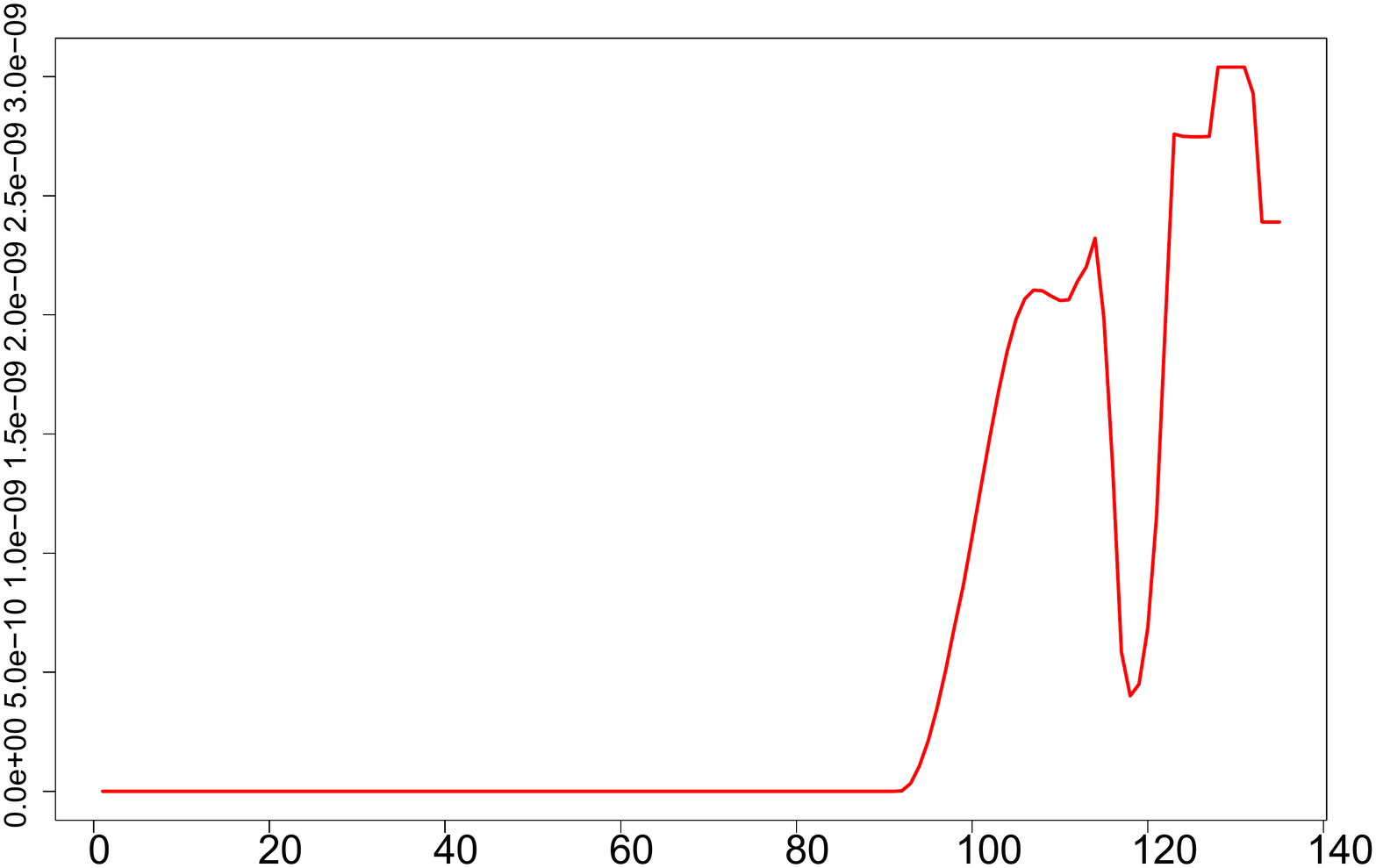}
\end{array}$
\caption{Synthetic LPPLS signal  (top) and corresponding TDA signal (bottom) for $m=0.7, 0.5, 0.1$ }\label{fig:LPPLS_model synthetic_3}
\end{figure}

We note   that when $m\approx 1$,  the main trend line of the log price is nearly linear.
The experiments displayed in Fig.~\ref{fig:LPPLS_model synthetic_3} show that as the frequency $m$ is decreased, the TDA signal gets shifted towards the end of the range.

\subsubsection{Dependence of positive/negative bubble parameter $B$}
\begin{figure}
\centering
$\begin{array}{cc}
\includegraphics[width=0.32\textwidth]{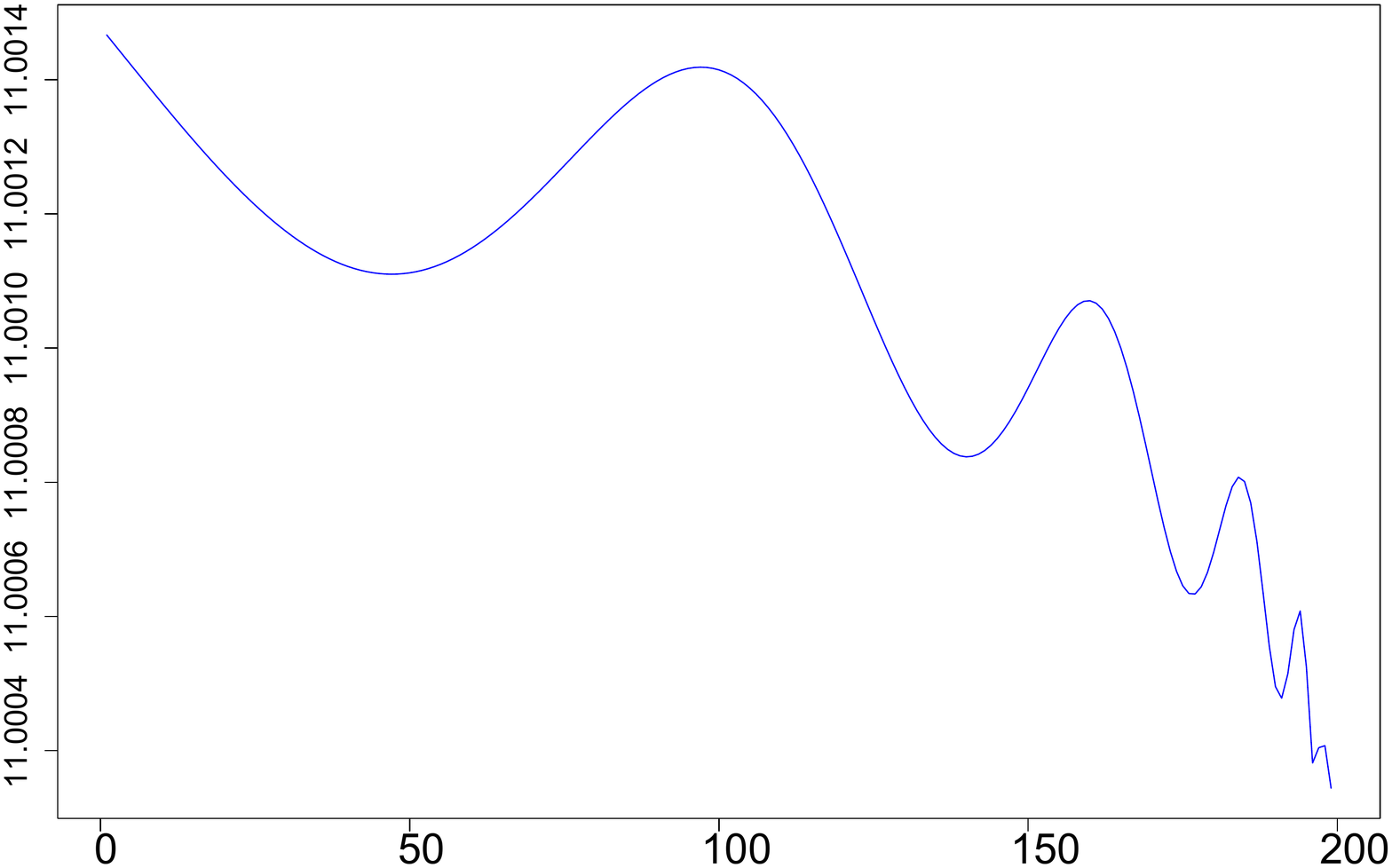}
&\includegraphics[width=0.32\textwidth]{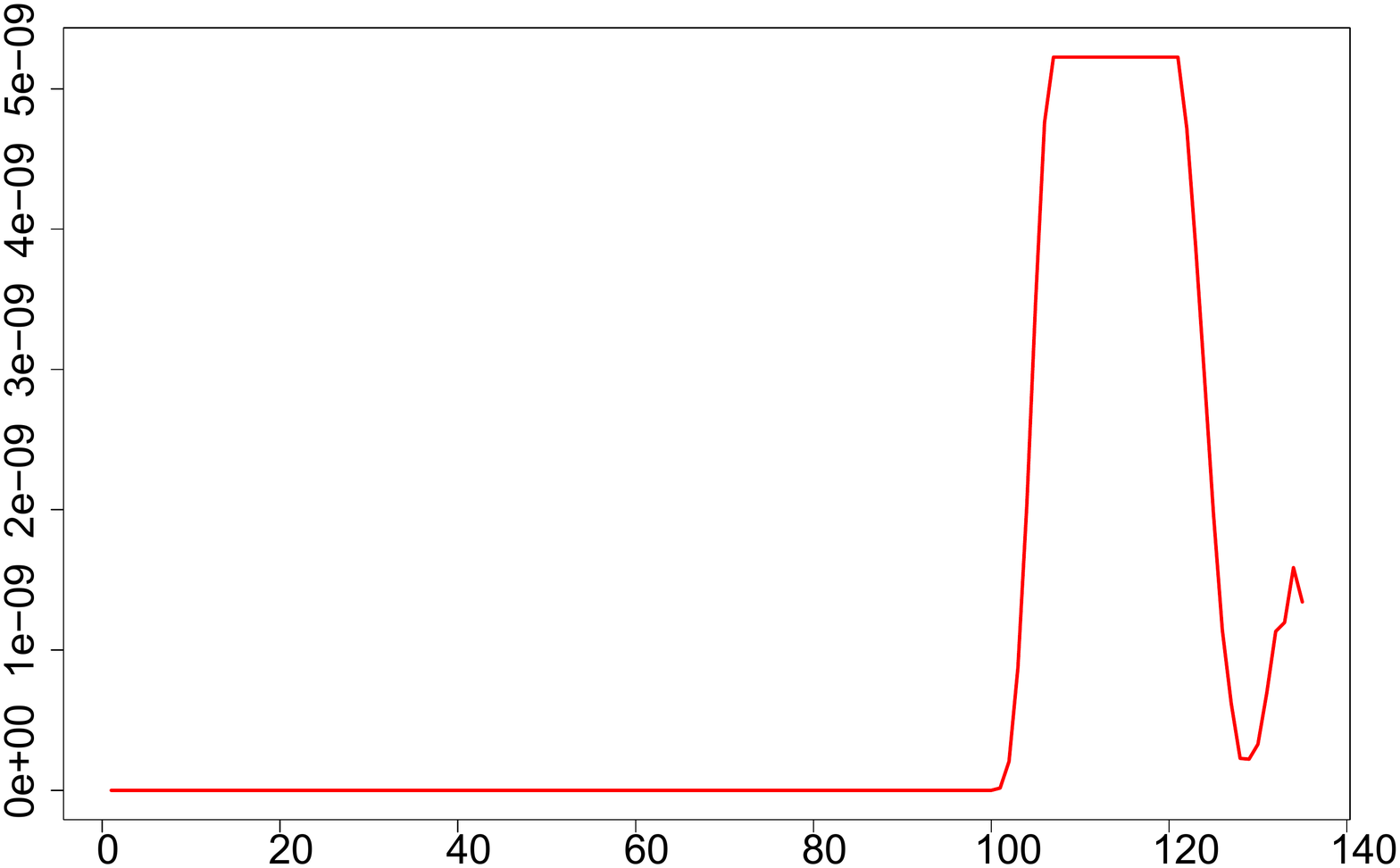}\\
\end{array}$
\caption{Synthetic LPPLS model and norm of persistence landscapes for a negative bubble $B>0$.}\label{fig:LPPLS_model synthetic_4}
\end{figure}

The sign of the parameter $b$ is associated with the bubble-type: $B<0$ corresponds to positive  bubble and $b>0$ corresponds to negative bubble.
In Figure \ref{fig:LPPLS_model synthetic_4} we show a  synthetic LPPLS model for a negative bubble. It is remarkable  that for both positive and negative bubbles the TDA signal spikes when approaching the critical time.

\subsubsection{Dependence on amplitude parameters $C_1,C_2$}
\begin{figure}
\centering
$\begin{array}{ccc}
\includegraphics[width=0.32\textwidth]{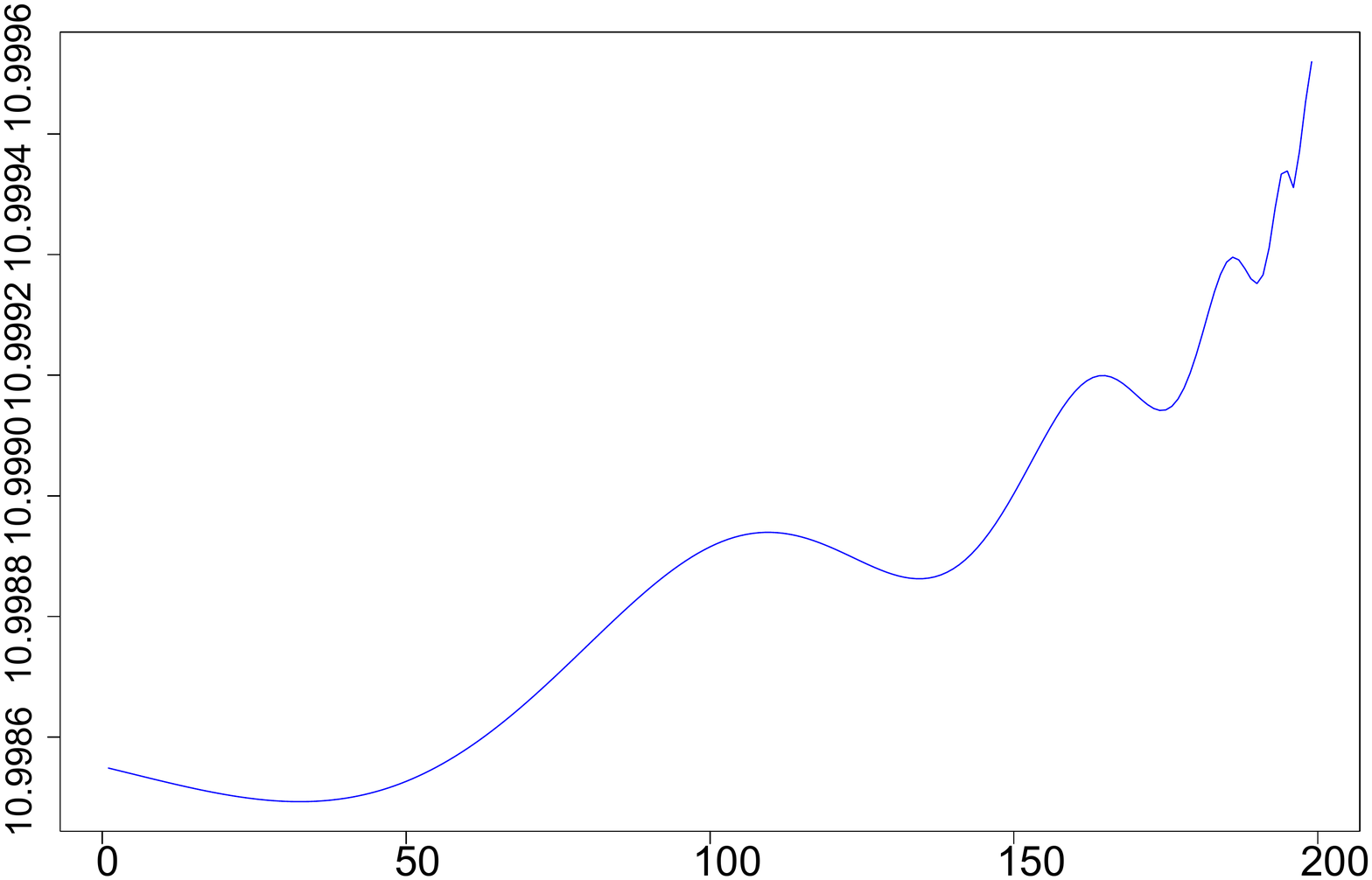}&\includegraphics[width=0.32\textwidth]{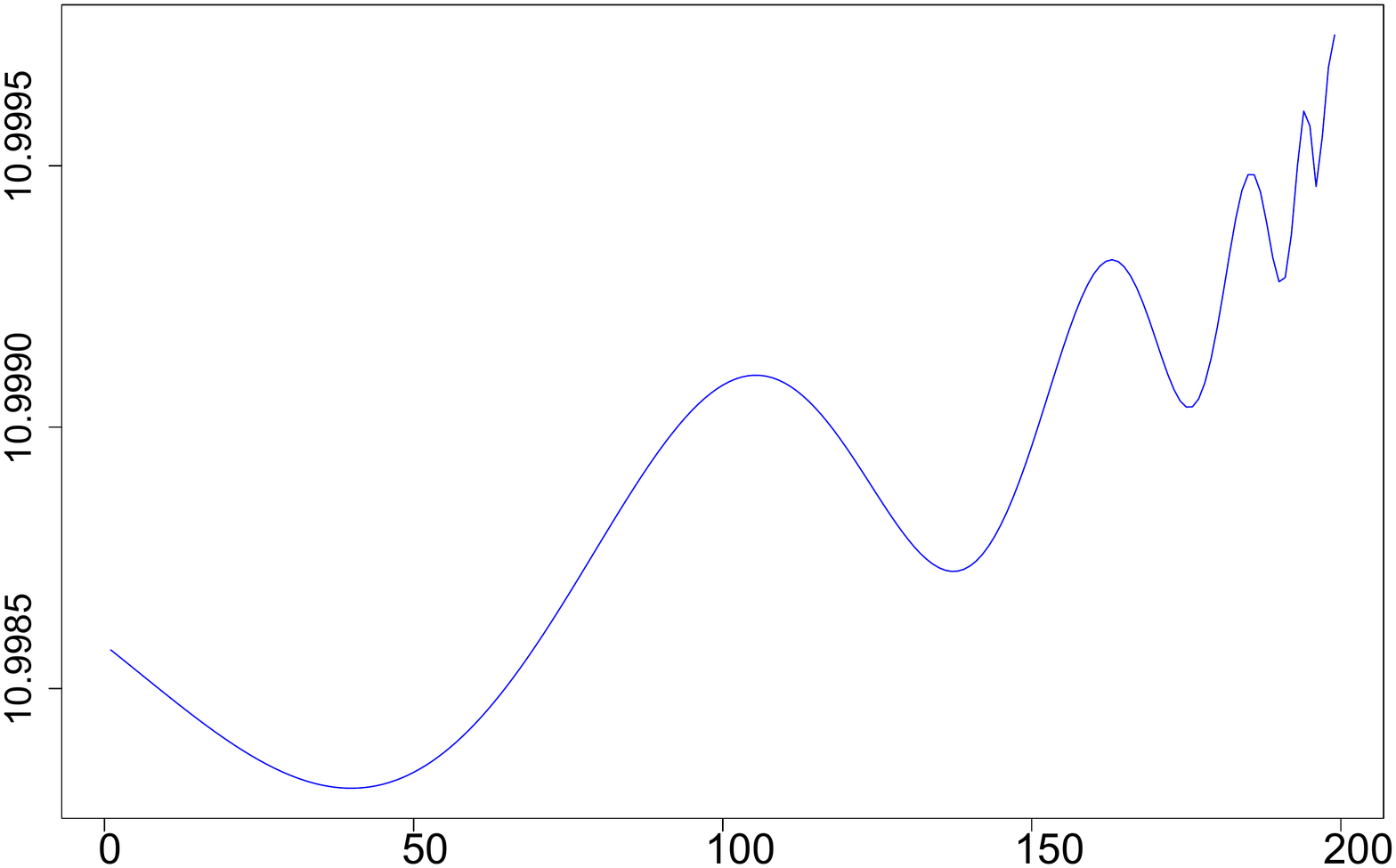}&\includegraphics[width=0.32\textwidth]{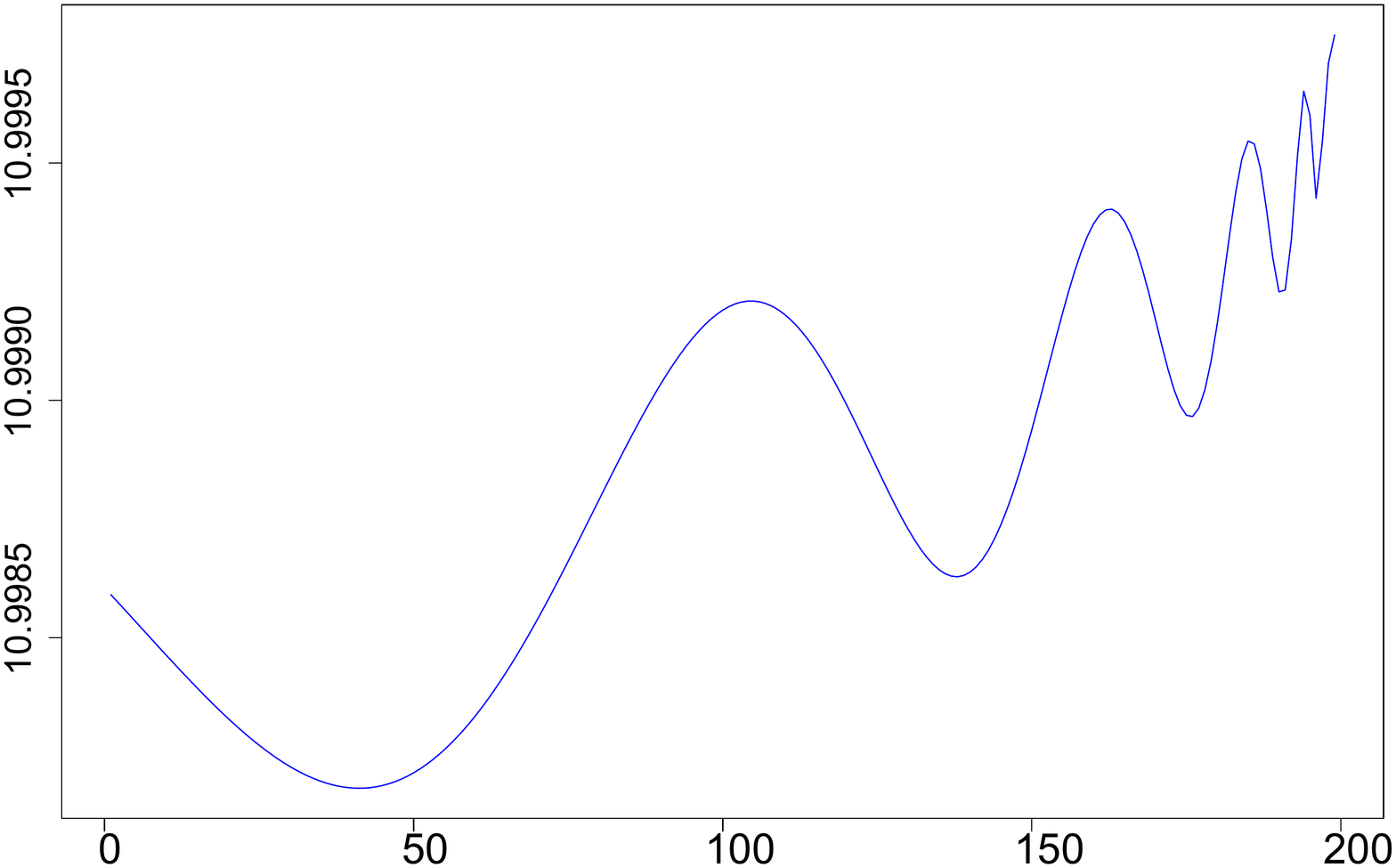}\\
\includegraphics[width=0.32\textwidth]{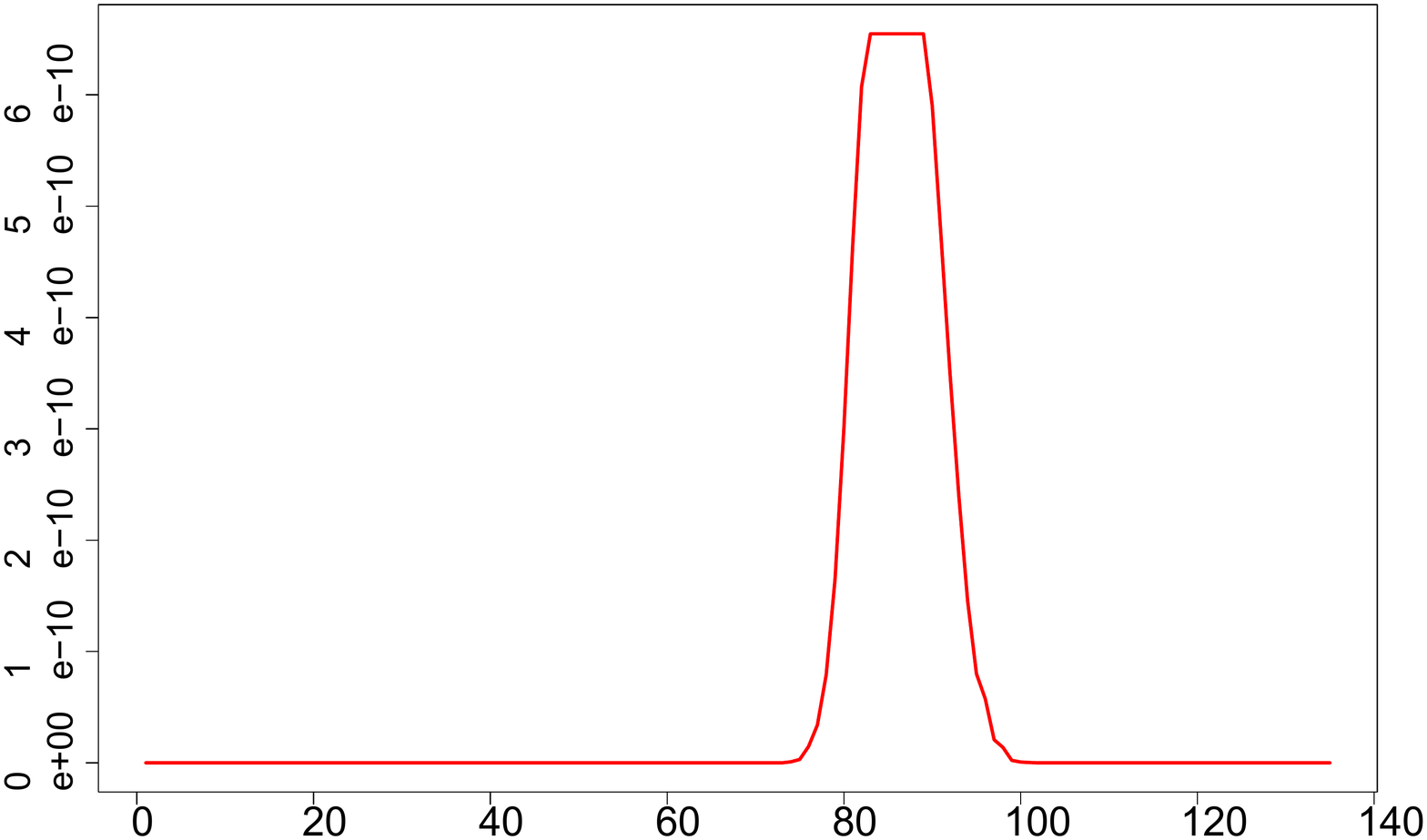}&\includegraphics[width=0.32\textwidth]{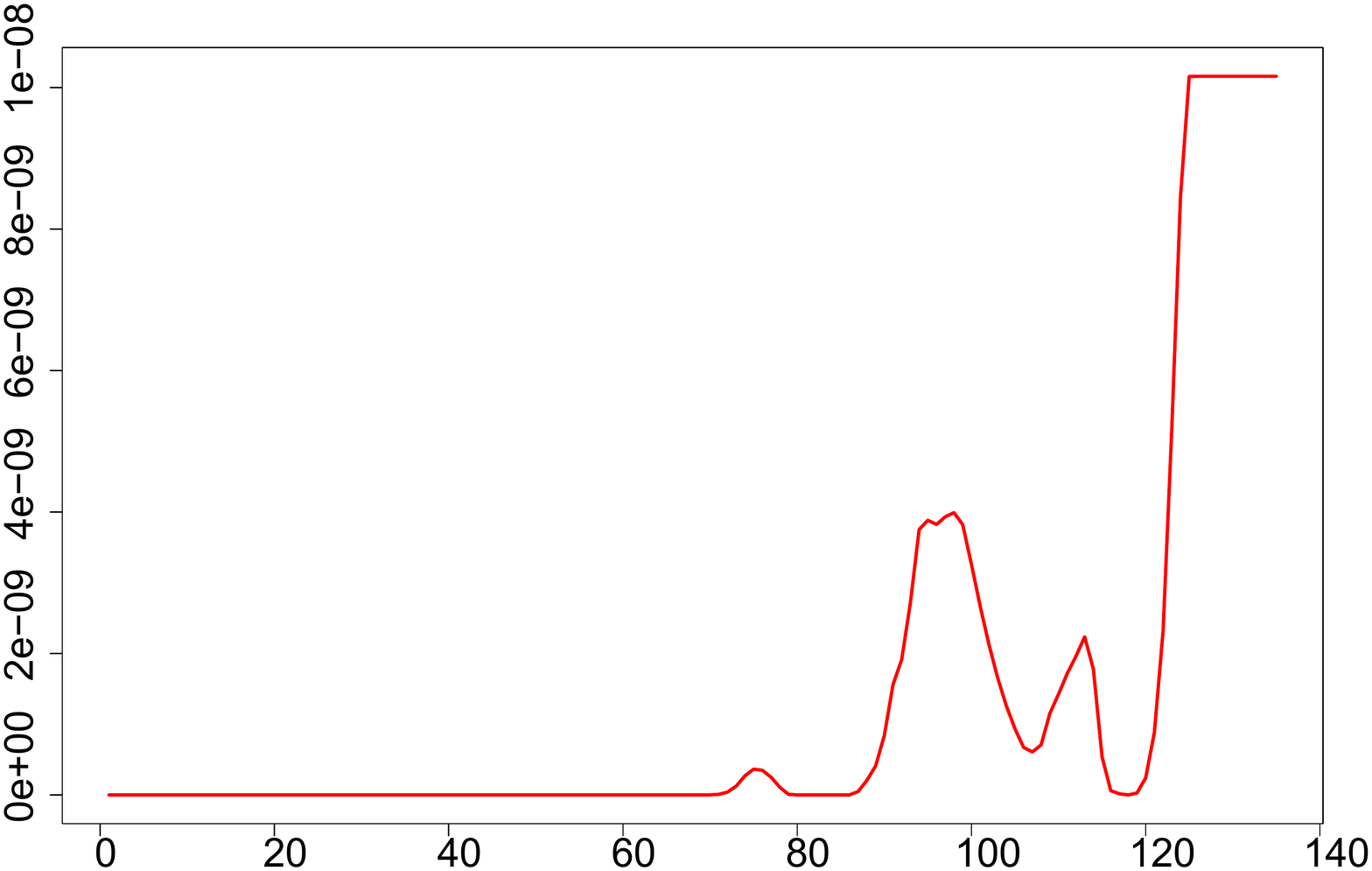}&\includegraphics[width=0.32\textwidth]{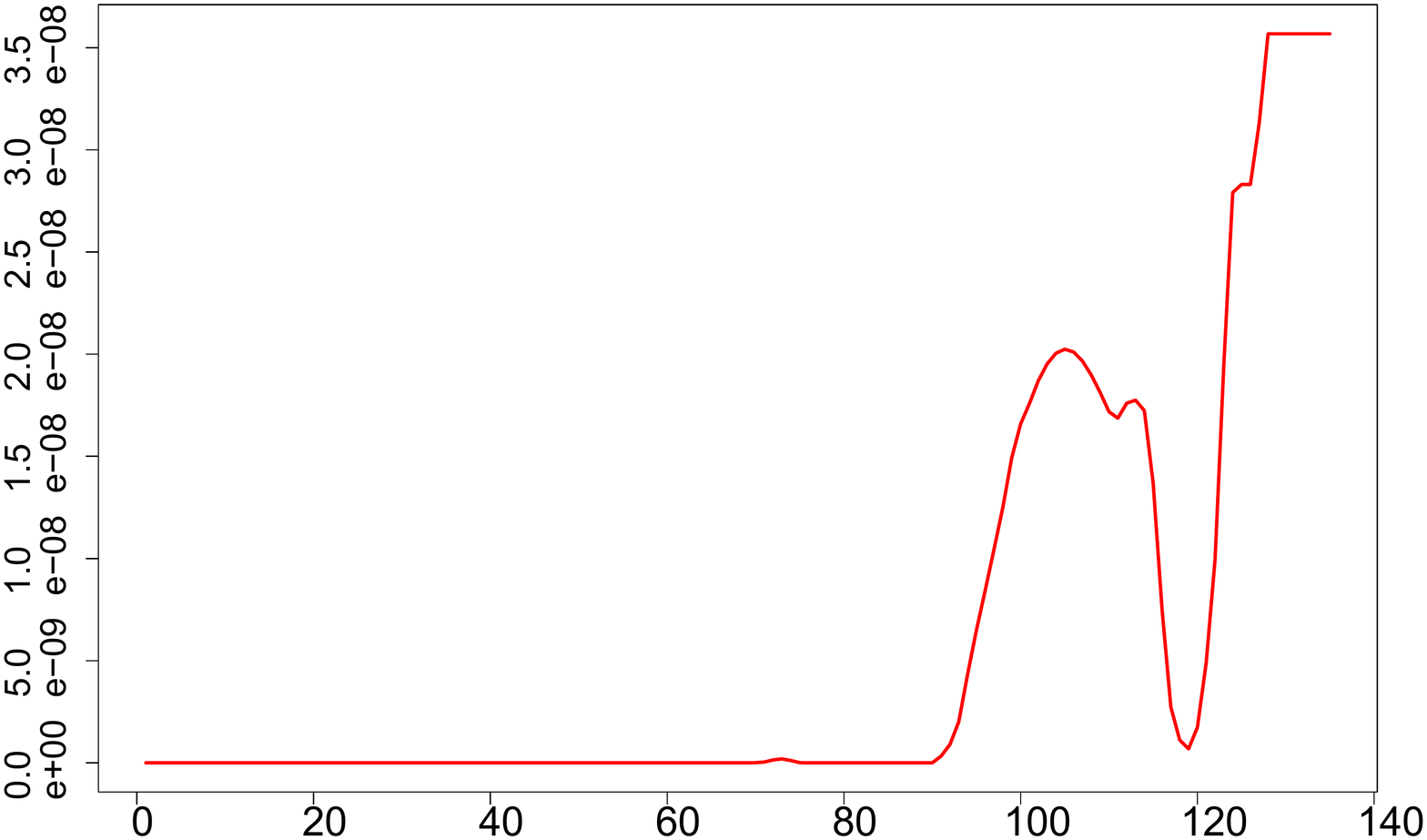}
\end{array}$
\caption{Synthetic LPPLS signal (top) and corresponding TDA signal (bottom) for $C_1=C_2=0.00002, 0.00005, 0.00007 $. }
\label{fig:LPPLS_model synthetic_5}
\end{figure}

We explore the dependence of the TDA signal on the parameters  $C_1,C_2$ which are responsible of the amplitude of the oscillations.
When the  amplitude of the signal is increased, the TDA signal grows in strength and becomes more prominent towards the end of the range.
See Fig.~\ref{fig:LPPLS_model synthetic_5}.

\section{Data segmentation into positive and negative bubbles}
\label{sec:data_segmentation}
In this section we describe a method  to segment a given financial time series into positive and negative bubbles.
For a positive bubble the main idea is there will be a  succession of positive
returns, which may be interrupted by negative returns no larger in amplitude than some pre-specified tolerance
level. An analogous description holds for negative bubbles.
We recall the data segmentation algorithm from \cite{gerlach2018dissection},
and some variants \cite{shu2020detection,shu2020real,yamashita2020segmentation}.

Let $x(t_i)$ be the time-series of the  price of an asset, where $t_{i}=t_{i-1}+\Delta t$.
The log-return of the asset at time $t_i$ is \[p_i=\ln x(t_i)-\ln x(t_{i-1}).\]
Assume that some initial time $t_{i_0}$ is the beginning of a up- (down-) trend.
The cumulative return at time $i$ is
\[p_{i_0,i}=\sum_{i=i_0+1}^{i} (\ln x(t_i)-\ln x(t_{i-1}))=\ln x(t_i)-\ln x(t_{i_0}).\]
The largest deviation $\delta_{i_0,i}$ of the price trajectory from a previous minimum (maximum)
is given by
\begin{equation}\delta_{i_0,i}=\left\{
                   \begin{array}{ll}
                     \max_{i_0\leq k\leq i} p_{i_0,k}-p_{i_0,i}, & \hbox{\textrm{ for an upward-trend}}, \\
                     p_{i_0,i}-\min_{i_0\leq k\leq i} p_{i_0,k}, & \hbox{\textrm{ for a  downward-trend}.}
                   \end{array}
                 \right.\label{eqn:delta}\end{equation}
To find the end of the upward-trend, we first compute $i_{\textrm{cross}_1}$ the first time $i>i_0$ when $\delta_{i_0,i}$ crosses a certain tolerance level $\eps_i$,
\begin{equation}\delta_{i_0,i}-\eps_i>0\label{eqn:crossing}\end{equation}
and then define the end of the upward- (downward-) trend as the time
\begin{equation} t_{i_1}=
\left\{
                   \begin{array}{ll}
                     \underset{i_0\leq k\leq i_{\textrm{cross}_ 1}}{\textrm{arg max}}  p_{i_0,k}, & \hbox{\textrm{ for an upward-trend}}, \\[0.5em]
                    \underset{{i_0\leq k\leq i_{\textrm{cross}_1}}}{\textrm{arg min}} p_{i_0,k}, & \hbox{\textrm{ for a  downward-trend}.}
                   \end{array}
                 \right.
\label{eqn:end_time}
\end{equation}

The time $t_{i_1}$ is considered as the beginning of the next downward (upward) trend. Through this procedure, the data is successively  segmented in
upward  and downward trends, which alternate one after the other.
The procedure is carried out for the whole time series and yields a sequence of peaks
\[\{t_{u,1}, t_{u,2},\ldots  \},\]
and  a sequence of  troughs
\[\{t_{d,1}, t_{d,2},\ldots  \},\]
such that peak is always followed by a  trough and vice-versa.

At the end of this procedure, each upward-trend corresponds to a positive bubble, and each downward-trend corresponds to a negative bubble.

The tolerance level $\eps_i$ can be chosen in multiple ways.
One option is to choose \[\eps_i=\eps,\] where $\eps>0$ is some preset constant.
Another option is to choose a time-dependent tolerance level $\eps_i$ proportional to the volatility (standard deviation) $\sigma_i=\sigma_i(w_0)$ at time $i$
estimated over a preceding time window of some fixed size $w_0$, i.e.,
\begin{equation}\eps_i=\eps_0\sigma_i,\label{eqn:threshold}\end{equation}
where $\eps_0$ is some scaling parameter. Thus the tolerance is  more permissive
when the market presents high volatility and is  stricter during calmer periods.
There are other ways to choose the tolerance level (see, e.g., \cite{yamashita2020segmentation}).

The segmentation procedure in \cite{gerlach2018dissection} does not  fix a single scaling parameter $\eps_0$ and a single window-size $w_0$ for the computation of the standard
deviation, but rather  considers some range of scaling parameters $\eps_0\in[\eps_1,\eps_2]$  and of window-sizes $w_0\in[w_1,w_2]$, each sampled
with some frequency, and computes a collection of peaks and a collection of troughs for each $\eps_0$ and $w_0$.   Finally, it selects the most frequent peaks and troughs that appear
over the whole range of $\eps_0,w_0$.

\begin{figure}
\centering
$\begin{array}{cc}
\includegraphics[width=0.32\textwidth]{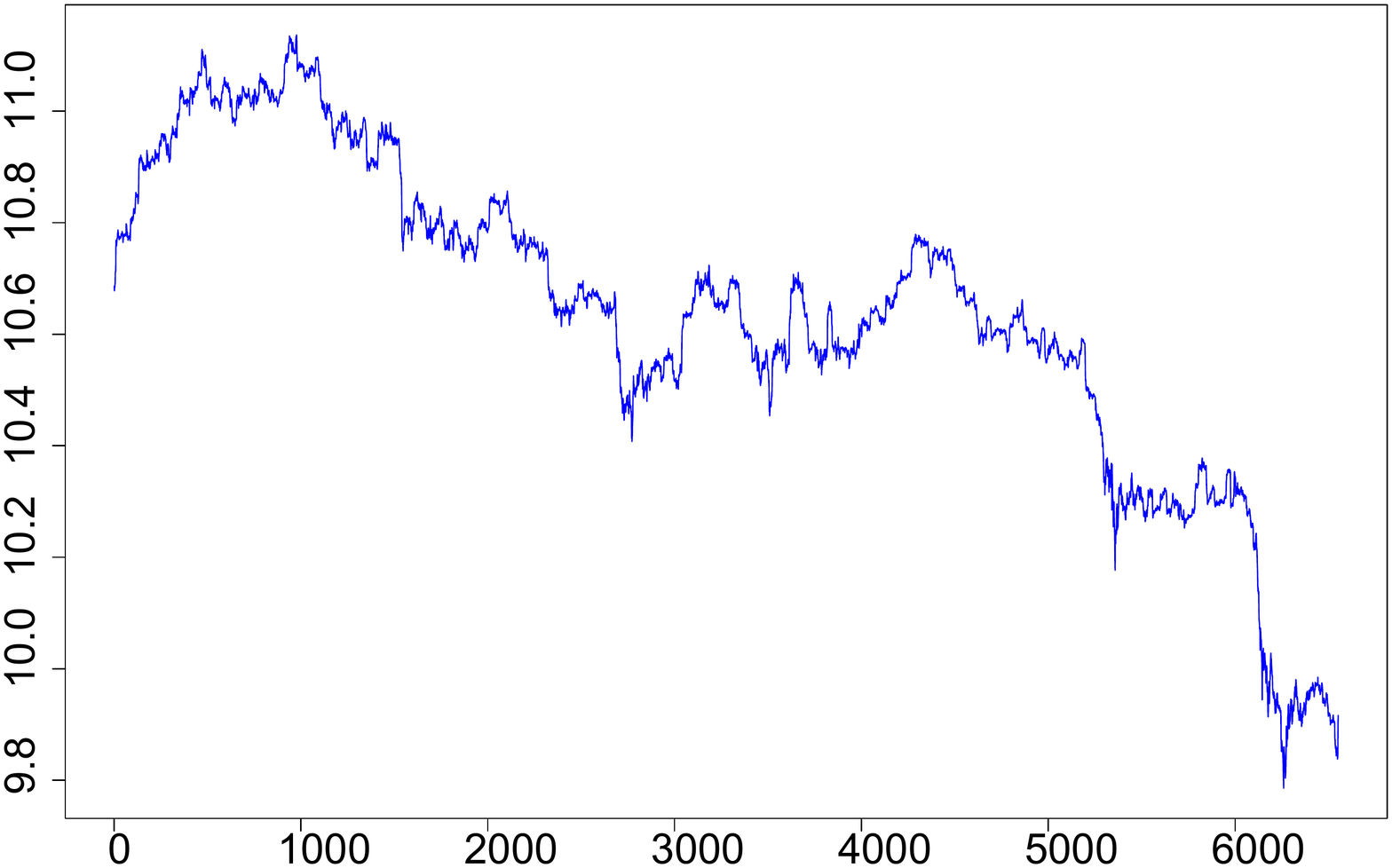}
&\includegraphics[width=0.32\textwidth]{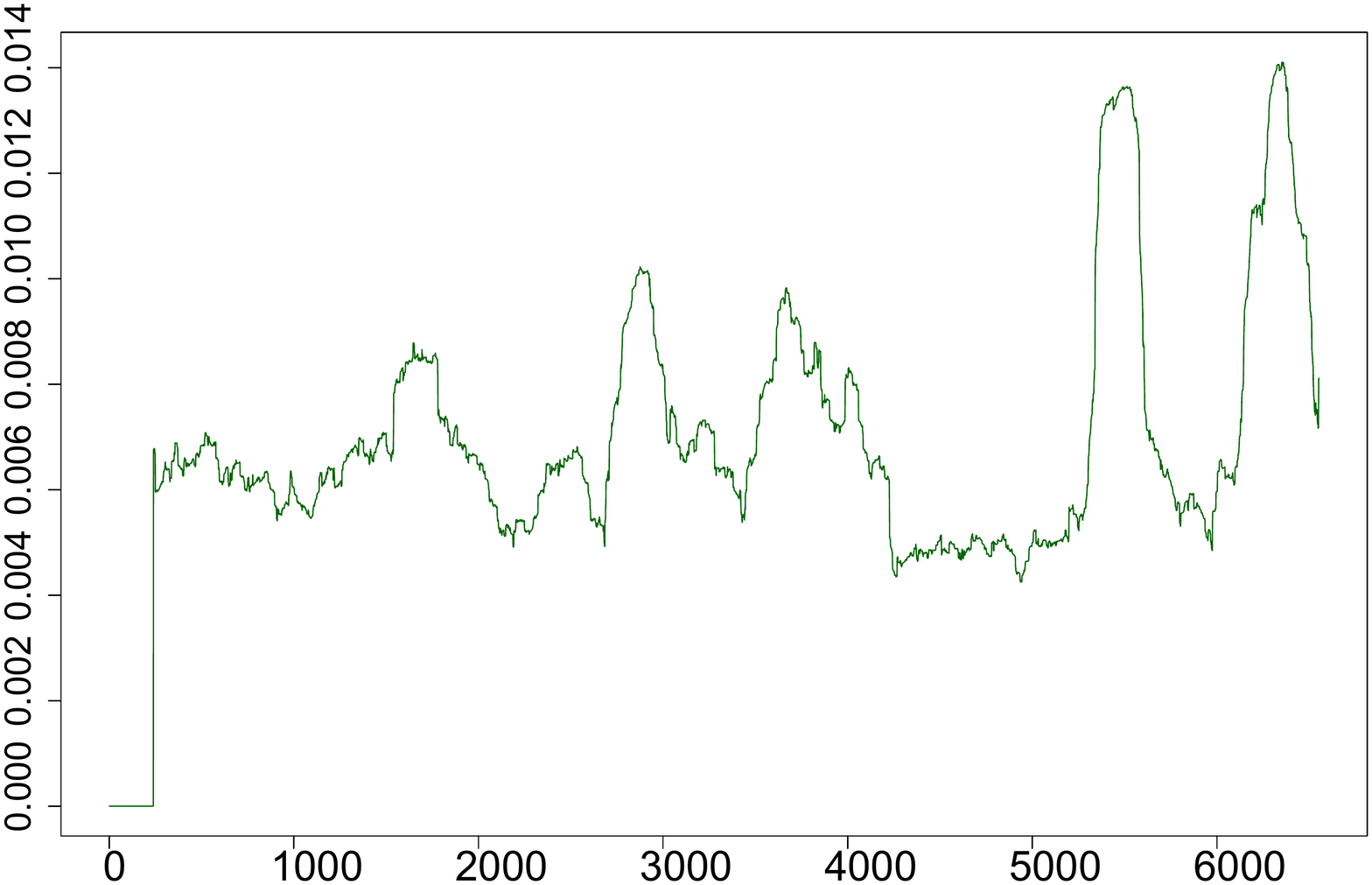}
\end{array}$
\caption{Log-price and standard deviation of Bitcoin over 10/01/2021-07/01/2022}
\label{fig:Logprice_bitcoin_Standard_deviation_bitcoin}
\end{figure}

In Fig.~\ref{fig:Logprice_bitcoin_Standard_deviation_bitcoin} we show the log-price of Bitcoin over the period 10/01/2021-07/01/2022 sampled at 1 hour intervals
(resulting in 6552 data points), and the standard deviation measured over a preceding time window of size $w_0=240$.

In the examples in Section \ref{sec:TDA_Bitcoin}, as we are interesting in `large' bubbles, we will fix the scaling parameter at some large value of $\eps_0$.

\section{Application to detection of Bitcoin bubbles}\label{sec:TDA_Bitcoin}
In this section we present a series of numerical experiments on hourly Bitcoin price during the period October 1st 2021 and July 1st  2022.
In Section \ref{sec:TDA_segmentation} we segment the data into positive bubbles and negative bubbles following the methodology described in Section \ref{sec:data_segmentation}.
In Section \ref{sec:LPPLS_fit} we fit the LPPLS model to each of the segments.
In Section \ref{sec:TDA_Bitcoin} we apply the TDA procedure to each segment, following the methodology described in  Sections \ref{sec:persistent} and \ref{sec:TDA_time_series}.

\subsection{Data segmentation}
\label{sec:TDA_segmentation}

The hourly Bitcoin price data set between October 1st 2021 and July 1st  2022 consists of 6552 data points.
We segment the data following the procedure described in Section \ref{sec:data_segmentation}.
To compute the standard deviation (volatility) $\sigma_i$ we use a fixed window size of $w_0=240$ data points.
We also use a  fixed scaling parameter $\eps_0=15.0$.
Thus, the tolerance level is taken as $\eps_i=\eps_0\sigma_i$, where $\sigma_i$ is computed over a sliding window of length $w_0$.

In the table below we identify the peak-time (P) and the crossing-time (C) for each upward-trend.

\smallskip

\noindent{\bf Table 2. Upward-trend peak and crossing times:}
\\
\begin{tabular}{|l|l|l|l|l|l|l|l|l|l|l|l|}
  \hline
P&471 & 977& 2105& 3184& 3662& 3831& 4290& 4861& 5447& 5824& 6443\\
  \hline
C &538& 1104& 2130& 3355 &3718 &3934& 4370 &4887 &5729& 5894 &6536\\
  \hline
\end{tabular}

\smallskip

In the table below we identify the  trough-time (T) and the crossing-time (C) for each downward-trend.

\smallskip

\noindent{\bf Table 3. Downward-trend troughs and crossing times:}
\\
\begin{tabular}{|r|l|l|l|l|l|l|l|l|l|l|l|}
  \hline

T&  648 & 1873& 2772& 3509 &3787 &3934 &4780 &5358 &5729&6260\\
  \hline
C& 783 &2010& 2821 &3531 &3831 &4075 &4835& 5601& 5804&6415\\
  \hline
\end{tabular}
\\

\begin{figure}
\centering
\includegraphics[width=0.5\textwidth]{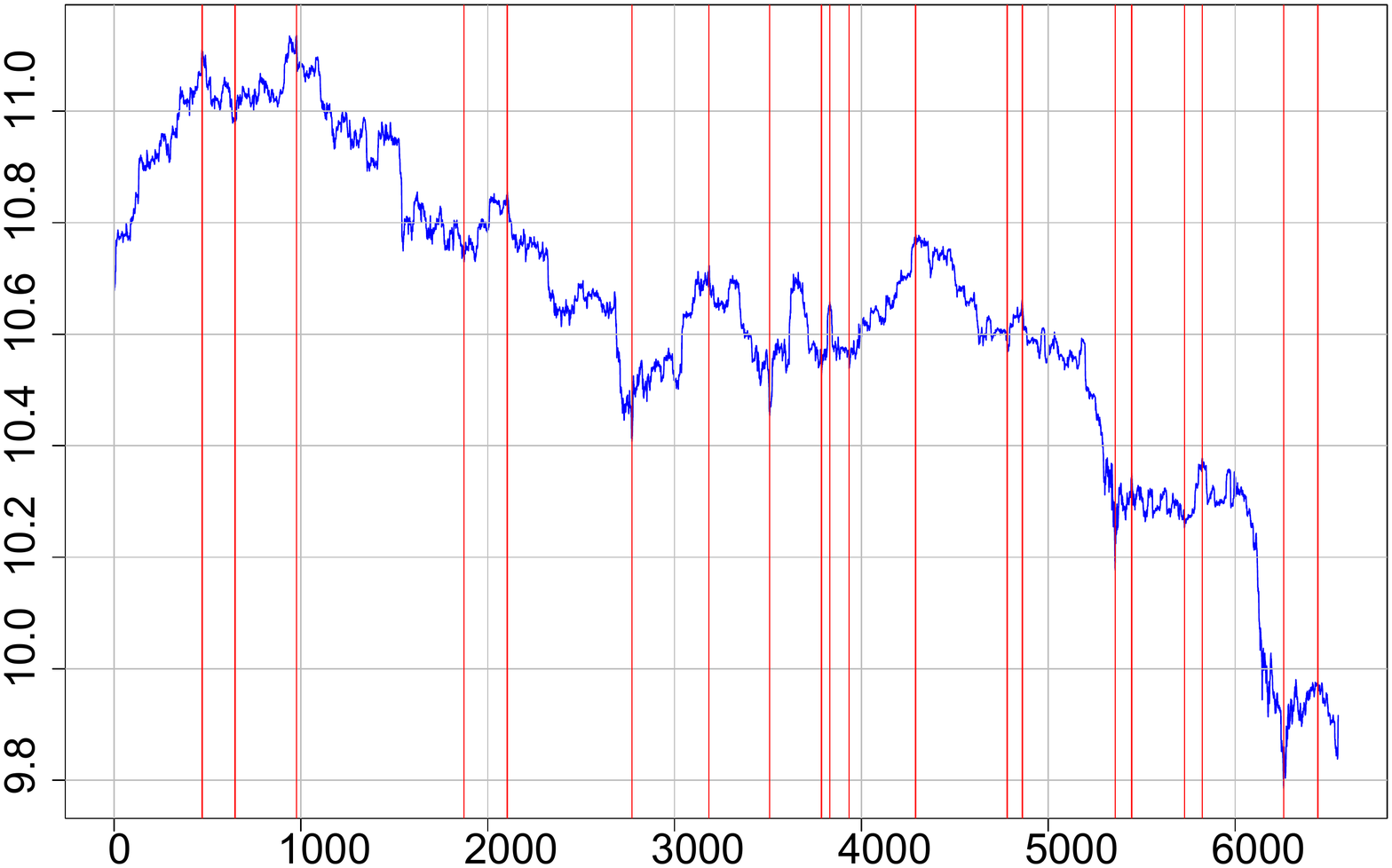}
\caption{Peaks and troughs}
\label{fig:peaks_and_troughs}
\end{figure}

The combined peaks and troughs from Tables 1 and 2 are shown in Fig.~\ref{fig:peaks_and_troughs}.

Then we define upward-trend (U) and downward-trend (D) segments to be fitted with LPPLS models as well as analyzed with  TDA.

First, each segment is determined by the following criterion.  For an upward- (downward-) trend, the end-point is a peak (trough) from the list and the starting-point of the
trend  is the nearest crossing time in the past from the previous downward- (upward-) trend.

\smallskip

\noindent{\bf Table 4. Upward-trend and downward-trend segments}
\\
\begin{tabular}{|r|l|l|l|l|l|l|}
  \hline
U      & 783-977&2010-2105&2821-3184& 3531-3831& 4075-4290& 4835-4861\\
\hline
U & 4835-5447&5804-5824&6415-6443& & &  \\
\hline
D & 538-648& 1104-1873& 2130-2772&3355-3509&3718-3787&3831-3934\\
\hline
D &4370-4780& 4370-5358&4887-5729& 5894-6260&  &\\
  \hline
\end{tabular}
\\

Second, we perform some additional adjustments.
We examine each of the resulting segments and if we observe  significant peaks or troughs inside  a segment  we do an ad-hoc adjustment of the segment: we cut the portion from the beginning of the segment to the last observed a peak/through.
This adjustment is occasionally needed since  using the segmentation procedure with fixed $w_0$ and fixed $\eps_0$
can miss some peaks/troughs. 

Third, from the remaining segments we  dismiss those that are very short, since  applying the LPPLS fitting and the TDA procedure to them is unreliable.
Thus, the segments $4835-4861$,  $5804-5824$ and $6415-6443$,  all corresponding  to upward-trends, were deemed as too short to perform the analysis.
We also disregard the first segment $1-471$ which was used to initialize  the process.

We retain the following:

\smallskip

\noindent{\bf  Table 5. Adjusted upward-trend and downward-trend segments}
\\
\begin{tabular}{|r|l|l|l|l|l|}
  \hline
U     & 783-977 &2010-2105&3020-3184& 3531-3831& 4175-4290\\
\hline
U  & 5358-5447& & & & \\
\hline
D & 538-648& 1104-1873& 2500-2772&3355-3509&3718-3787\\
\hline
D &3831-3934 & 4370-4780& 5150-5358&4887-5729& 5894-6260  \\
  \hline
\end{tabular}

\subsection{Fitting the LPPLS model to Bitcoin time-series}
\label{sec:LPPLS_fit}

\begin{figure}
\centering
\includegraphics[width=.5\textwidth]{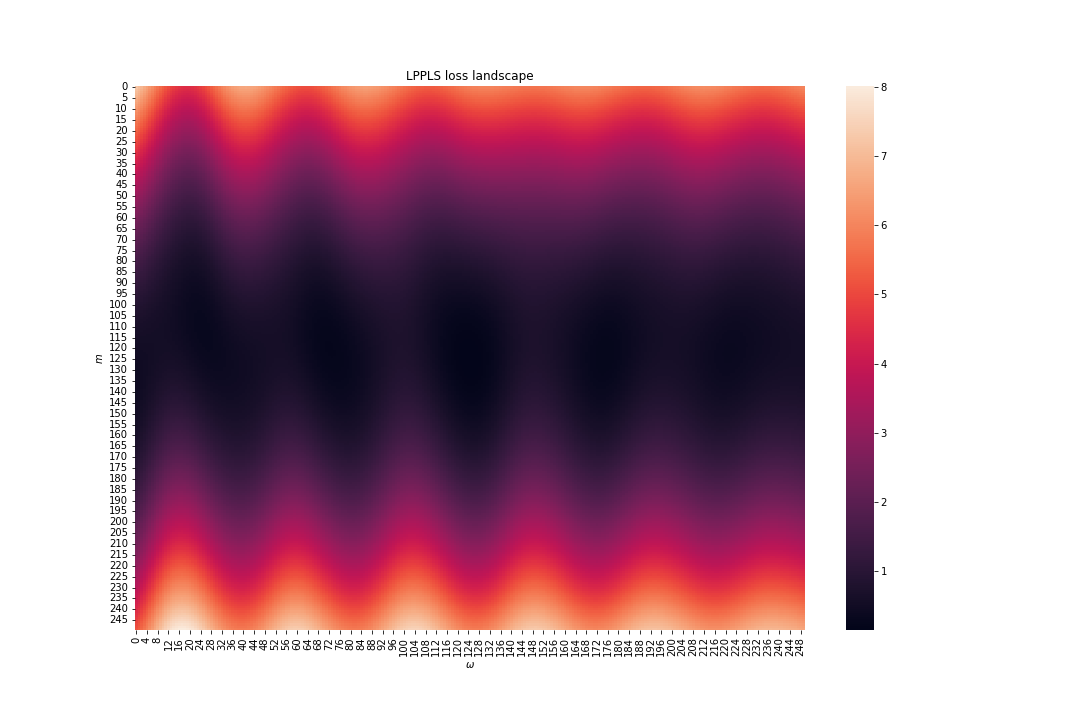}
\caption{Fitness function in the domain $[m, \omega]$, for constant values of the remaining parameters. The objective function displays strong non-convexity.}\label{fig:loss_land}
\end{figure}

Once a specific segment is defined, we address the optimization problem, in which we aim at best fitting the LPPLS model to the time series.
More formally, using \eqref{eq:lppls} to obtain a vector of estimates of the expected log-price $\hat{y} = LPPLS(y)$ from the segment $x$, the goal is to solve the
unconstrained, non-convex (Figure \ref{fig:loss_land})  optimization problem:

\begin{equation}
	\min_{tc, m, \omega, A, B, C_1, C_2} \quad   ||\hat{y} - y ||^2_2
    \label{eq:lppls}
\end{equation}

\begin{figure}
\centering
\includegraphics[width=.5\textwidth]{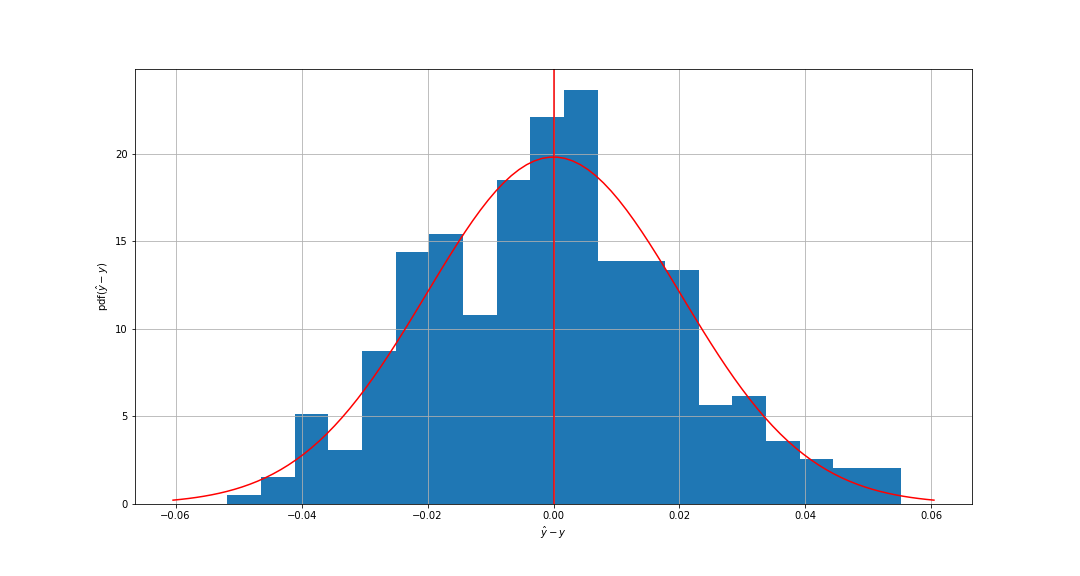}
\caption{Distribution of residuals for a fitted model.}
\label{fig:res}
\end{figure}

Given the assumption for the time series to be characterized by Gaussian noise, the choice of such fitness function is equivalent to maximum likelihood estimation as depicted in Figure \ref{fig:res}.

For this procedure,  together with Gaussianity, we are assuming time-independent variance; in the case in which heteroscedasticity is taken into account, a weighting scheme can be introduced in the definition of the fitness function.

Given the geometry of the problem, a global optimization scheme based on differential evolution \cite{storn1997differential} has been used. This method, as opposed to gradient-based local optimizers, is based on performing function evaluation on a set of sample points drawn from a large candidate space followed by a local optimizer at the end. While similar works propose the use of grid search in the space of initial conditions, each followed by local optimization, also \cite{shu2020real} makes use of evolutionary algorithms (specifically, covariance matrix adaptation evolution strategy).

In order to speed up convergence, the problem is reformulated as

\begin{equation}
	\min_{p = [tc, m, \omega]} \quad   ||\hat{y}(p, A(p), B(p), C_1(p), C_2(p)) - y ||^2_2
\end{equation}

In which $q = [A(p), B(p), C_1(p), C_2(p)]$ is determined, for each $p$, solving a linear least square problem:

\begin{equation}
	q = ( \textbf{X}^T \textbf{X} )^{-1} \textbf{X}^T \textbf{p}.
\end{equation}

The software developed in this context is built on top of the open-source package \cite{Boulder-Investment-Technologies}. The results of the LPPLS fitting are depicted in  Appendix~\ref {sec:appendix}.

\subsection{TDA applied to Bitcoin time-series and comparison to fitted LPPLS models}
\label{sec:TDA_Bitcoin}

We apply TDA to each  adjusted upward-trend and downward-trend segment from Table 5.
For each segments we have chosen a suitable set of TDA  parameters from the following range:
\begin{equation}
w=48,60,72,\, d=3,5,\, N=4.
\label{eqn:TDA_parameters_2}
\end{equation}
As noted in Section \ref{sec:dependence}, adjustments of the TDA parameter are needed because the segments have different lengths,
and the trends  of the corresponding time-series (reflected by the fitted LPPLS models) are different.   

In Appendix  \ref {sec:appendix} we show, side-by-side, the fitted LPPLS model and TDA signal for each upward-/downward-trend, in chronological order (i.e., alternating upward- and downward-trends).

We observe that most of the segments show a primary peak in the TDA signal towards the end of the range.  Some of the  segments  show some secondary peaks in the TDA signal (e.g.,  Appendix  \ref {sec:appendix} -- Figs.\ref{fig:Data set 783-977}, \ref{fig:Data set 1104-1873}, \ref{fig:Data set 2500-2772},  \ref{fig:Data set 3531-3831}, \ref{fig:Data set 5894-6260}), which seem  to capture some intermediate price jumps or drops.

In a few instances (e.g.,  Appendix  \ref {sec:appendix} -- Fig.  \ref{fig:Data set 4370-4780}, \ref{fig:Data set 4887-5729})) the primary peak in the TDA signal  occurs earlier than the end of the range.
We note that in these instance that an  apparent tipping point seems to occur not at the end of the range but closer to the middle. The fact that the segmentation procedure used in Section \ref{sec:data_segmentation} does not identify this apparent tipping point is  likely due to the dependence of the  threshold \eqref{eqn:threshold} on volatility; if the preceding time period is highly volatile, the threshold for  the crossing time \eqref{eqn:crossing} is higher, and if the volatility becomes lower at a later time, then the threshold will also become lower, and hence the crossing time will be registered at a later time.

We also note that some of the fitted LPPLS models do not seem to display the desired characteristics, that is, super-exponential growth (or decay)  superimposed with oscillations increasing in frequency and decreasing in amplitude. For instance,   Appendix  \ref {sec:appendix} -- Figs. \ref{fig:Data set 2010-2105}, \ref{fig:Data set 3831-3934}, \ref{fig:Data set 4370-4780}) show very few oscillations, or weak nonlinearity. Such LPPLS models should, in fact, be dismissed;  only  LPPLS models that pass certain filters should be selected as viable solutions (see \cite{gerlach2018dissection} and the references listed there). Nevertheless, the TDA signal is clearly distinguishable even in cases when the LPPLS fit is  flawed.

\section{Conclusions}

We provide  an empirical argument on why TDA is able to provide early warning signals for critical transitions in financial time series, for both positive and negative financial bubbles.  The premise of our investigation is that financial bubbles are reflected by the LPPLS model,
asserting that the price of an asset exhibits  super-exponential growth (or decay)  superimposed with oscillations that  increase in frequency and decrease in amplitude when approaching a critical transition. When the time series is transformed into a point cloud via time-delay coordinate embedding, these oscillations give rise to persistent $1$-dimensional homology generators (loops), which can be quantified via persistence landscapes.  By applying a sliding window to the time series, one can capture the changes in the oscillations through the growth in the norms of the corresponding persistence landscapes.

Our experiments show that, when the TDA procedure is applied to synthetic LPPLS time-series, the  norms of the persistence landscapes grow as predicted.
They also show an intricate dependence of the TDA signal on the parameters of the LPPLS model and of the TDA procedure.
Further work is needed to fully analyze this dependence and automatize the parameter selection.

As a practical application, we run the TDA procedure on a collection of positive  and negative bubbles of the Bitcoin price between October 1st 2021 and July 1st  2022. In most cases,  the TDA gives early warnings for tipping points, even when the fitted LPPLS appears to be flawed.

\appendix
\section{}
\label{sec:appendix}
\begin{figure}[H]
\centering
\begin{subfigure}[t]{0.7\linewidth}
    \centering
    $\begin{array}{cc}
    \includegraphics[width=0.52\linewidth]{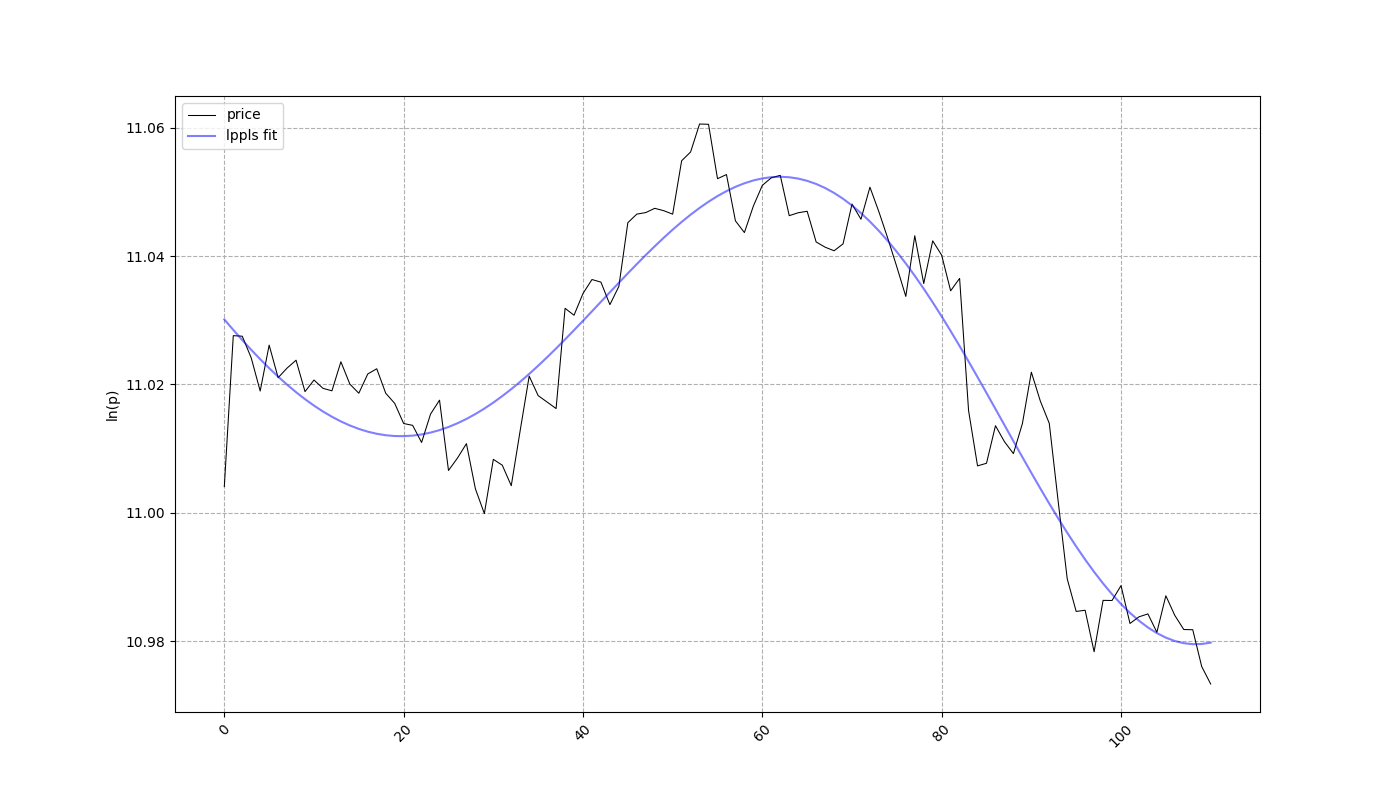}&
    \includegraphics[width=0.41\linewidth]{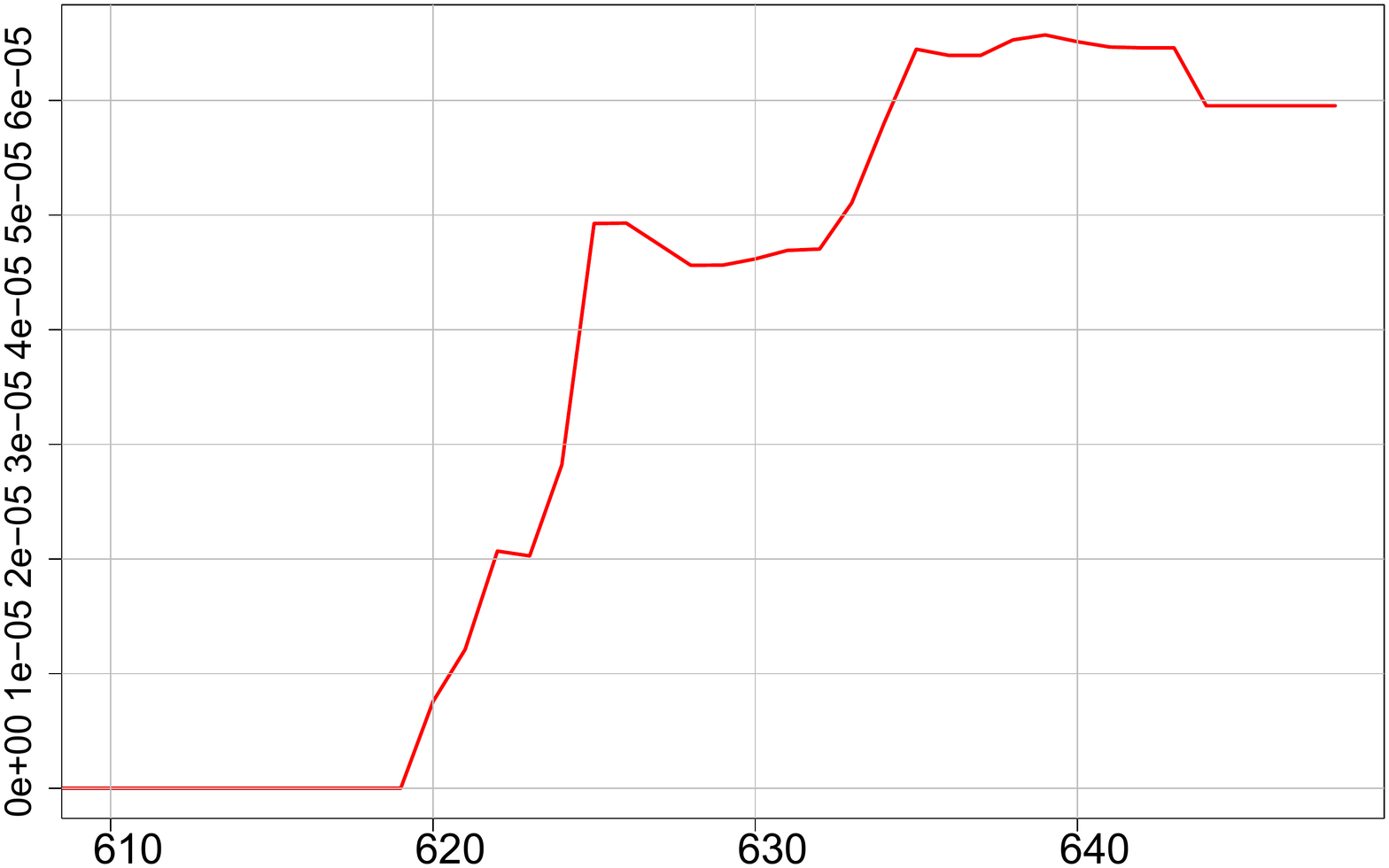}
    \end{array}$
    \caption{Data set 538-648}
    \label{fig:Data set 538-648}
\end{subfigure}
\end{figure}

\begin{figure}[H]
\ContinuedFloat
\centering
\begin{subfigure}[t]{0.7\linewidth}
    \centering
    $\begin{array}{cc}
    \includegraphics[width=0.52\linewidth]{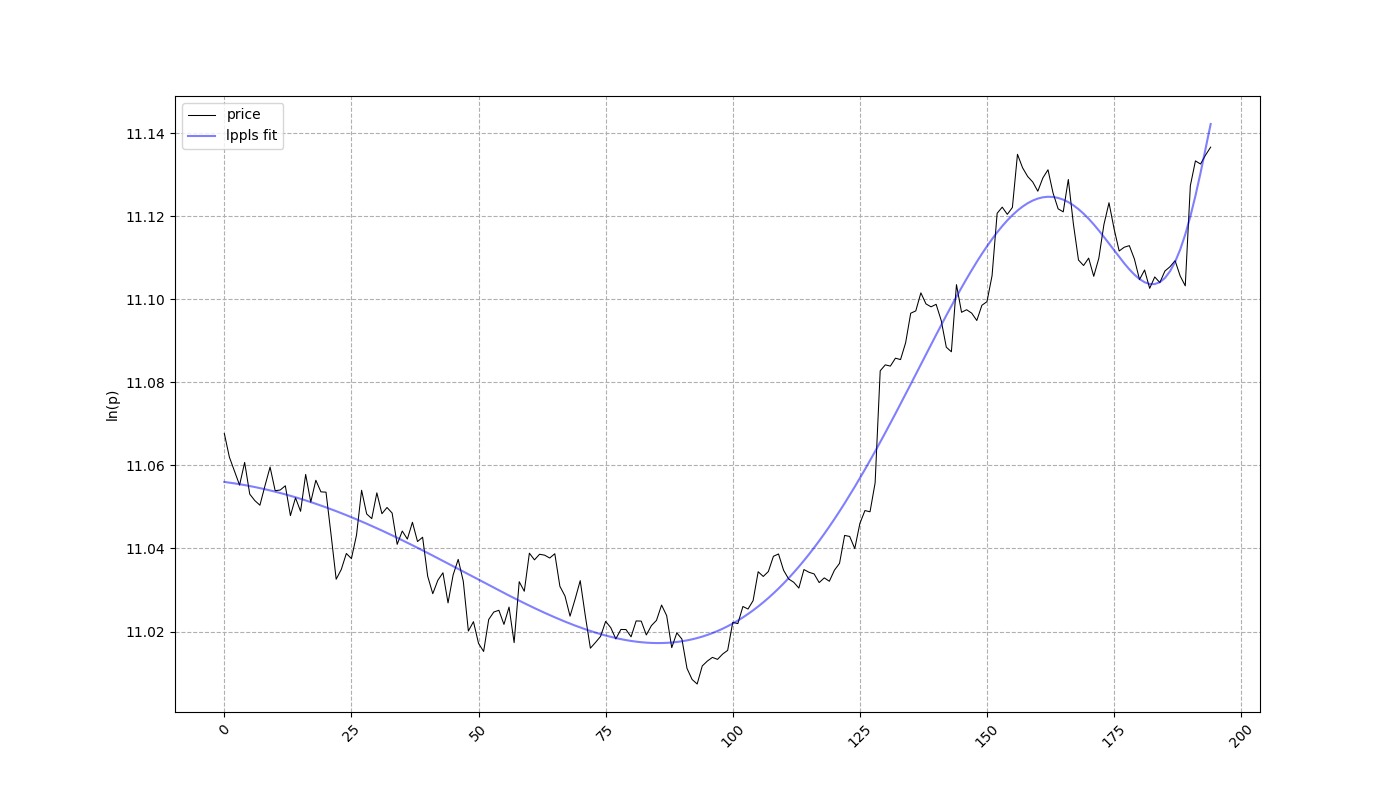}&
    \includegraphics[width=0.41\linewidth]{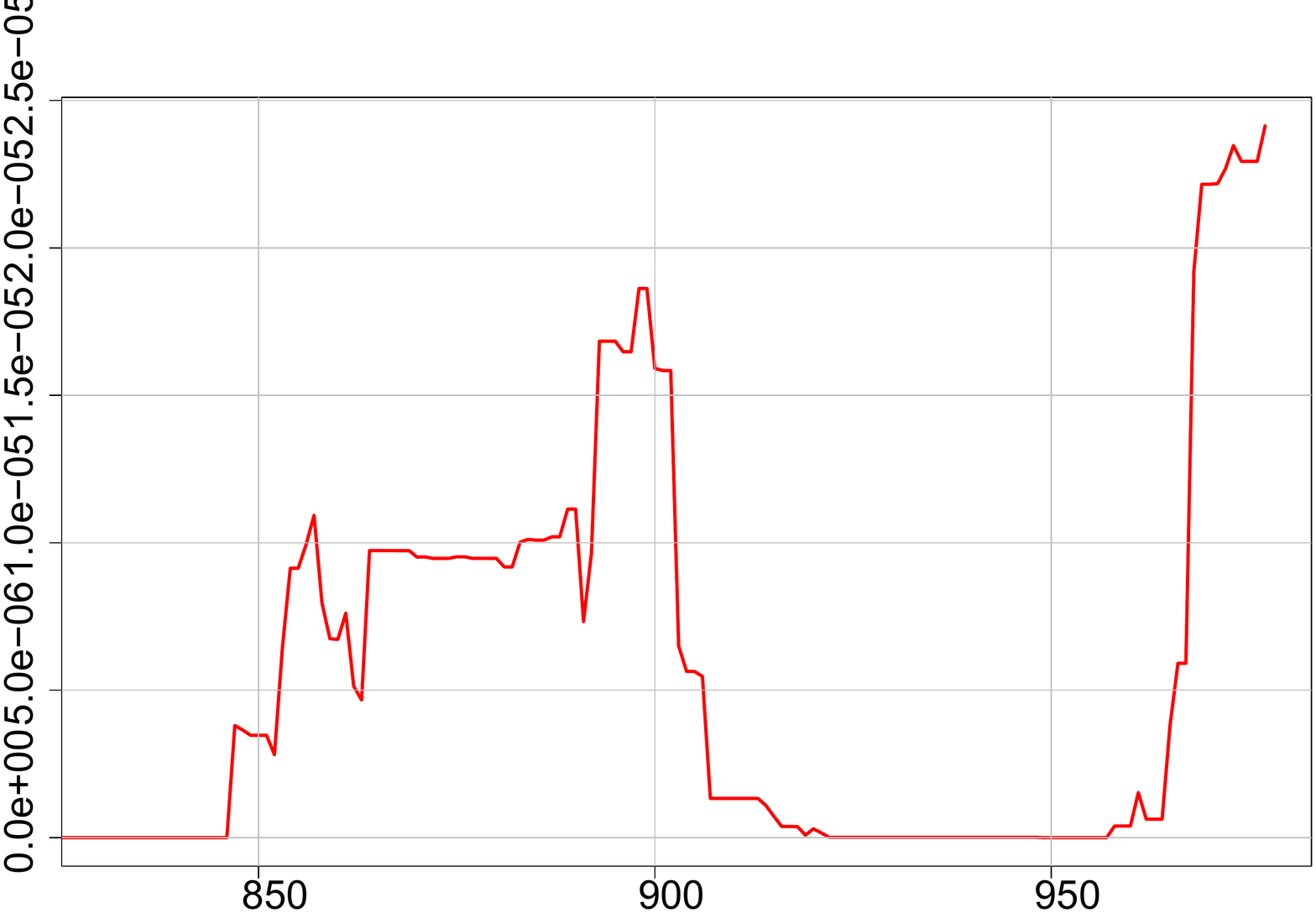}
    \end{array}$
    \caption{Data set 783-977}
    \label{fig:Data set 783-977}
 \end{subfigure}
\\
 \begin{subfigure}[t]{0.7\linewidth}
    \centering
    $\begin{array}{cc}
    \includegraphics[width=0.52\linewidth]{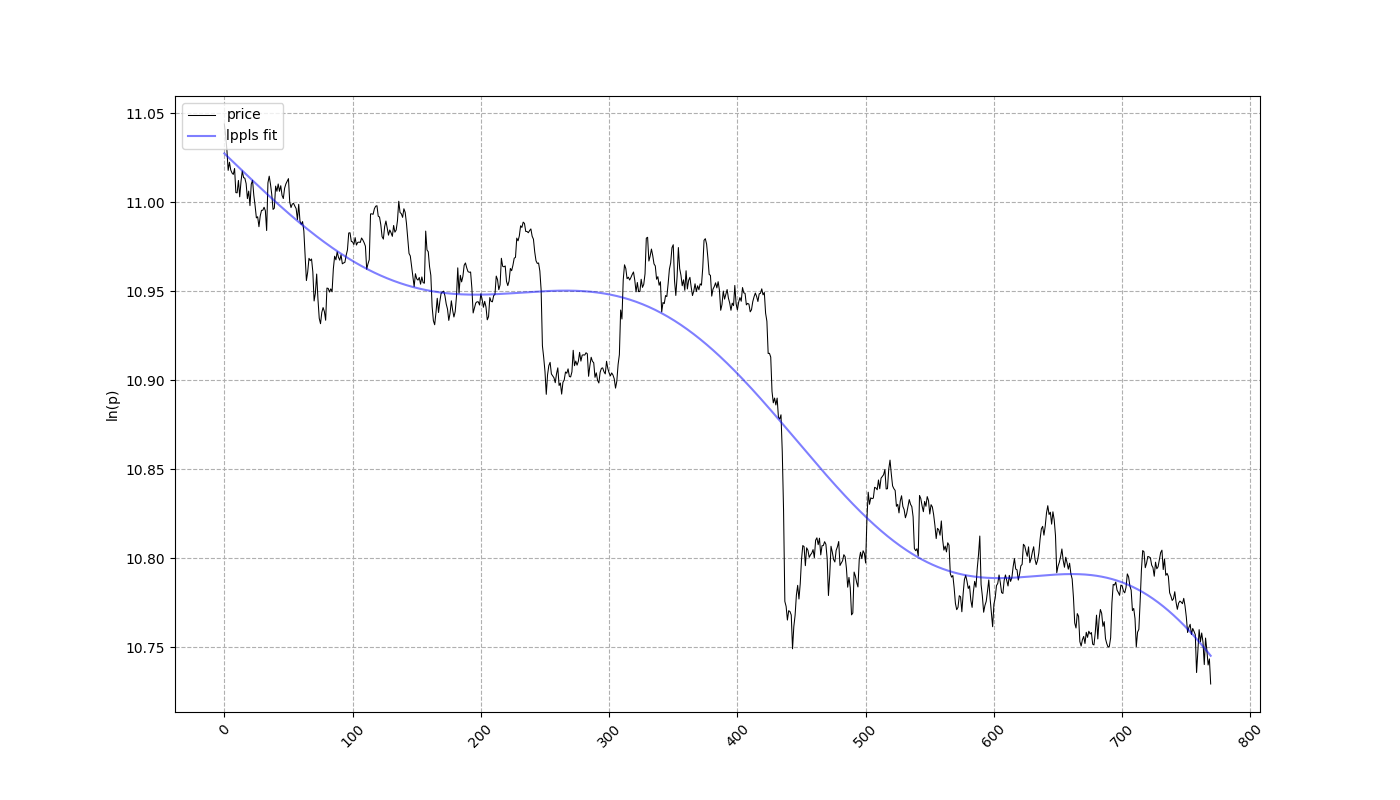}&
    \includegraphics[width=0.41\linewidth]{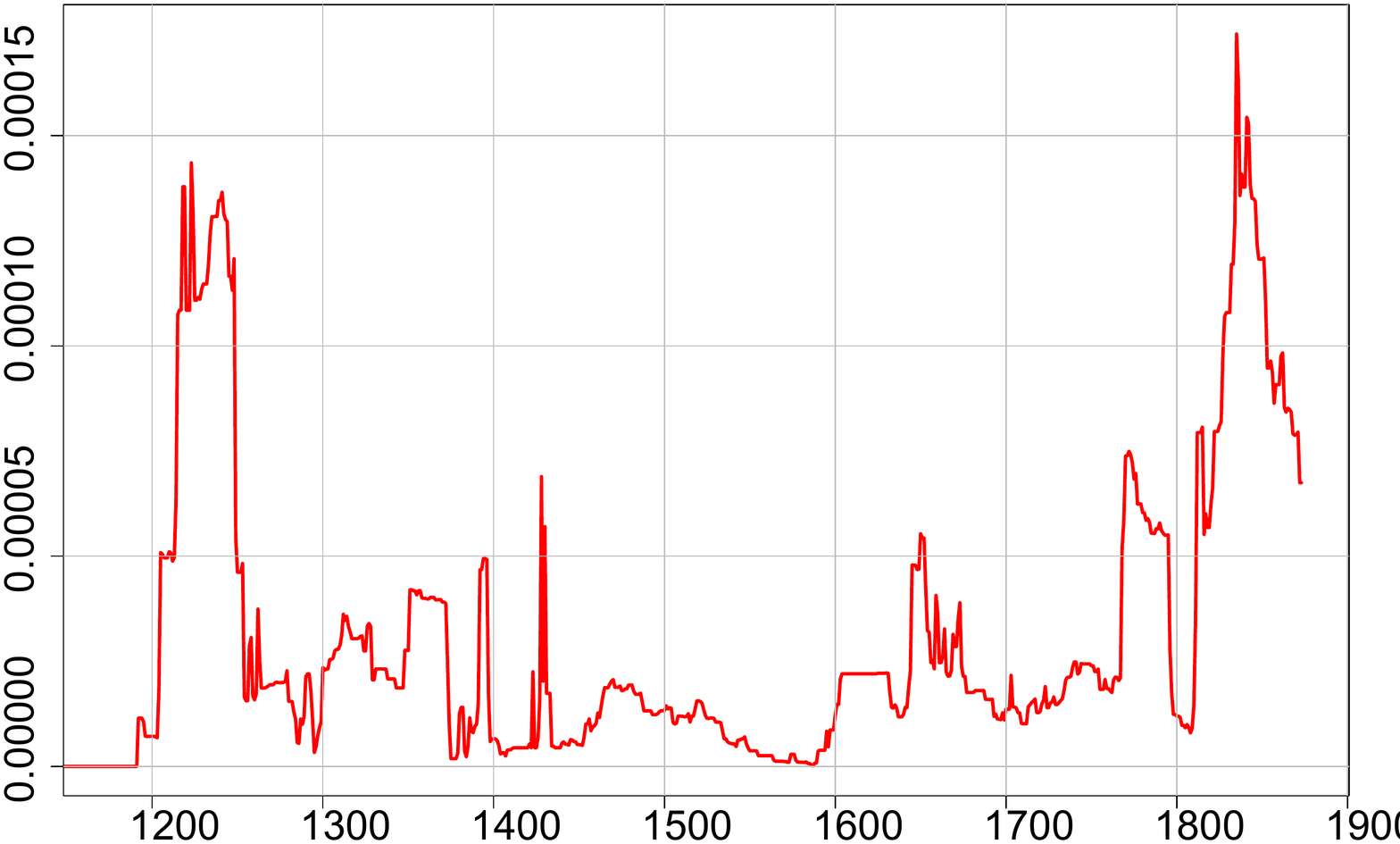}
    \end{array}$
    \caption{Data set 1104-1873}
     \label{fig:Data set 1104-1873}
\end{subfigure}
\\
\begin{subfigure}[t]{0.7\linewidth}
    \centering
    $\begin{array}{cc}
    \includegraphics[width=0.52\linewidth]{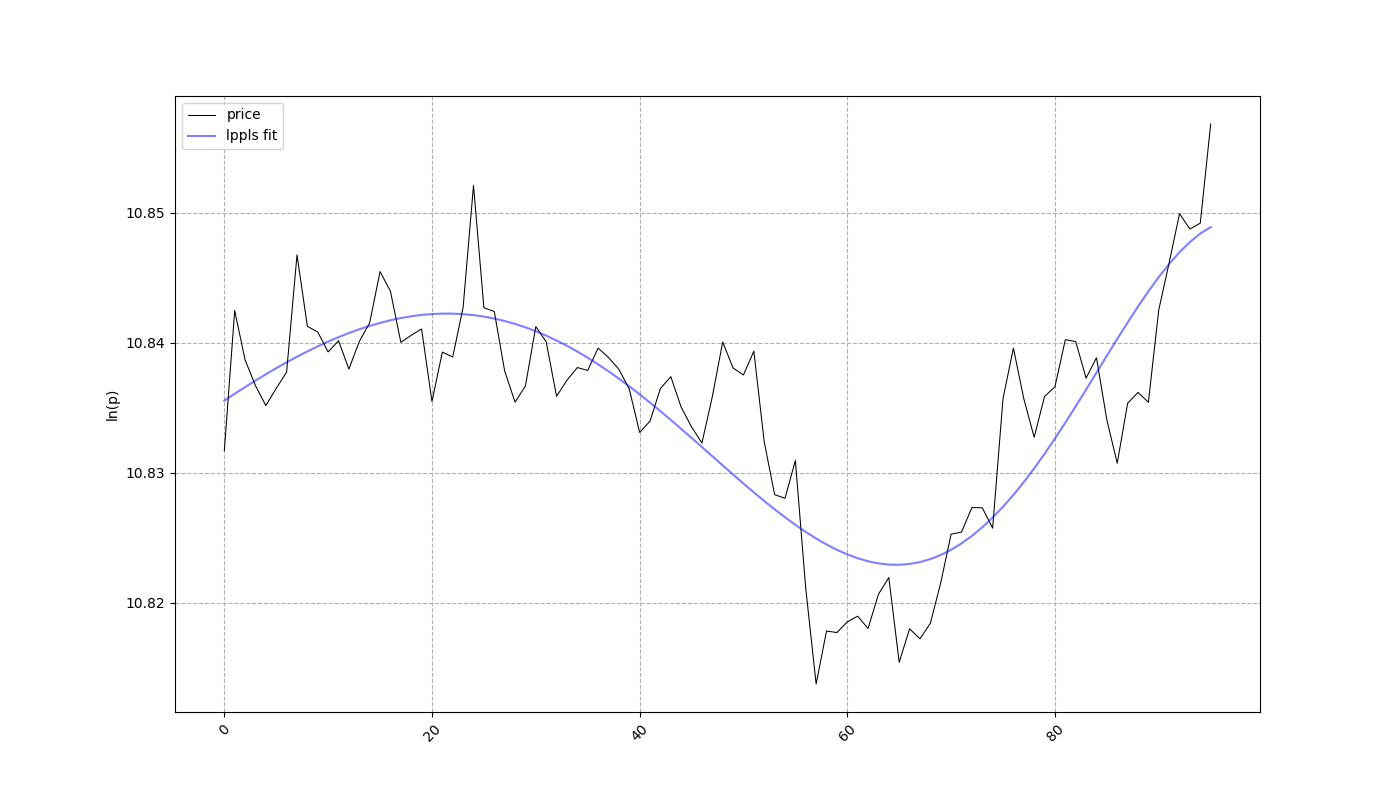}&
    \includegraphics[width=0.41\linewidth]{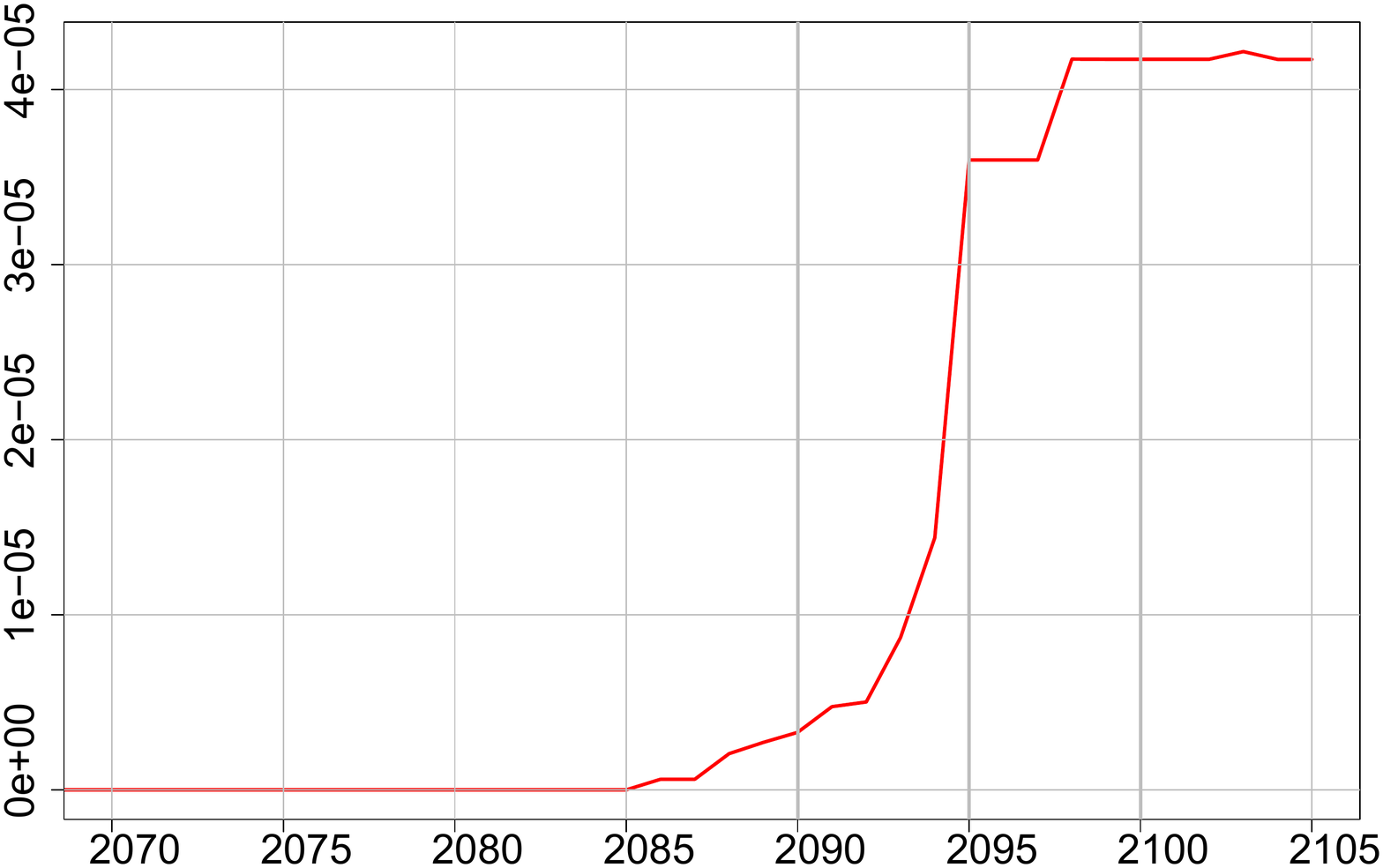}
    \end{array}$
    \caption{Data set 2010-2105}
     \label{fig:Data set 2010-2105}
\end{subfigure}
\\
\begin{subfigure}[t]{0.7\linewidth}
    \centering
    $\begin{array}{cc}
    \includegraphics[width=0.52\linewidth]{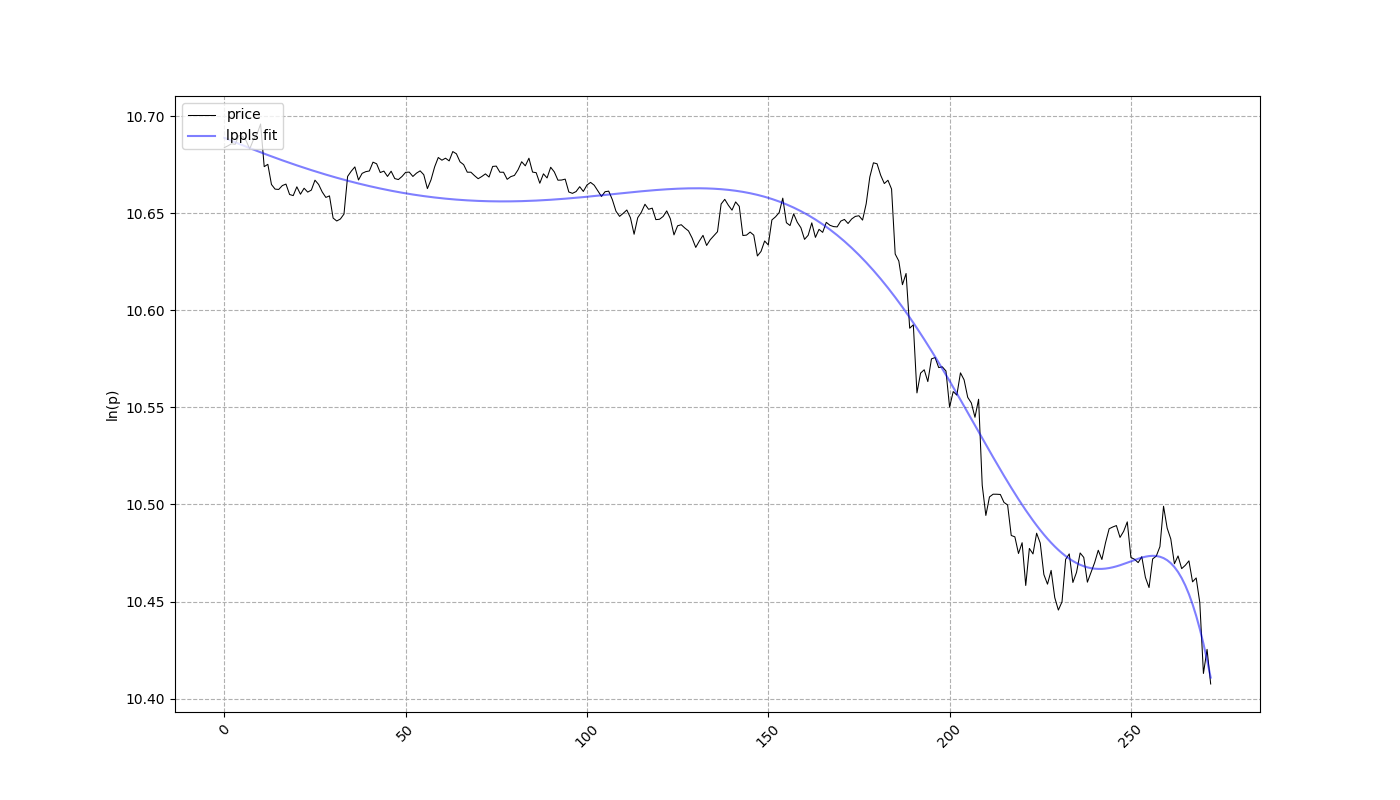}&
    \includegraphics[width=0.41\linewidth]{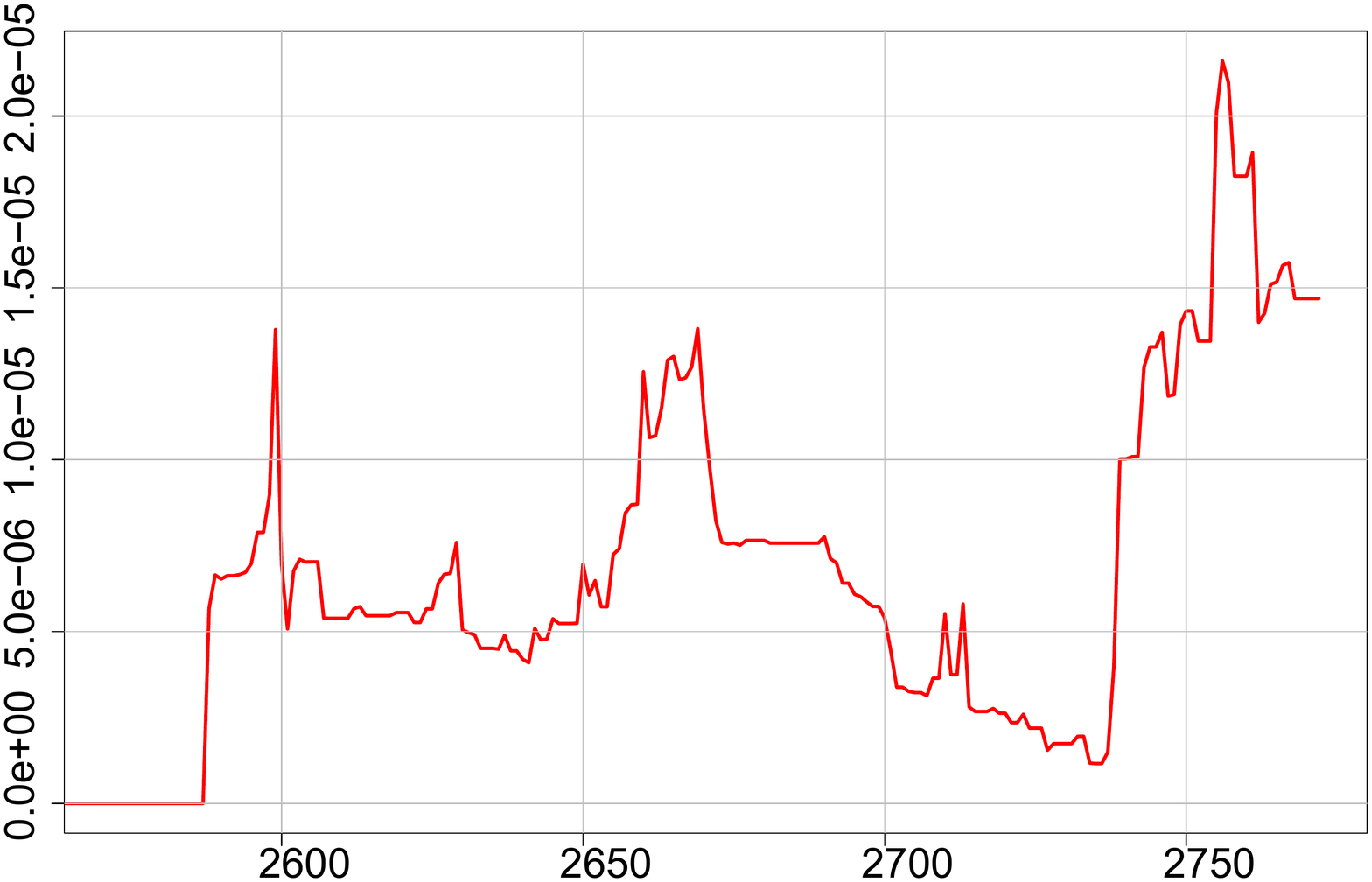}
    \end{array}$
    \caption{Data set 2500-2772}
     \label{fig:Data set 2500-2772}
\end{subfigure}
\\
\begin{subfigure}[t]{0.7\linewidth}
    \centering
 $\begin{array}{cc}
    \includegraphics[width=0.52\linewidth]{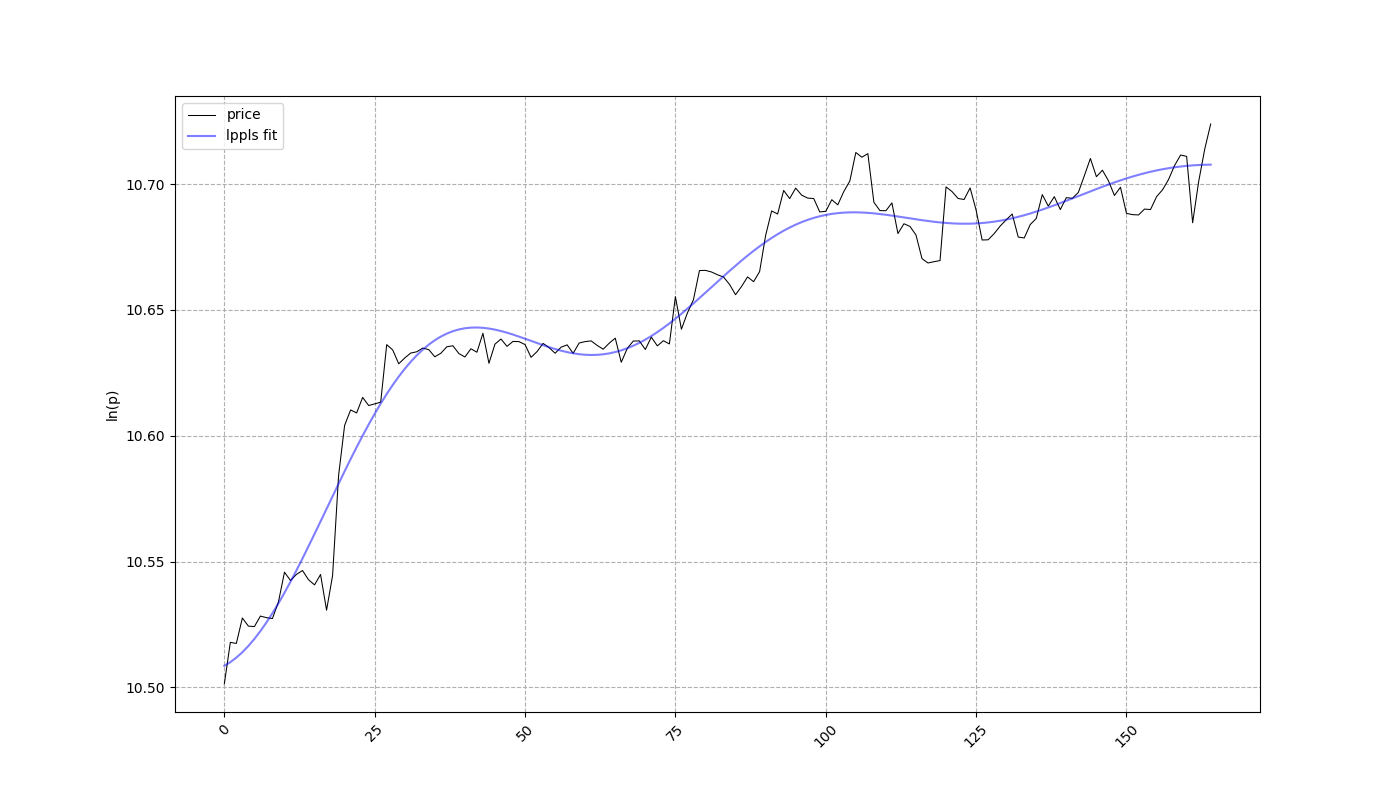}&
    \includegraphics[width=0.41\linewidth]{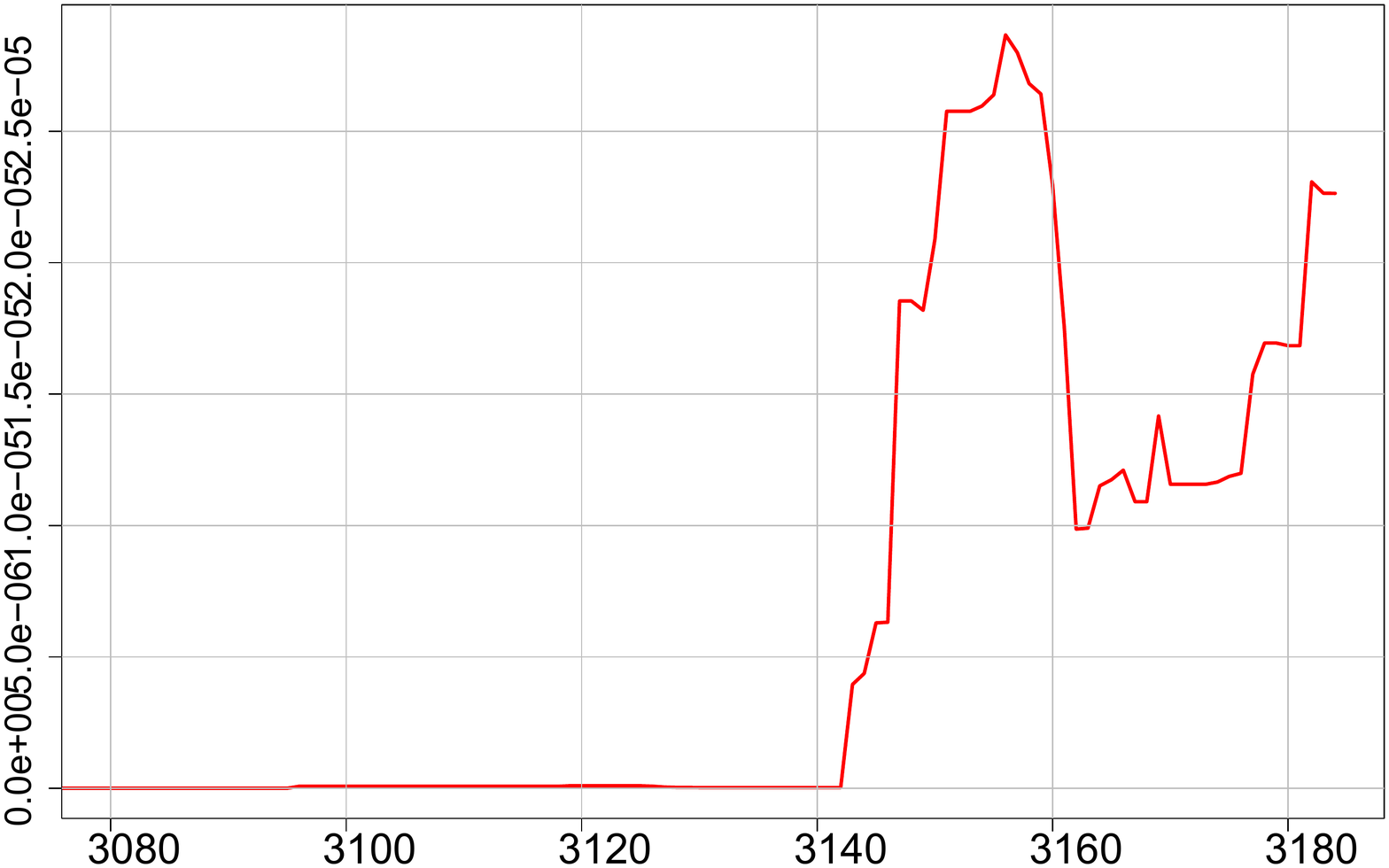}
    \end{array}$
    \caption{Data set 3020-3184}
     \label{fig:Data set 3020-3184}
 \end{subfigure}
 \\
  \begin{subfigure}[t]{0.7\linewidth}
    \centering
    $\begin{array}{cc}
    \includegraphics[width=0.52\linewidth]{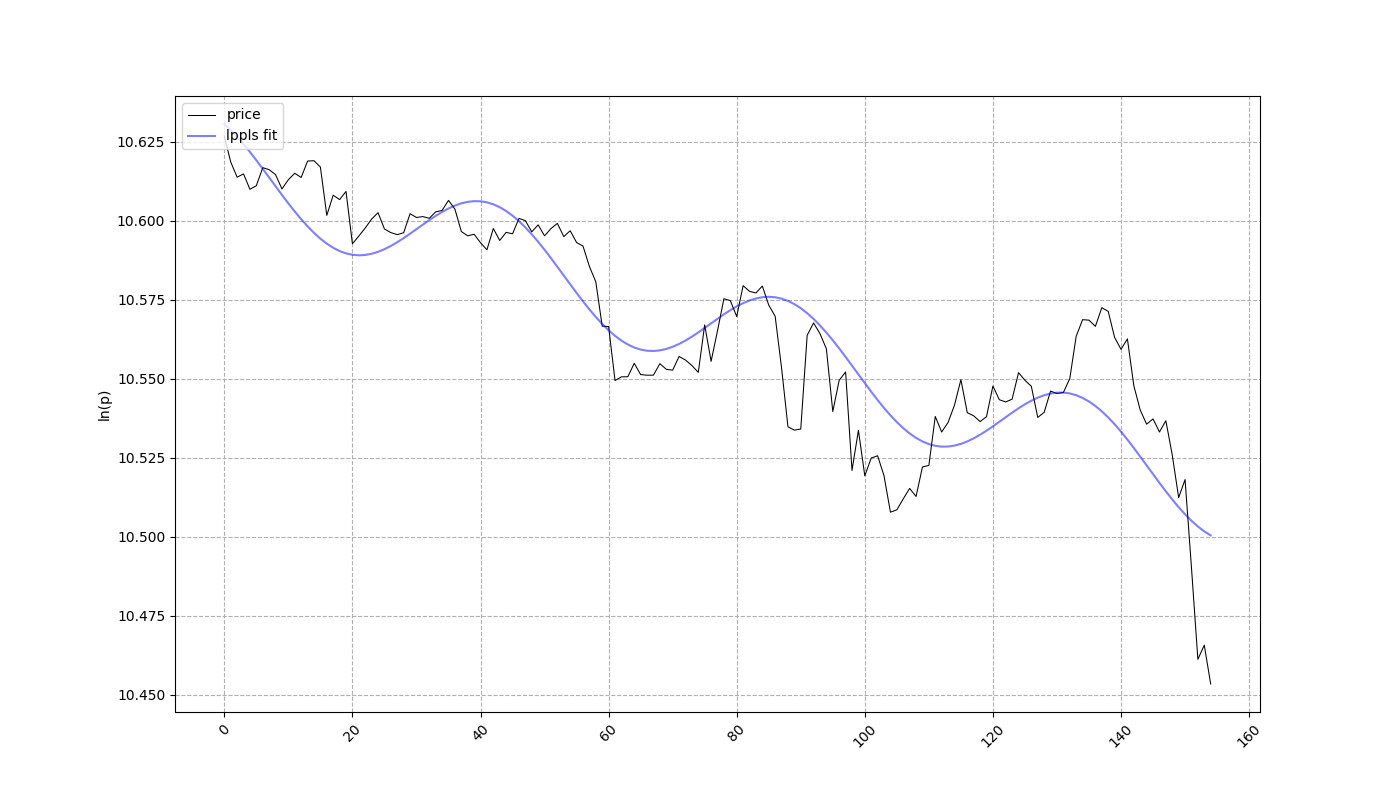}&
    \includegraphics[width=0.41\linewidth]{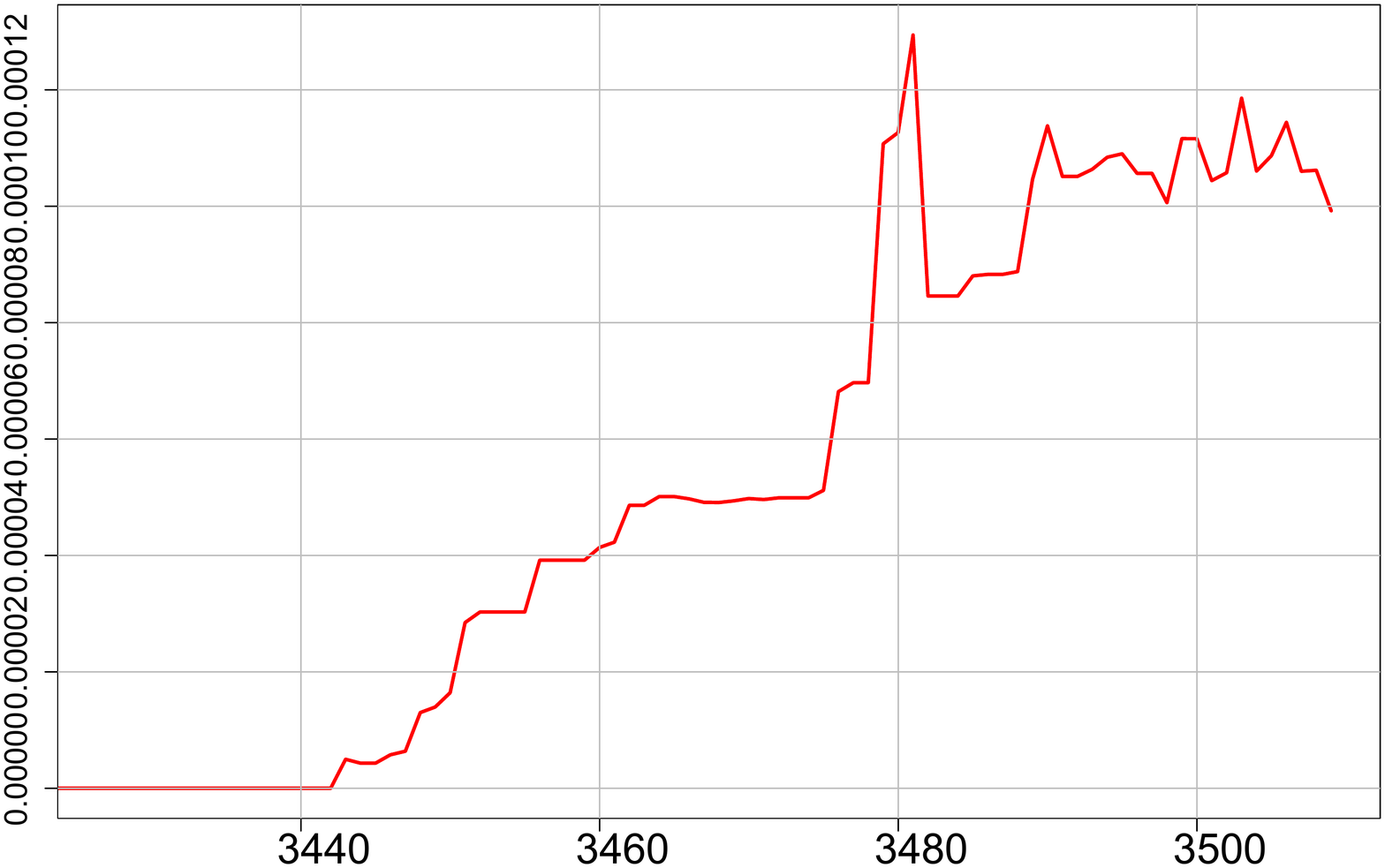}
    \end{array}$
    \caption{Data set 3355-3509}
     \label{fig:Data set 3355-3509}
 \end{subfigure}
 \end{figure}

 \begin{figure}[H]
\ContinuedFloat
 \begin{subfigure}[t]{0.7\linewidth}
    \centering
    $\begin{array}{cc}
    \includegraphics[width=0.52\linewidth]{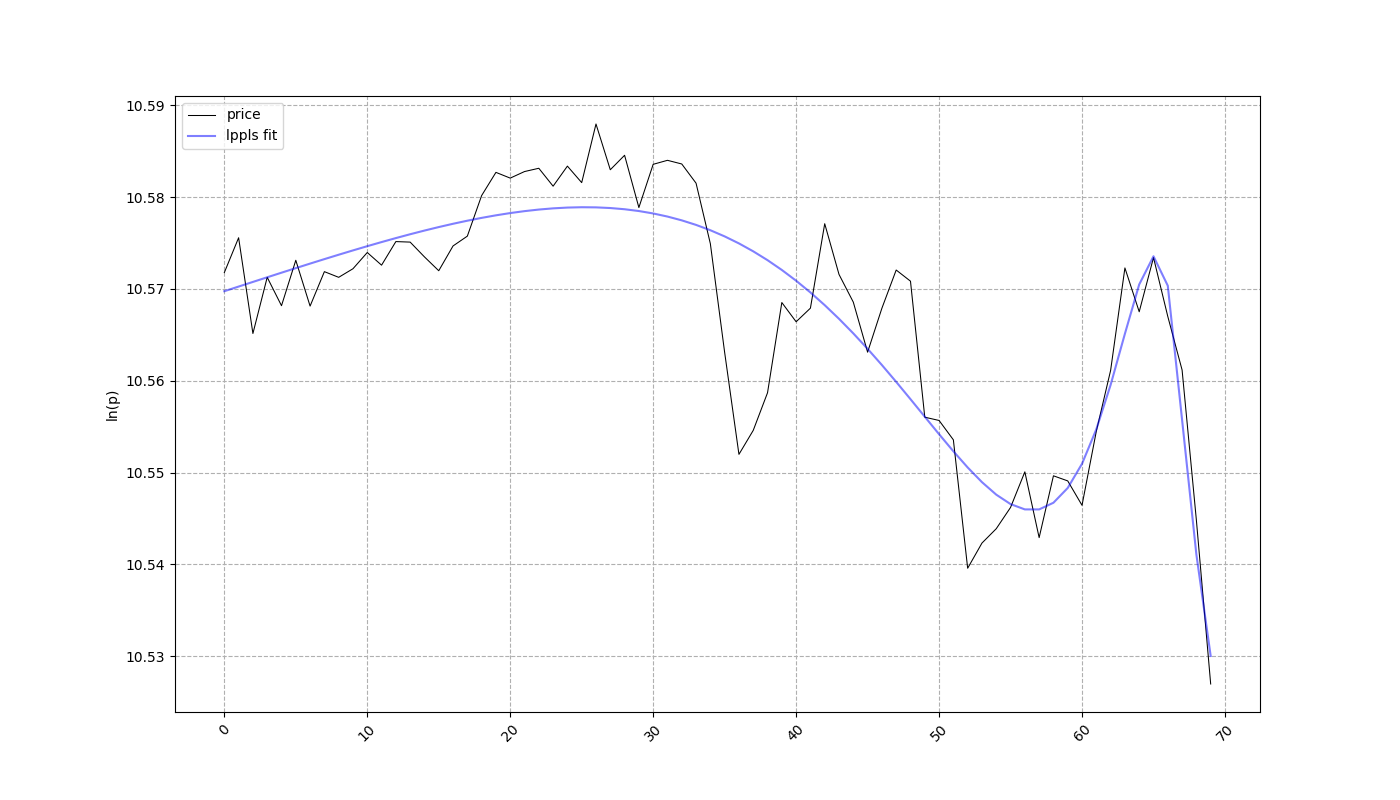}&
    \includegraphics[width=0.41\linewidth]{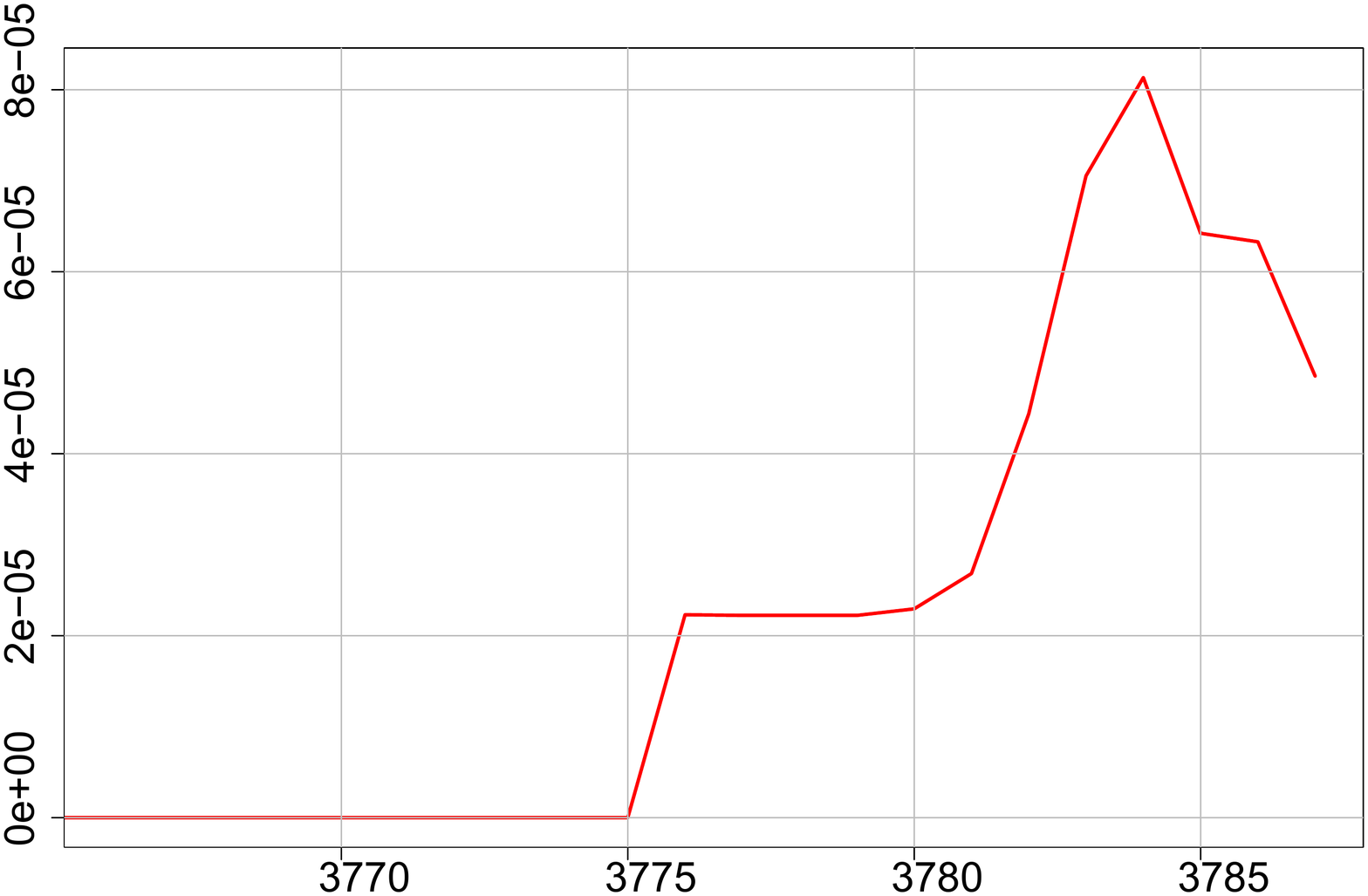}
    \end{array}$
    \caption{Data set 3718-3787}
     \label{fig:Data set 3718-3787}
\end{subfigure}
\\
\begin{subfigure}[t]{0.7\linewidth}
    \centering
    $\begin{array}{cc}
    \includegraphics[width=0.52\linewidth]{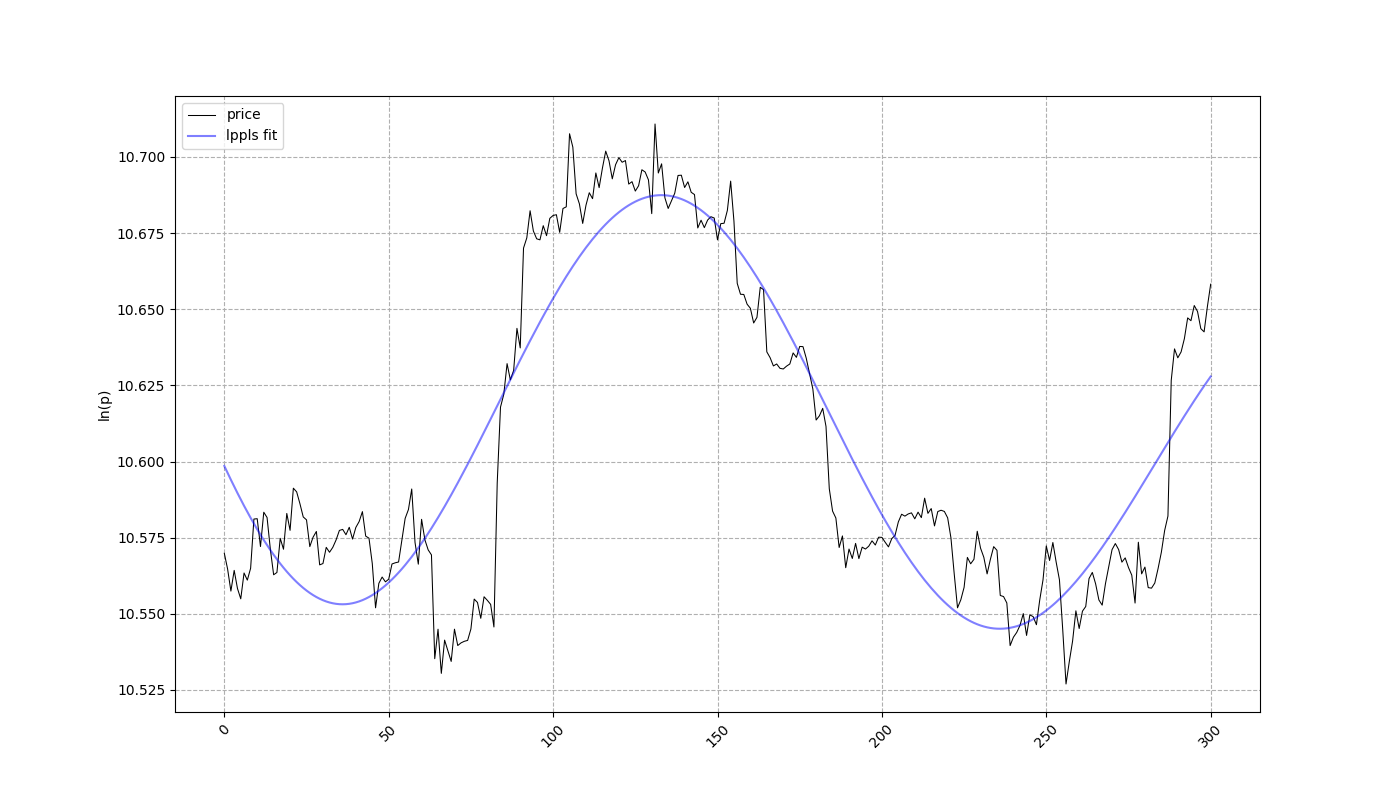}&
    \includegraphics[width=0.41\linewidth]{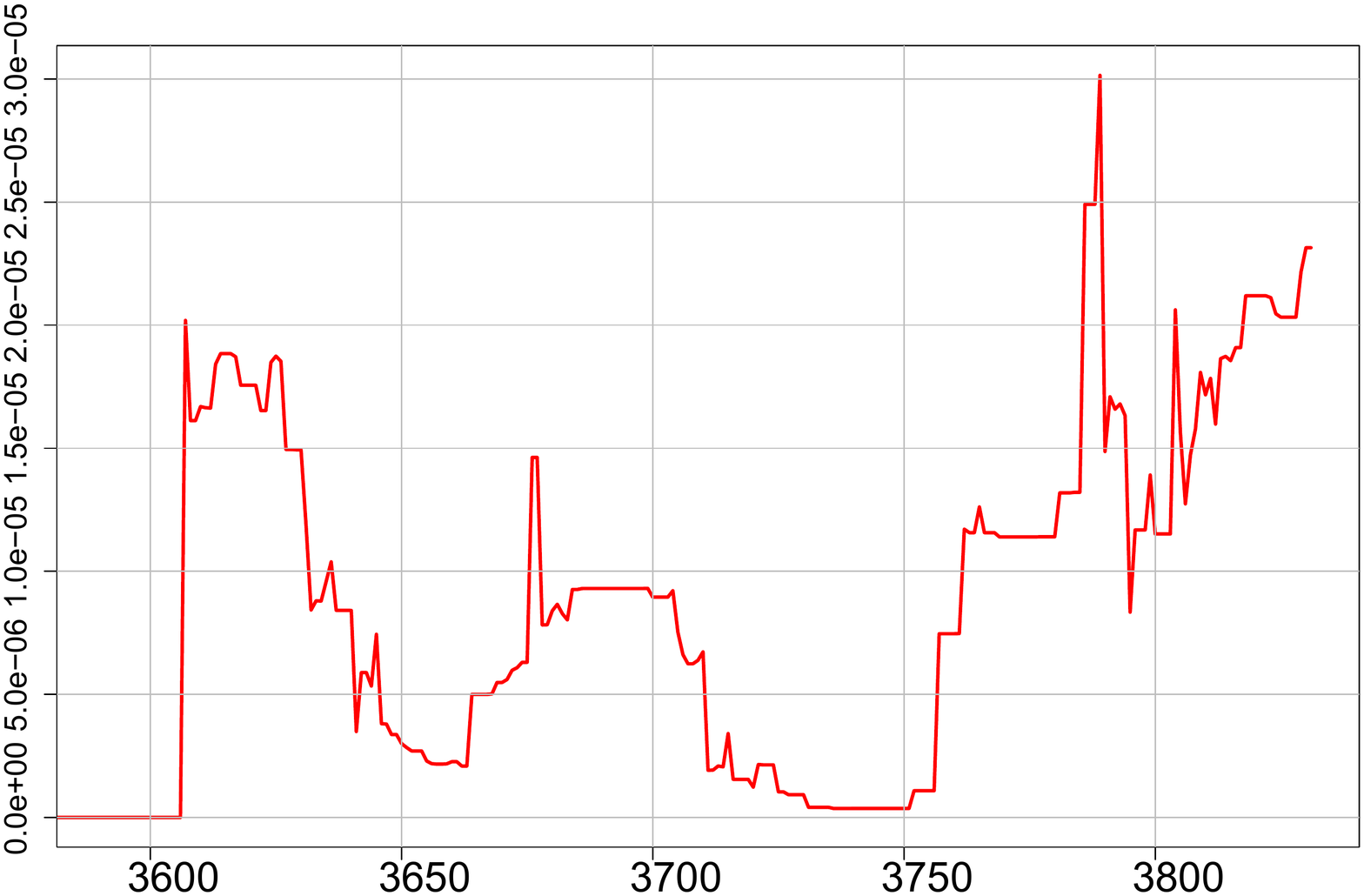}
    \end{array}$
    \caption{Data set 3531-3831}
     \label{fig:Data set 3531-3831}
 \end{subfigure}
 \\
 \begin{subfigure}[t]{0.7\linewidth}
    \centering
    $\begin{array}{cc}
    \includegraphics[width=0.52\linewidth]{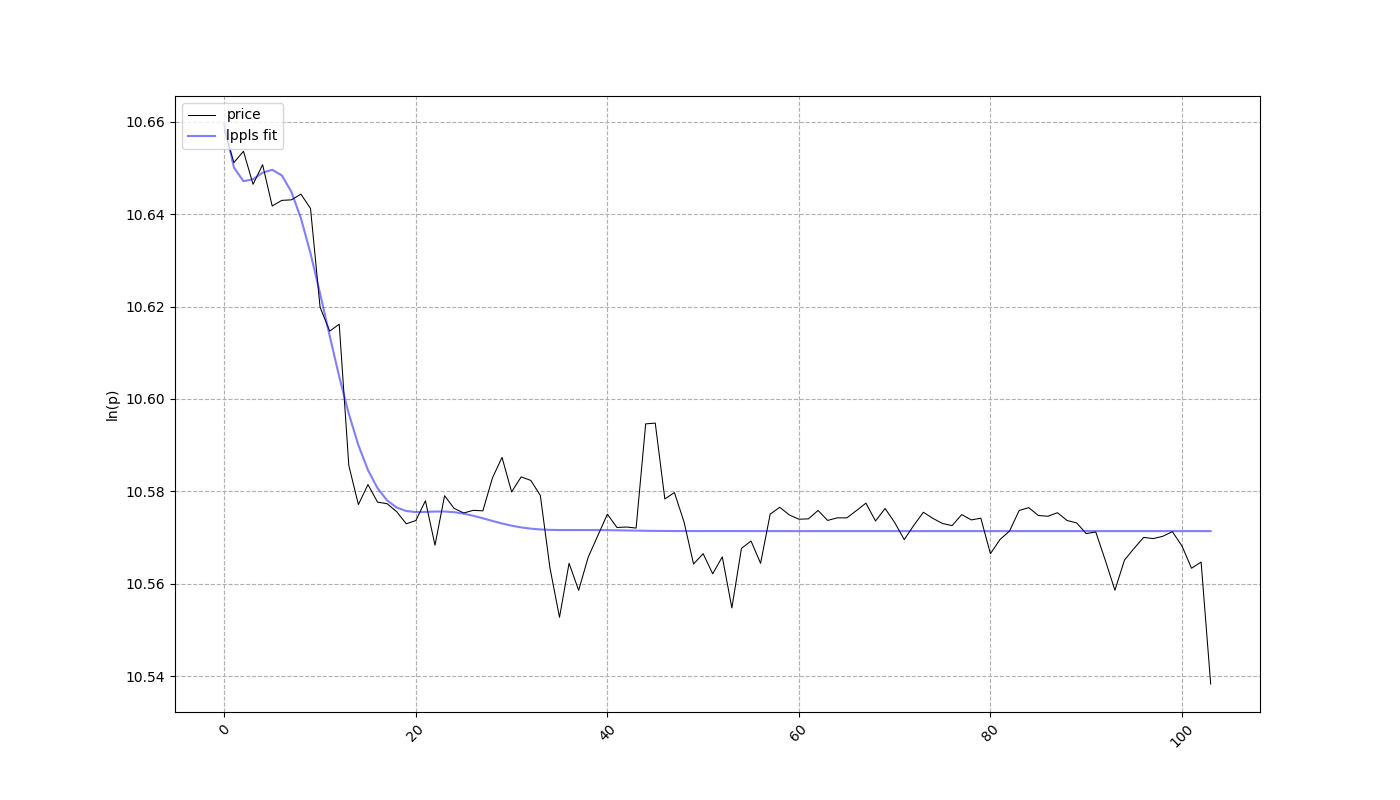}&
    \includegraphics[width=0.41\linewidth]{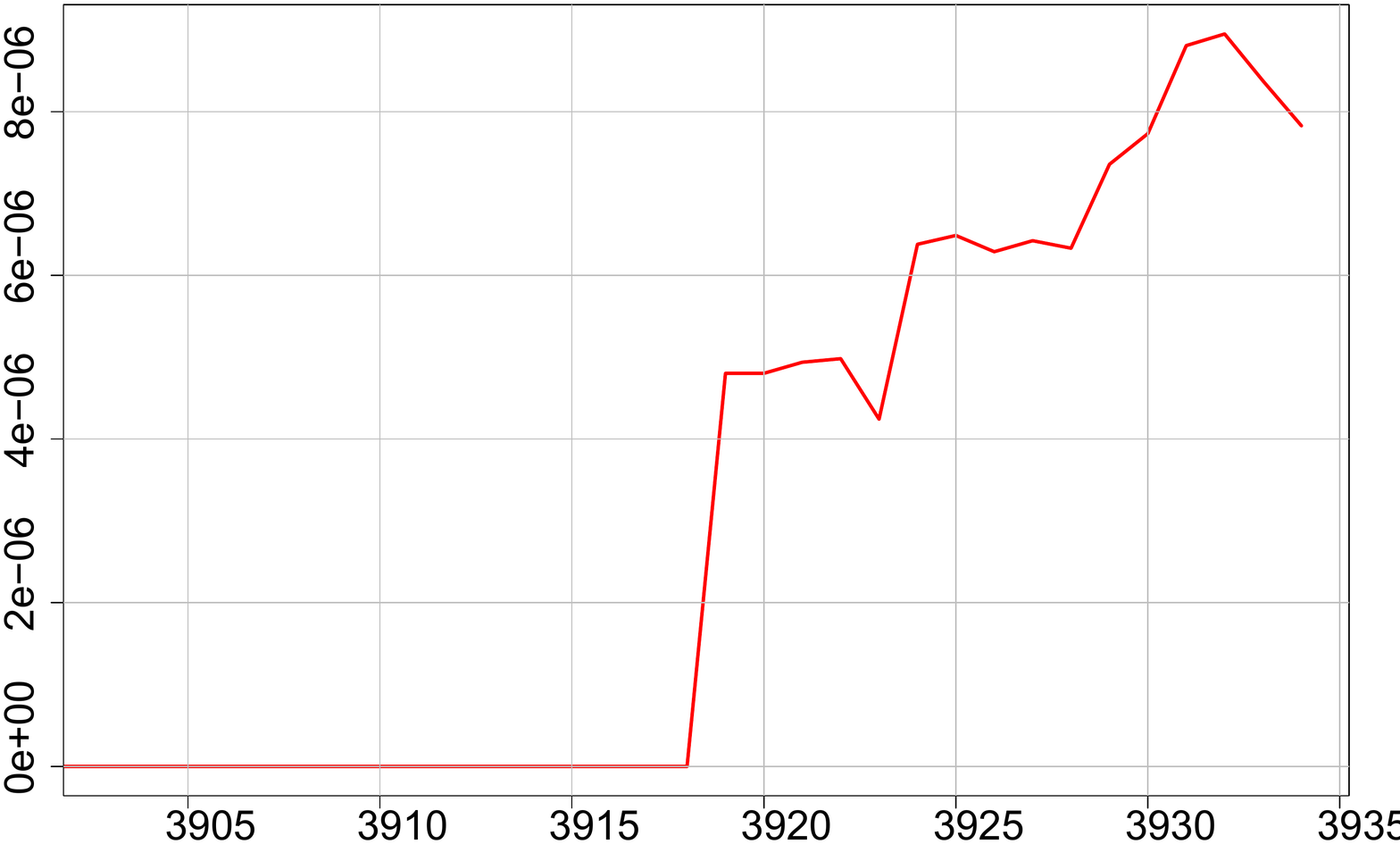}
    \end{array}$
    \caption{Data set 3831-3934}
     \label{fig:Data set 3831-3934}
 \end{subfigure}
 \\
 \begin{subfigure}[t]{0.7\linewidth}
    \centering
    $\begin{array}{cc}
    \includegraphics[width=0.52\linewidth]{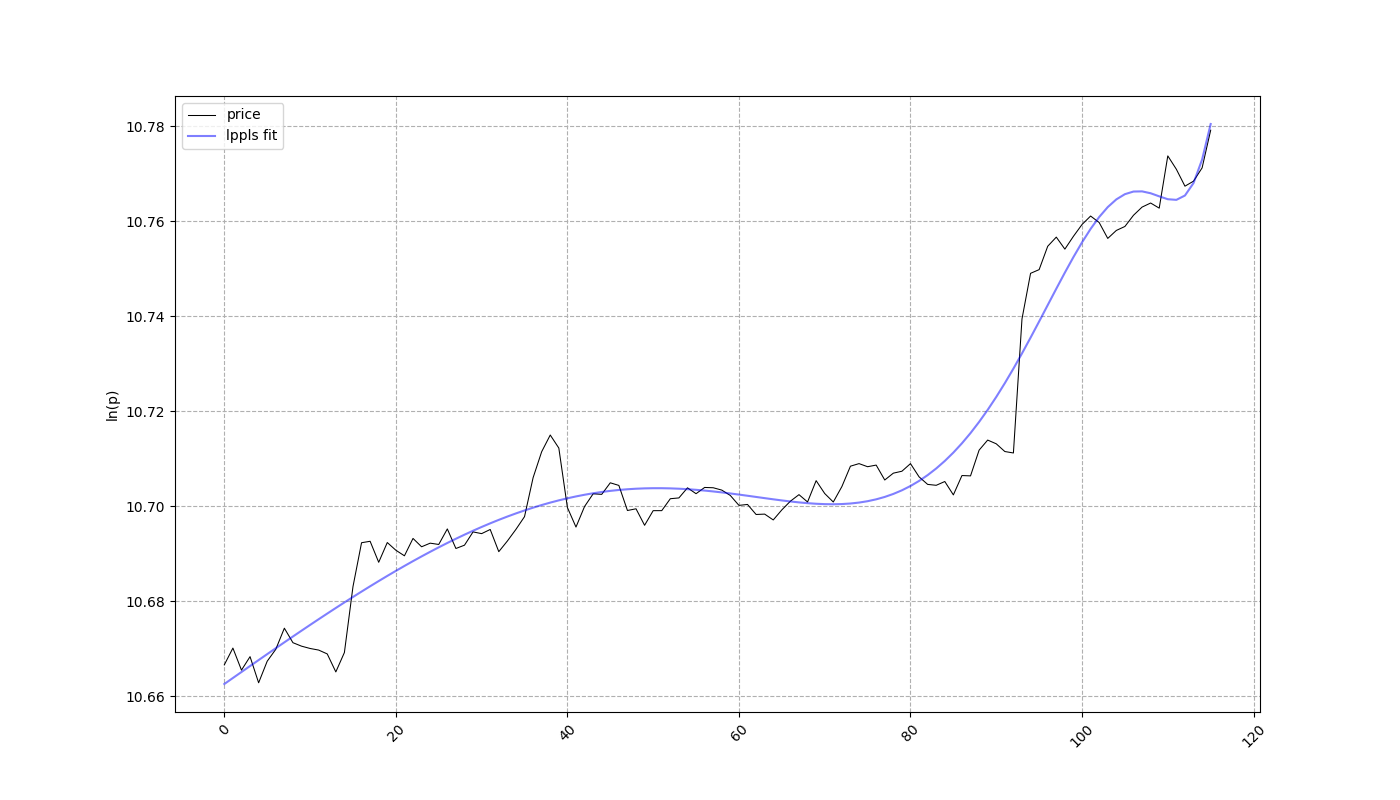}&
    \includegraphics[width=0.41\linewidth]{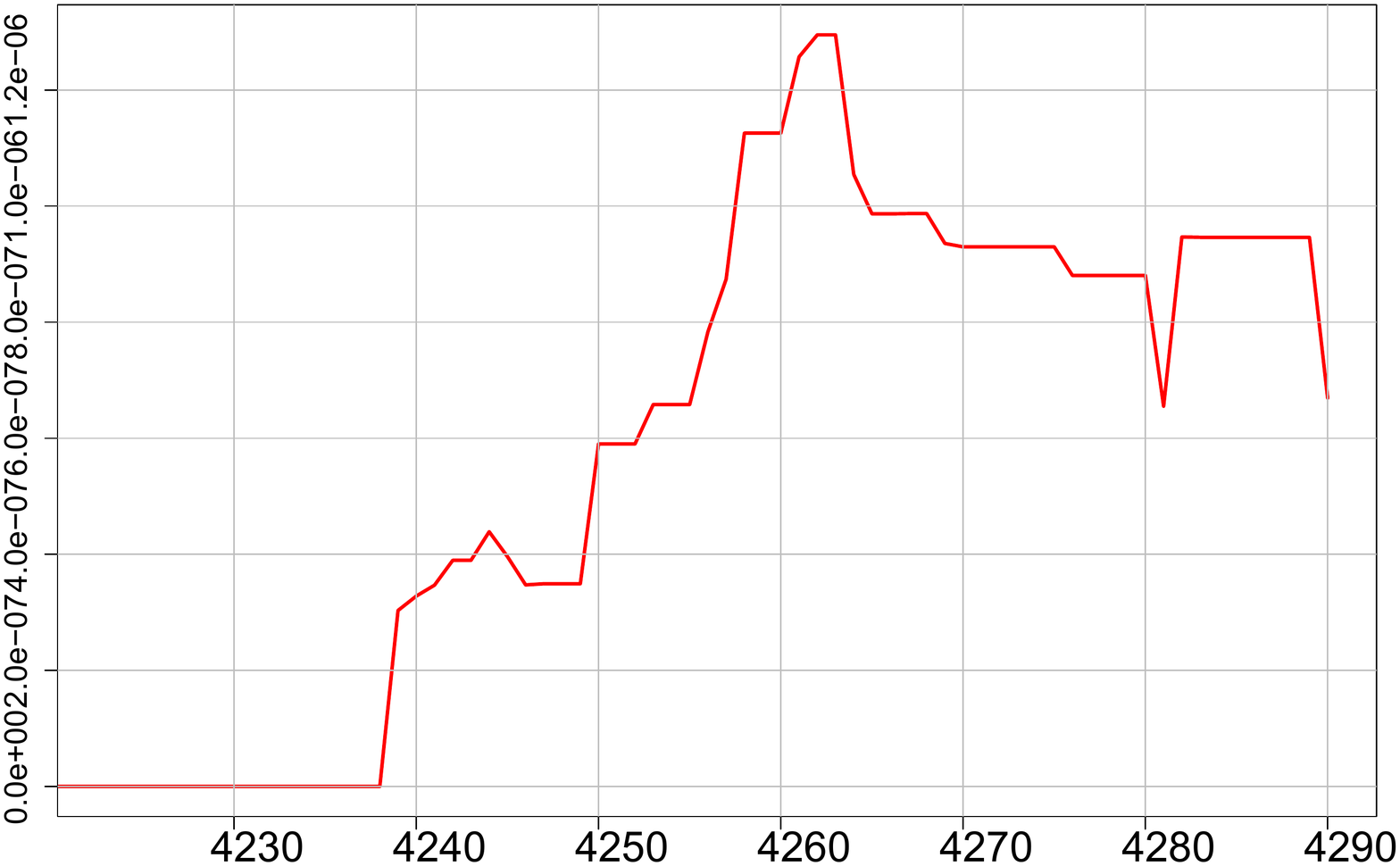}
    \end{array}$
    \caption{Data set 4175-4290}
     \label{fig:Data set 4175-4290}
 \end{subfigure}
 \\
 \begin{subfigure}[t]{0.7\linewidth}
    \centering
    $\begin{array}{cc}
    \includegraphics[width=0.52\linewidth]{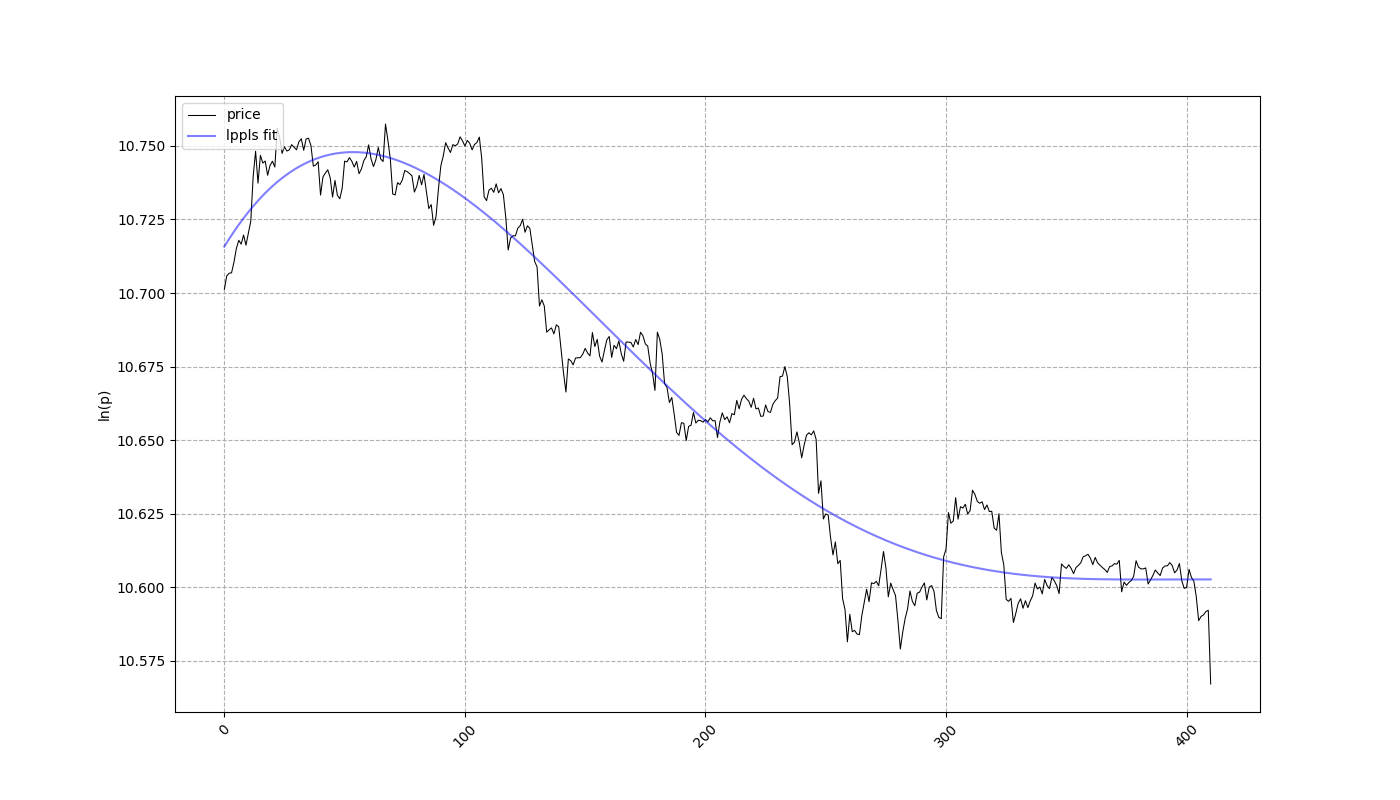}&
    \includegraphics[width=0.41\linewidth]{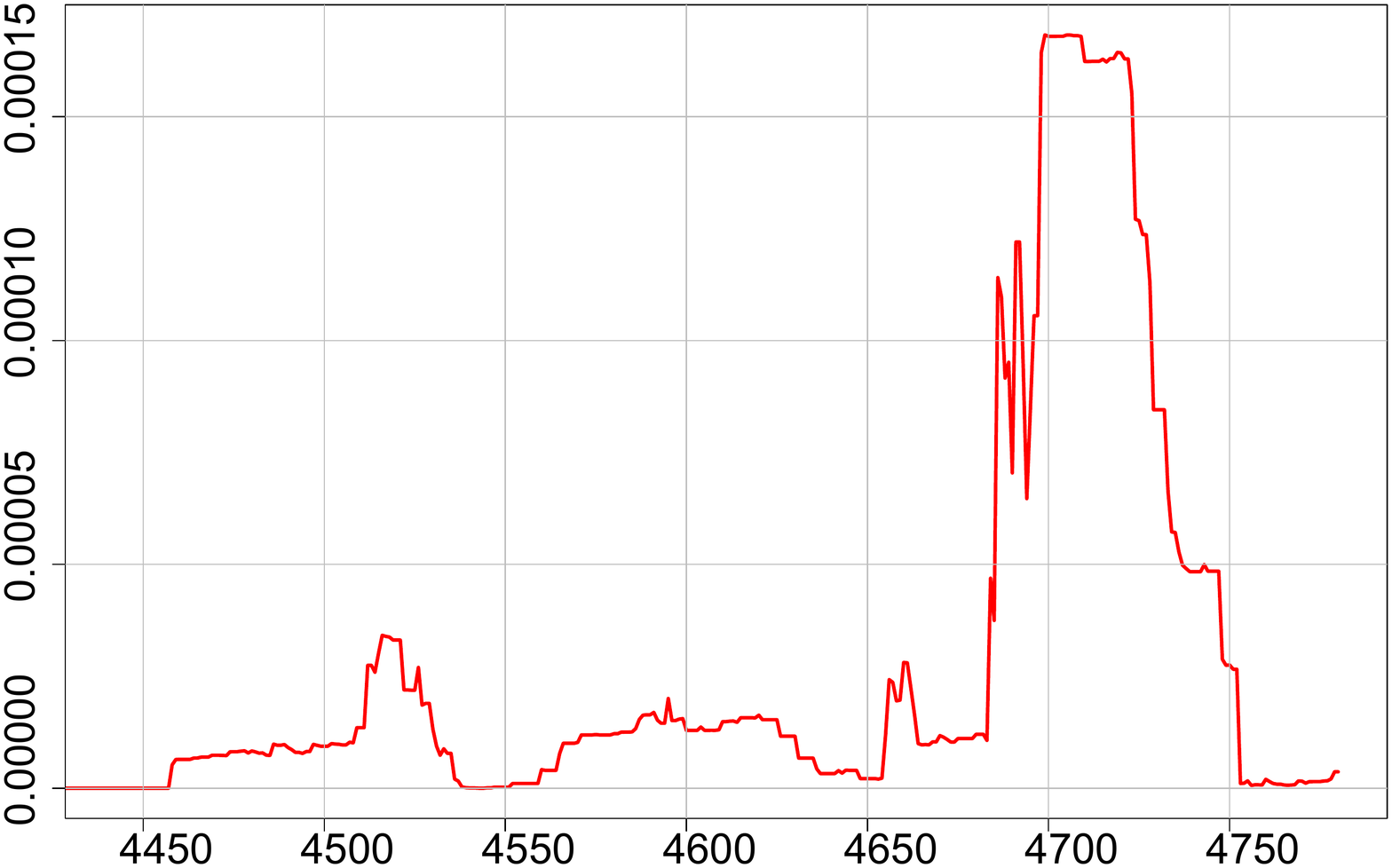}
    \end{array}$
    \caption{Data set 4370-4780}
     \label{fig:Data set 4370-4780}
\end{subfigure}
\\
\begin{subfigure}[t]{0.7\linewidth}
    \centering
    $\begin{array}{cc}
    \includegraphics[width=0.52\linewidth]{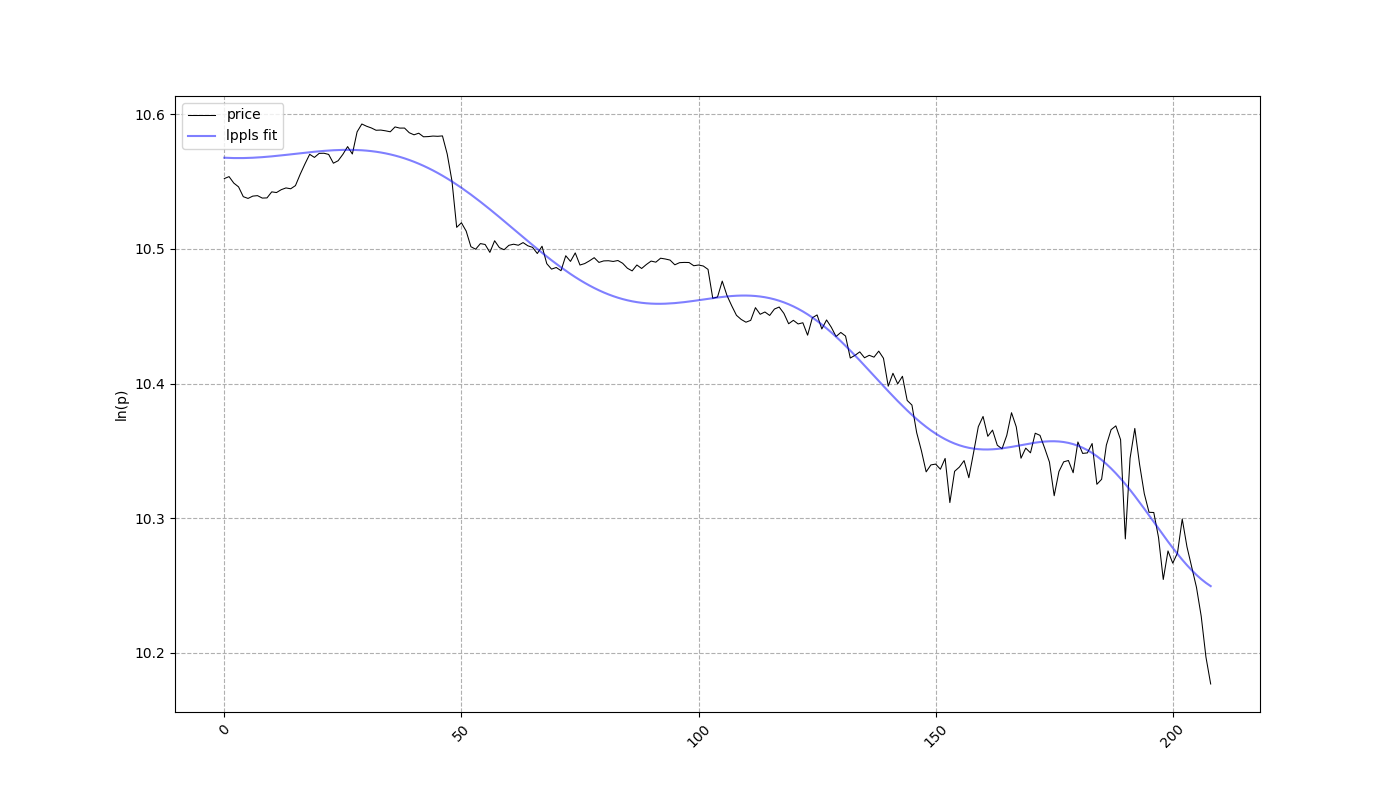}&
    \includegraphics[width=0.41\linewidth]{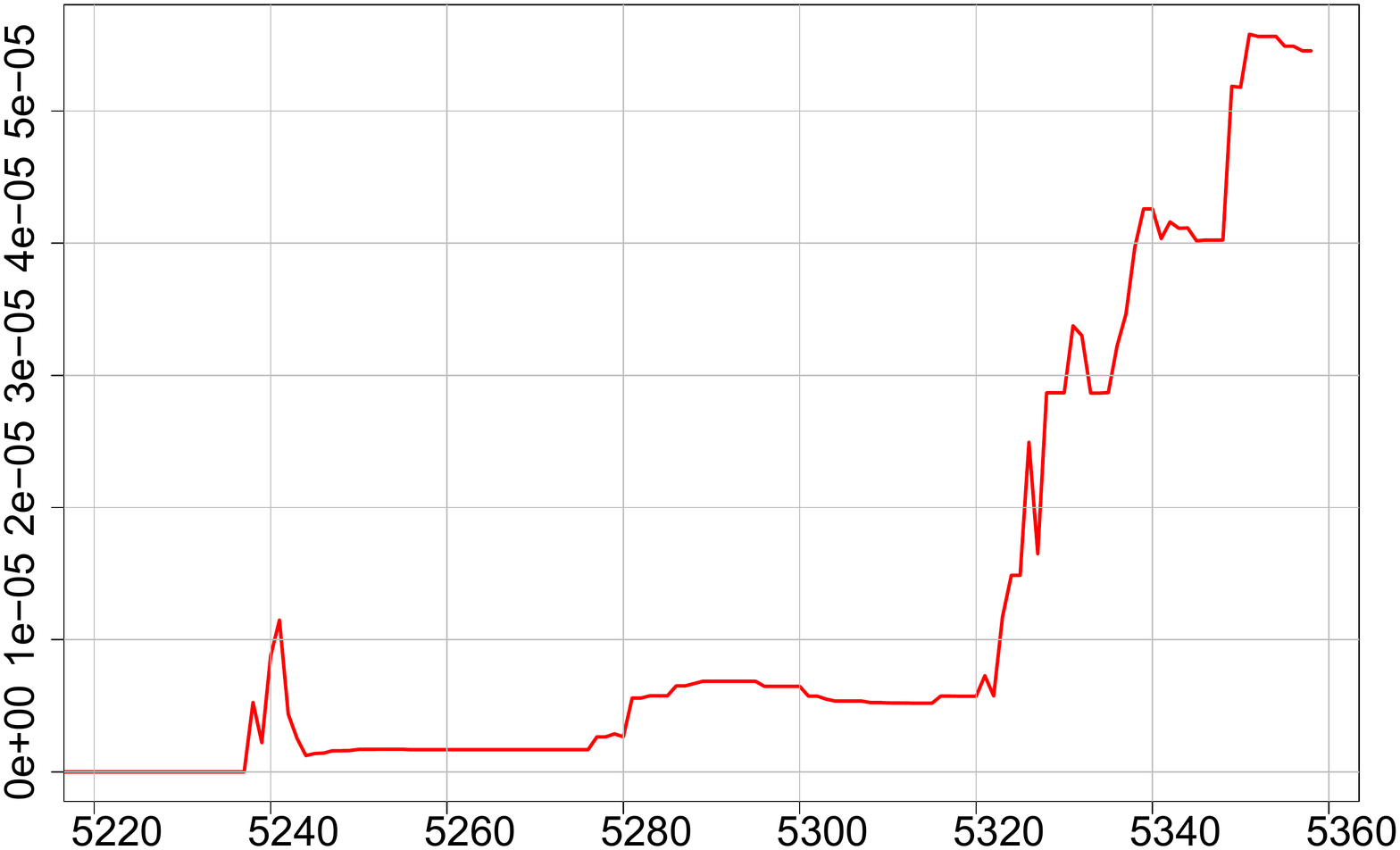}
    \end{array}$
    \caption{Data set 5150-5358}
     \label{fig:Data set 5150-5358}
\end{subfigure}
\end{figure}

\begin{figure}[H]
\ContinuedFloat
\centering
\begin{subfigure}[t]{0.7\linewidth}
    \centering
    $\begin{array}{cc}
    \includegraphics[width=0.52\linewidth]{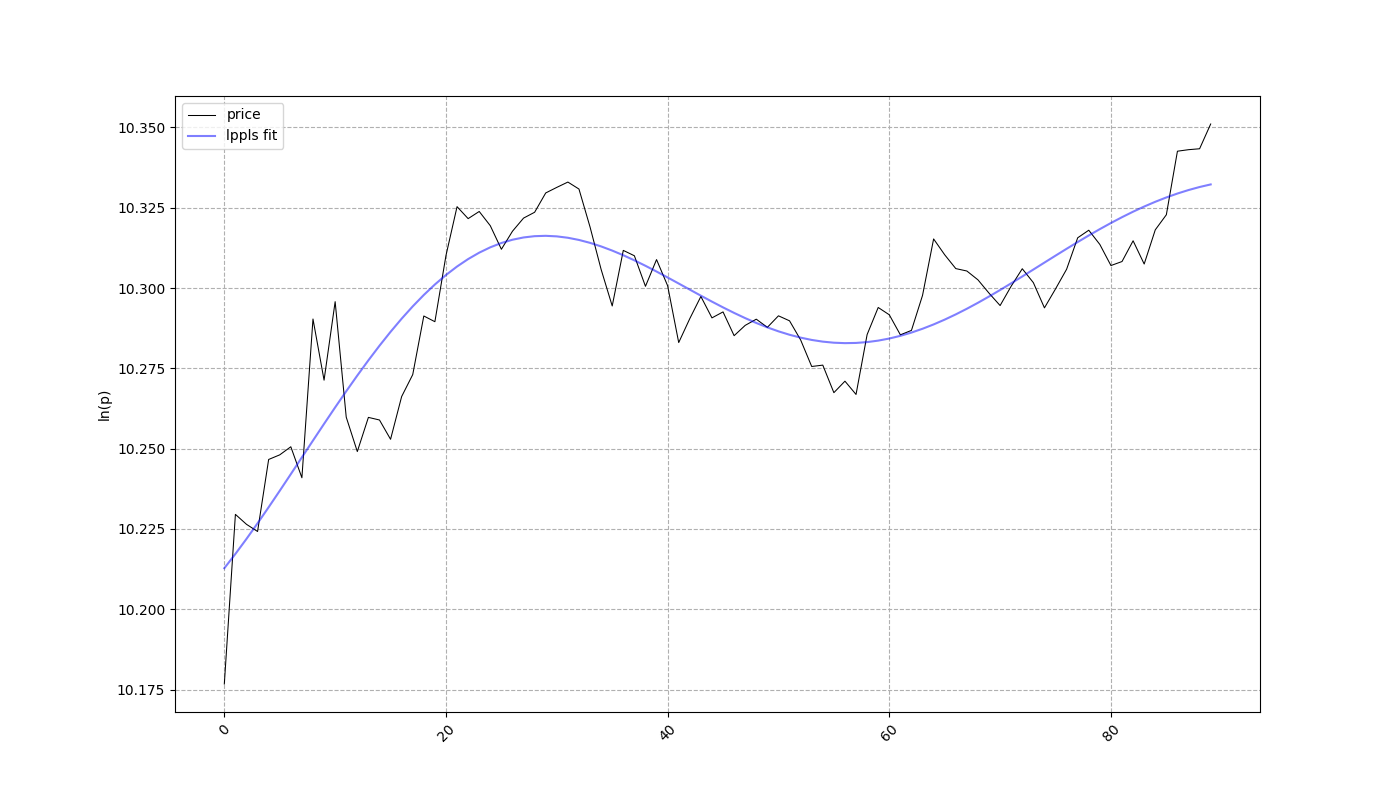}&
    \includegraphics[width=0.41\linewidth]{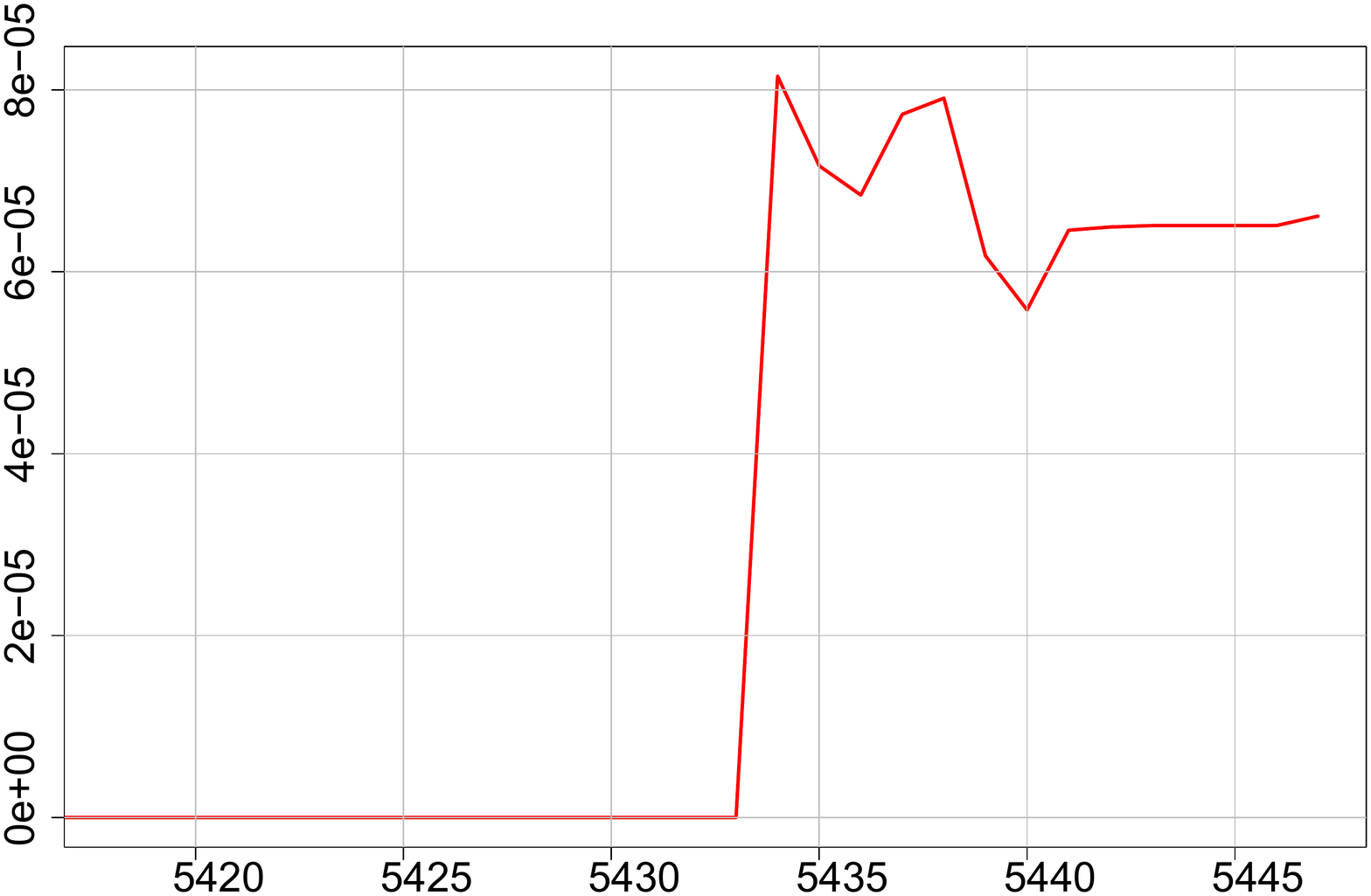}
    \end{array}$
    \caption{Data set 5358-5447}
     \label{fig:Data set 5358-5447}
 \end{subfigure}
 \\
\begin{subfigure}[t]{0.7\linewidth}
    \centering
    $\begin{array}{cc}
    \includegraphics[width=0.52\linewidth]{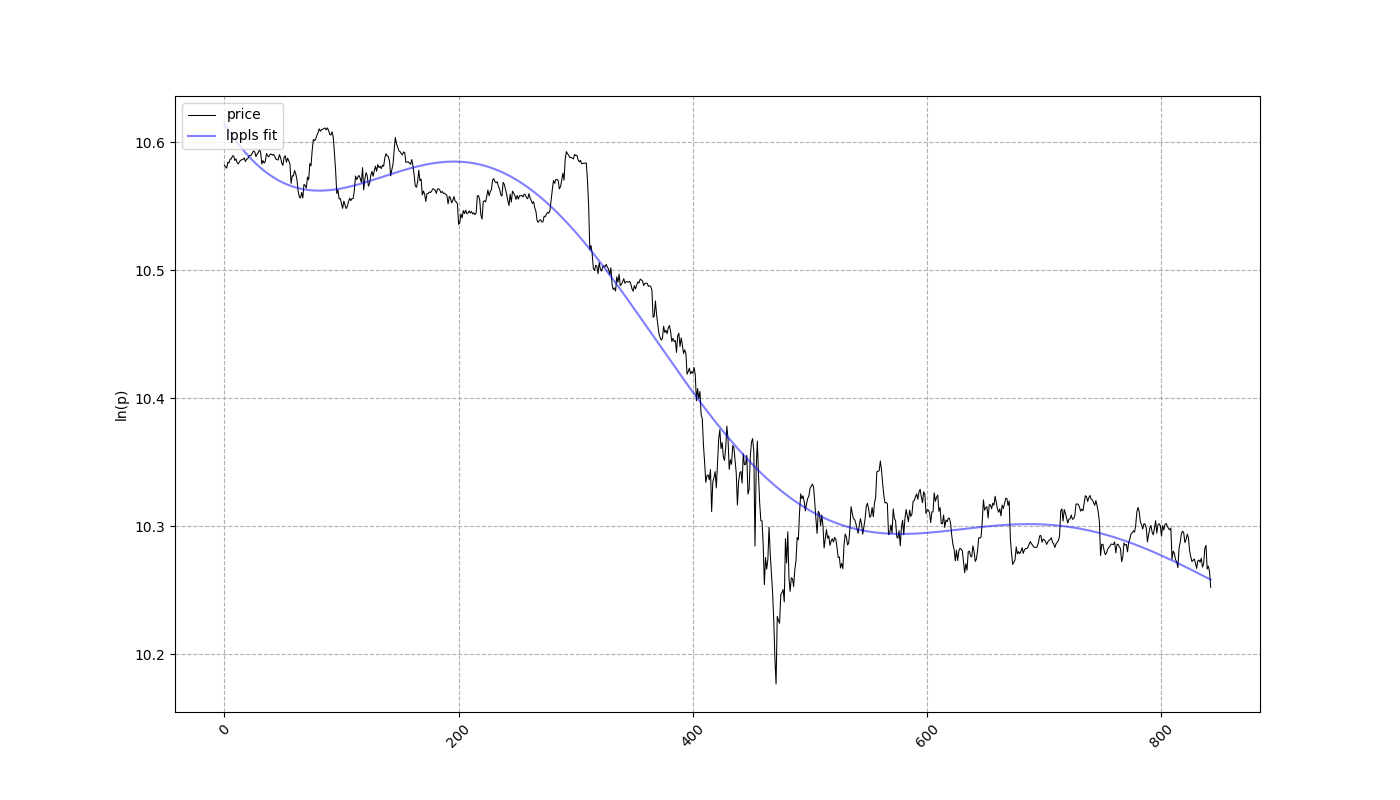}&
    \includegraphics[width=0.41\linewidth]{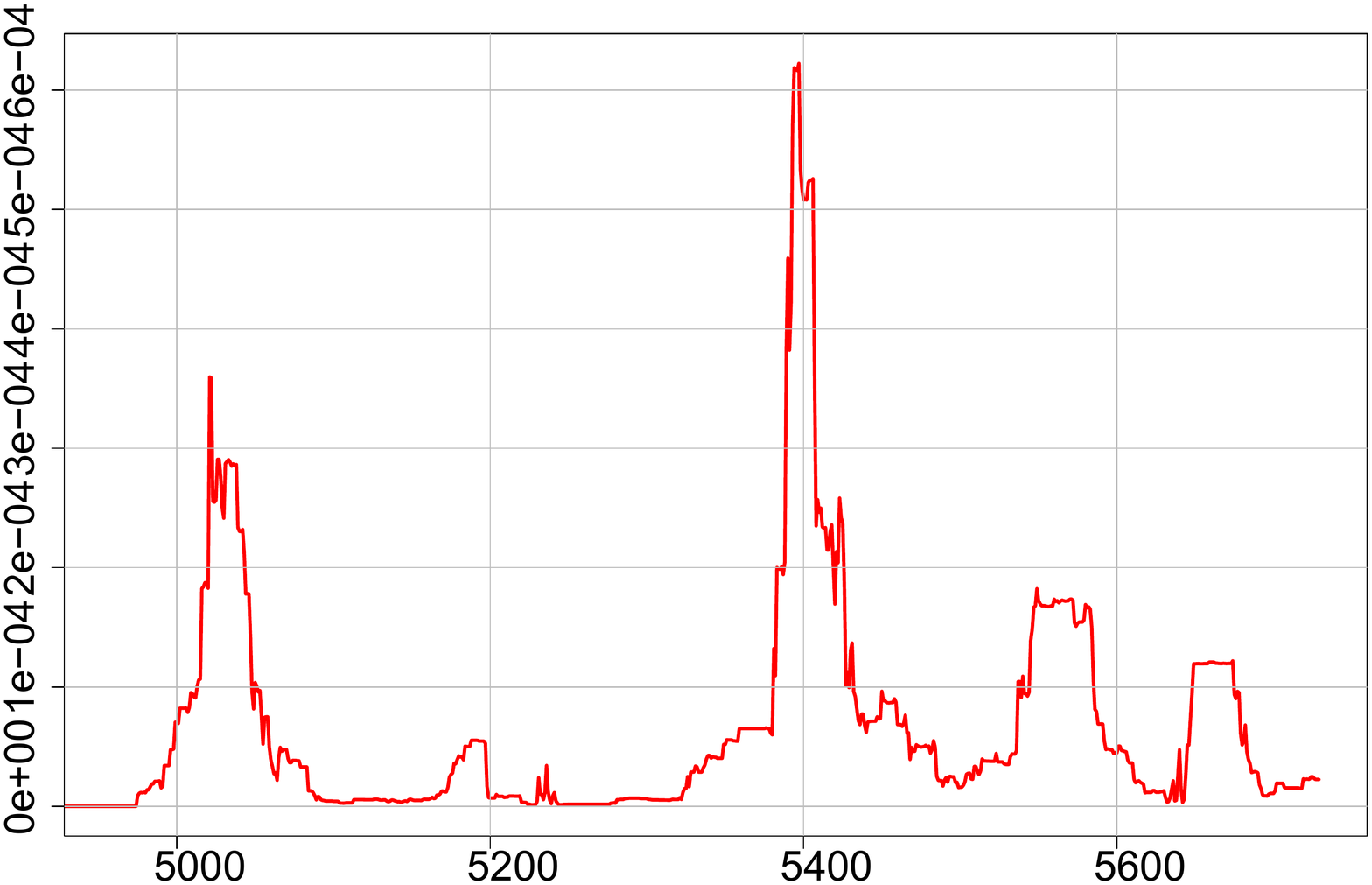}
    \end{array}$
    \caption{Data set 4887-5729}
     \label{fig:Data set 4887-5729}
\end{subfigure}
\\
\begin{subfigure}[t]{0.7\linewidth}
    \centering
    $\begin{array}{cc}
    \includegraphics[width=0.52\linewidth]{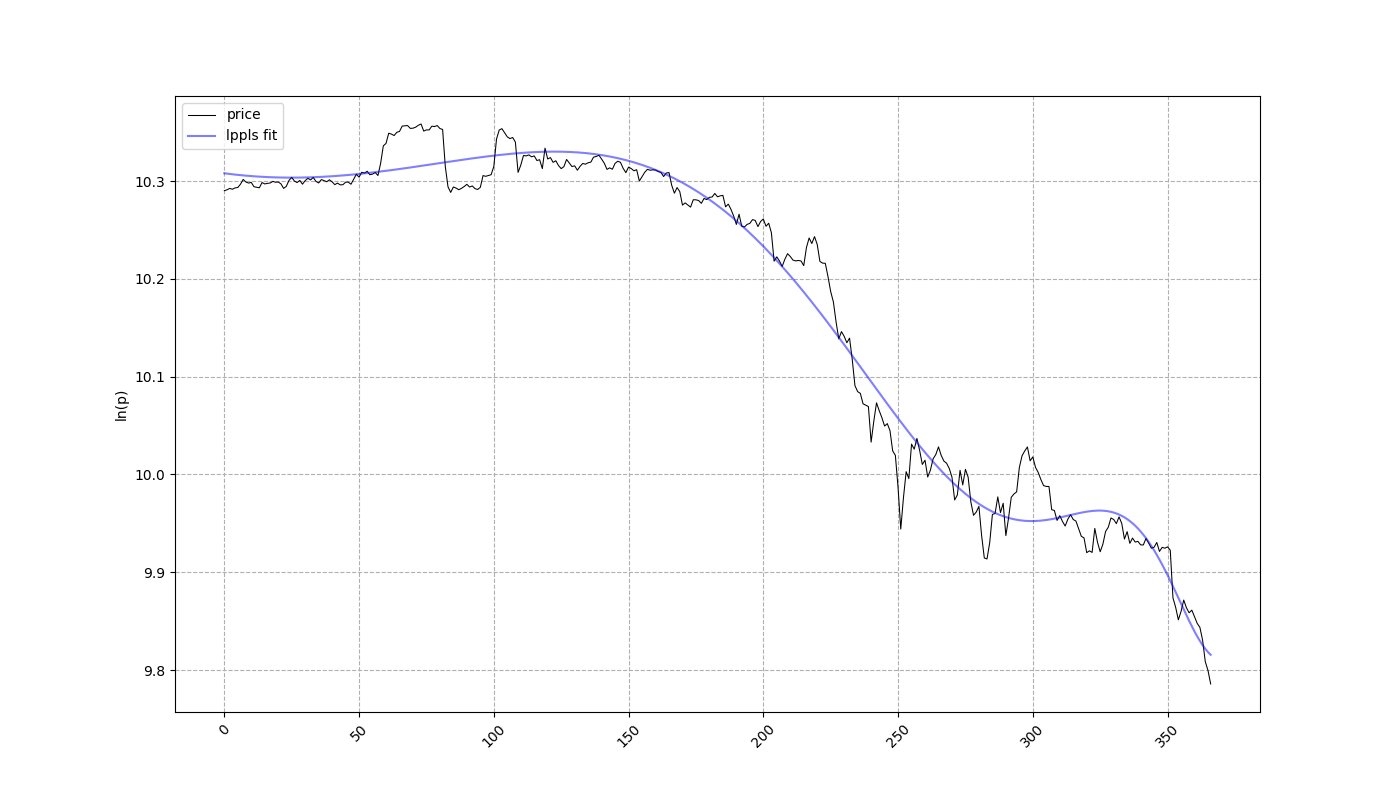}&
    \includegraphics[width=0.41\linewidth]{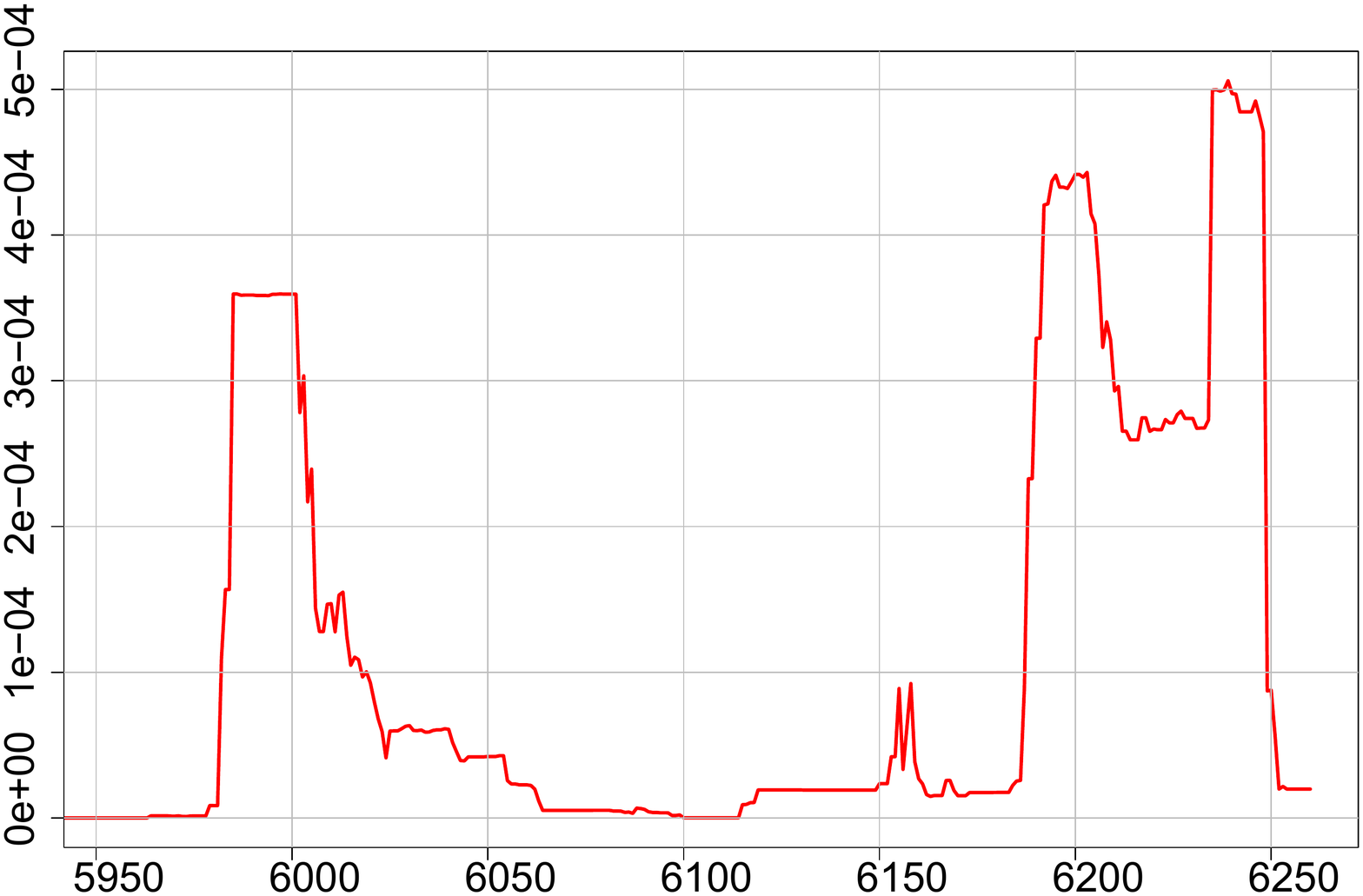}
    \end{array}$
    \caption{Data set 5894-6260}
     \label{fig:Data set 5894-6260}
 \end{subfigure}
 \caption{Fitted LPPLS models and norms of persistence landscapes for the segments from Table 5}
\end{figure}

\bibliographystyle{alpha}
\bibliography{TDA_references}

\end{document}